\documentclass{ws-ijmpe}
\usepackage[super,compress]{cite}
\usepackage{multirow}

\providecommand{\rbtrr}{\rule[-0.8ex]{0ex}{3.2ex}}
\begin{document}

\markboth{Authors' Names}{Instructions for typing manuscripts (paper's title)}

\catchline{}{}{}{}{}

\title{Investigation of particle production in $h{\text -}A$ collisions using statistical distributions.}

\author{A.~Kaur}

\address{Post Graduate Govt. College for Girls, Sector-11, Chandigarh-160 011, India.\\
anter.18@gmail.com}

\author{M.~Kaur}
\address{Physics Department, Panjab University, Chandigarh, India\\
manjit@pu.ac.in \footnote{corresponding author}}
\author{R.~Aggarwal}           
\address{Department of Technology, Savitribai Phule Pune University, Pune-411 007, India.\\
ritu.aggarwal1@gmail.com}

\maketitle

\begin{history}
\received{Day Month Year}
\revised{Day Month Year}
\end{history}

\begin{abstract}
Study of the characteristic properties of charged particle production in hadron-nucleus collisions at high energies, by utilising the approaches from different statistical models is performed.~Predictions from different approaches using the Negative Binomial distribution, shifted Gompertz distribution, Weibull distribution and the Krasznovszky-Wagner distribution are utilised for a comparative study of the relative successes of these models.~These distributions derived from a variety of functional forms are based on either phenomenological parameterizations or some model of the underlying dynamics.~Some of these have have also been used to study the data at the LHC for both proton-proton and nucleus-nucleus collisions.~Various physical and derived observables have been used for the analysis.

\end{abstract}


\ccode{PACS numbers: 5.90.+m, 12.40.Fe, 13.66.Bc}

\section{Introduction}
Investigation of charged particle production in collisions of high energy particles has always been of interest.~Gradual increase in small number of such particles produced in fixed target experiments by manifolds in collider experiments has not diminished this interest.~The increase in numbers follows logarithmic rise in the average values with the increasing energy of collision in center of mass system (c.m.s.).~Several theoretical and phenomenological studies have explored the dynamics of particle production.~Numerous statistical models predicting the trends have been proposed, studied and used to successfully describe the particle production in hadron${\text -}$hadron~($pp$ and $\overline{p}p$), lepton${\text -}$lepton~($e^{+}e^{-}$), hadron${\text -}$nucleus~($hA$) and nucleus${\text -}$nucleus~($AA$) collisions over a wide range of c.m.s. energy.~Some of the statistical distributions and their modified forms which have been widely used are Negative Binomial distribution~(NB) [\refcite{NBD}], Gamma distribution [\refcite{Gam1}], Tsallis distribution  [\refcite{TS1,TS2}], Weibull distribution (WB) [\refcite{WB1},~\refcite{WB2},~\refcite{WB3},~\refcite{WB4}] etc.~In one of our previous papers, we introduced shifted Gompertz distribution (SG) for the description of particle production and have shown that it explains very well the multiplicity distributions in different kinds of collisions.~The modified version of SG distribution described the data extremely well as shown in the references [\refcite{SGD1},~\refcite{SGD2}].~Another distribution which remarkably well described the data on hadron${\text -}$hadron and hadron${\text -}$nucleus interactions, from a large number of experiments was Krasznovszky-Wagner distribution~(KW)~[\refcite{KW1},~\refcite{KW2},~\refcite{KW3}].

In an attempt to describe the energy-independence of the multiplicity distribution, it was proposed by Koba et al. [\refcite{Kob},~\refcite{KNO}] that the probability distribution function $P(n)$ expressed as a function of $n$ shows an asymptotic behaviour of the shape at very high energy:
\begin{equation}
P(n)=\frac{1}{\langle{n}\rangle}\psi(z,s)= \lim_{s \to \infty}  \frac{1}{\langle{n}\rangle}\psi(z) \hspace*{4mm}  where  \hspace*{2mm}   z=\frac{n}{\langle{n}\rangle} 
\label{eqKNO}
\end{equation}

This behaviour was known as KNO-scaling.~The KNO scaling has been widely studied in different types of collisions.~Deviation from KNO-scaling was first observed at ISR energies [\refcite{ISR}] in proton${\text -}$proton collisions at $\sqrt{s}$\, =\,30.4 to 62.2~GeV.~With the collider data being available in different pseudorapidity windows and in diferent classes of events such as inelastic, non-single diffractive etc., the scaling behaviour is still being studied in data in small as well as large intervals  at the highest energy colliders, the Large Electron Positron Collider~(LEP), Large Hadron Collider~(LHC) and the Relativistic Heavy Ion Collider~(RHIC) [\refcite{ALICE},~\refcite{LHCb},~\refcite{CMS},~\refcite{OPAL}] etc. ~After the observation of KNO-scaling violation, negative binomial distribution became the most widely used distribution, by every collider experiment around the years since 1985. 

In this paper, the first study of multiplicity distributions in hadron-nucleus interactions from fixed target experiments in terms of four distributions namely, NB, SG, WB and KW is reported.~The data used were collected by different experiments, almost four decades ago from different projectile-target combinations at different energies.~In one of our earlier papers [\refcite{SM}] we analysed these data on hadron-nucleus interactions to study the Tsallis non-extensive entropy model.~We reanalyse the data in the light of new concepts.~In addition to the multiplicity distributions, we study other standard physical observables such as the normalized moments $C_{q}$, the normalized factorial moments $F_{q}$ and $H_{q}$ moments.~The ratio $K_{q}/F_{q}$, known as the $H_{q}$ moments have also been studied.~An outline of the models used is described in the following sections:

\section{Probability functions and the particle multiplicity}

The probability of producing $n{\text -}$charged particles in an interaction can be understood in terms of a probability distribution function (PDF), expressed as $P(n|\tilde{\mu})$, where $\tilde{\mu}$ represents a set of parameters which influence the shape and scale of the PDF.~The number of particles produced are distributed according to a PDF which depending upon $\tilde{\mu}$ produces different types of distributions.~We discuss four such PDFs in the subsections below.

\subsection{The Negative Binomial~(NB) distribution}
Van Hove and Giovannini showed that the equation~(\ref{eqNB}) describing a negative binomial distrtibution of probability, could arise from a partially stimulated emission or from a cascade mechanism [\refcite{Hove}];
 
\small
\begin{equation}
P_{n}\big(n|\langle n \rangle,k\big)=\frac{\Gamma(n+k)}{\Gamma(n+1)\Gamma(k)}\frac{(\langle n \rangle/k)^n}{(1+\langle n \rangle/k)^{n+k}} \hspace{0.8cm} 
\label{eqNB}
\end{equation} 
\\
\normalsize
Later on the results from the NA22 collaboration [\refcite{NA22}] ruled out the stimulated-emission explanation.~While the cascade mechanism with its large clans decaying as per the logarithmic distribution [\refcite{clan}] continued to attract attention.~The negative binomial distribution has been studied for almost every experiment, at all the colliders such as LEP~(Electron Positron collider) and LHC at CERN, RHIC at BNL, Tevatron at Fermilab etc.~The experimental results are understood in terms of distributions of negative binomial form both for total multiplicities and multiplicities in restricted pseudorapidity windows.~A plausible explanation of the experimental findings leads to a concept of cluster formation and an insight into the properties of clusters.~The negative binomial distribution is characterised by two free parameters, which determine the mean $\langle{n}\rangle$ and shape $k$ of the distribution.
\subsection{Shifted Gompertz~(SG) distribution}
We introduced the shifted Gompertz distribution to study its applicability for describing the multiplicity data from $pp$, $\overline{p}p$ and $e^{+}e^{-}$ collisions obtained by various experiments at different energies [\refcite{SGD1},~\refcite{SGD2},~\refcite{SGD3}].~The detailed studies have shown that SG distribution explains the data trends very well and the precision is better than NB in several cases.~It is defined in terms of the probability density function~(PDF) of two non-negative parameters, $b$ the scale parameter and $\eta$ the shape parameter.~Following equations define the PDF and the mean of the distribution:

\small
\begin{equation}
P_{n}\big(n|b,\eta\big) = b e^{-bn}e^{-\big(\eta e^{-bn}\big)}\big[1+\eta(1-e^{-bn}\big)\big]\hspace{0.3cm} for\hspace{0.15cm} n > 0 \, 
\label{eqSG}
\end{equation}
Mean of the distribution:
\begin{equation}
 \Big(-\frac{1}{b}\Big)\Big(E\big[ln(\zeta)\big]-ln(\eta)\Big) \hspace*{0.3 cm} where \hspace*{0.2 cm} \zeta = \eta e^{-bn}
\label{eqSGm}
\end{equation}
and
\small
\begin{equation}
\begin{split}
 E\big[ln(\zeta)\big] = \Big[1 + \frac{1}{\eta}\Big]\int_{0}^{\infty}e^{-\zeta}\big[ln(\zeta)\big]d\zeta \\ 
 - \frac{1}{\eta}\int_{0}^{\infty}\zeta e^{-\zeta}\big[ln(\zeta)\big]d\zeta\
\end{split}
\label{eqSGm2}
\end{equation}
\normalsize
Though the SG distribution has been studied recently, with very good results in hadronic and leptonic interactions, this is our first attempt to analyse the $p/\pi^{-}{\text -}$nucleus data from the fixed target experiments, in terms of shifted Gompertz distribution.
\subsection{The two parameter Weibull~(WB) distribution}
The Weibull distribution was originally proposed as a model for studying the material breaking strength.~It is commonly used in many fields such as engineering sciences, biology etc. to assess product reliability, analyze life data and model failure times.

Two different probability distribution functions (PDF) of Weibull exist; one having three free parameters and another having two free parameters.~Weibull distribution can also be fitted to non-symmetrical data.~The two parameter Weibull has been used during the last few years to describe the collision data from high energy experiments [\refcite{WB2},~\refcite{WB3},~\refcite{WB4}].\\

The PDF of a Weibull random variable is:
\small
\begin{equation}
 P_n\big(n|\lambda,K\big) =
 \begin{cases}
  \frac{K}{\lambda} \Big(\frac{n}{\lambda}\Big)^{(K-1)} exp^{-(\frac{n}{\lambda})^{K}} & \text{\it{n} $\geq$ 0}    \\
  0 & \text{\it{n} $<$ 0 }
 \end{cases}
\label{eqWB}
\end{equation}
\normalsize
 The standard Weibull has a scale parameter $\lambda >0$ and a shape parameter $K>0$ for its two free parameters.~The mean of the distribution function is given by: 
\begin{equation}
\bar{n} = \lambda \Gamma\bigg(\frac{K+1}{K}\bigg)
\label{eqWBm}
\end{equation}

\subsection{The Krasznovszky-Wagner~(KW) distribution}
This distribution based on the generalized geometrical-optical model in the impact parameter representation [\refcite{KW1},~\refcite{KW2}] has three free parameters; a mean $\langle{n}\rangle$ and two others, $m$ and $A$:

The PDF of the distribution is:

\small 
\begin{equation}
P_{n}\big(n|\langle{n}\rangle,m,A\big) = \frac{mF(A)^{A}z^{mA-1}e ^{-F(A)z^{m}}}{\langle{n}\rangle\Gamma(A)} \hspace{.0cm} 
\label{eqKW}
\end{equation} 
where
$z = n/\langle{n}\rangle$
and 
\small
\begin{equation}
F(A) = \frac{\Gamma^{m}(A+1/m)}{\Gamma^{m}(A)}
\label{eqKW2}
\end{equation}
\normalsize
 The distribution have been used for studying particle production in $e^{+}e^{-}$, $pp$ and $\pi{\text -}p$ interactions mainly in the context of KNO scaling [\refcite{KW1},~\refcite{KW2},~\refcite{KW3}]. 
\section{Moments of multiplicity distributions}
The shape of the charged particle multiplicity distribution depends on the mechanism of particle production and hence serves as a fundamental tool to study the production dynamics.~It is well established that in the case of independent emission of single particles, the shape is Poissonian.~Any deviation from this shape points to the presence of correlations [\refcite{WO}].~To study such deviations and the particle correlations, normalised moments $C_{q}$, normalised factorial moments $F_{q}$, normalised factorial cumulants $K_{q}$ and $H_{q}$ moments are used, which are defined as: 
\small
\begin{gather}
C_{q} = \frac{\sum_{n=1}^{\infty}n^{q}P(n)}{(\sum_{n=1}^{\infty}nP(n))^q} \label{eqMom}\\
F_{q} = \frac{\sum_{n=q}^{\infty}n(n-1).......(n-q+1)P(n)}{(\sum_{n=1}^{\infty}nP(n))^q} \label{eqFacMom}\\
K_{q} = F_{q}-\sum_{m=1}^{q-1}\frac{(q-1)!}{m!(q-m-1)!}K_{q-m}F_{m}\\
H_{q} = K_{q}/F_{q}\label{eqHq}\
\end{gather}

\normalsize
The multiplicity distribution shape is rather non-trivial in terms of its dependence on the interaction energy.~A scaling behaviour predicted by Koba${\text -}$Nielsen-Olesen (KNO)~[\refcite{Kob}] showed that probability distributions at all energies fall on to one curve when plotted as a function of a variable $z=n/\langle{n}\rangle$.~Study of moments of distribution reveals that if the scaling hypothesis holds, the moments $C_q$, equation (\ref{eqMom}) are independent of energy.~And if the moments depend upon energy, the KNO scaling breaks down.~Also the moments $F_q$ scale if the KNO scaling holds.~In addition equation (\ref{eqFacMom}) shows correlations in the production of up to $q$ particles.~For the particle distribution following Poissonian shape, all $F_q$ are equal to unity.~The distribution becomes broader if the particles are correlated, and narrower if they are anti-correlated.~In case of a positive(negative) correlation the $F_q$ are more(less) than unity.~A special oscillation pattern for the ratio of cumulants to factorial moments $H_{q}=K_{q}/F_{q}$ was predicted with first minimum occuring around $q_{min}\backsim{5}$ as determined from the quantum chromodynamical considerations [\refcite{DR},\refcite{Cap}].

\section{Results and Discussion}
The following sections describe the data analysed.~The above mentioned distributions have been used to study the interaction dynamics by fitting the data from different varieties of collider data such as $e^+e^-$ at LEP, $pp$ and heavy ion data at the Large Hadron Collider (LHC).~The study of nuclear interactions has been recently done by using different types of heavy ions such as $Au$, $Pb$ etc. at the LHC by the ALICE collaboration.
\subsection{The data used}
In the present work we extend the analysis to check the validity of the mentioned distributions in hadron-nucleus interactions in fixed target experiments using nuclear emulsion as a target.~Nuclear emulsion acts as a photographic film, capable of recording three-dimensional tracks of charged particles with submicron spatial resolution, which can be studied under a microscope.~The emulsion consists of silver bromide ($AgBr$) crystals dispersed in a gelatin layer which contains carbon, nitrogen and oxygen atoms.~The projectile can undergo interaction with any one of these nuclei.~On the average the interaction of the projectile is estimated as an interaction of the projectile with a nucleus of atomic weight 72.~The latest usage of emulsion technology has been in the OPERA experiment [\refcite{OP}], which aims at a direct observation of tau neutrino appearance in a pure muon neutrino beam.~The data used for the present study, are though old, and at relatively lower center of mass energies (roughly between 100 to 350 GeV), carry valuable information suitable for testing new models.~In addition, for comparison, we also use the conventional NB distribution which has never been tested with these data although it has been the most frequently tested for almost all types of interactions at the modern day colliders.
We present analysis of the following inelastic interactions from the fixed target experiments using $p$ or $\pi^{\pm}$ as the projectiles.\\
(i) A set of $p{\text -}Em$ interactions at $P_{Lab}$ = 27, 67, 200, 300, 400 and 800 GeV [\refcite{Barb},~\refcite{Babecki},~\refcite{Hebert},~\refcite{Boos},~\refcite{Abdu}].\\
(ii) $p{\text -}Au$ interactions at $P_{Lab}$ = 200 GeV [\refcite{Brick}].\\
(iii) A set of $\pi^{-}{\text -}Em$ interactions at $P_{Lab}$ = 50,~200,~340 and 525 GeV [\refcite{Kumar},~\refcite{Anzon},~\refcite{Nadi},~\refcite{Cherry}].
(iv) $\pi^{\pm}{\text -}Ne$ interactions at $P_{Lab}$ = 30 and 64 GeV [\refcite{Rees}].

The data for 50 GeV and 340 GeV $\pi^{-}{\text -}Em$ interactions are our own data, collected using nuclear emulsion stacks [\refcite{Kumar},~\refcite{Nadi}].~In one case, the emulsion stack was irradiated to a single pulse of negative pions of momentum $p$\,=\,50~GeV ($\Delta{p}/p \sim 1\%$) under the effect of a strong pulsed magnetic field of intensity 180 KOe and is of a unique kind.~The magnetic field was used for identifying the charge of the particles from the curvatures of the tracks produced and calculating the momenta.~The stack also carried a 9~GeV proton beam in the perpendicular direction.~This enabled us to make corrections towards thickness distortions in emulsion while determining the momenta of secondary charged particles.

\section{Comparison of PDFs of different distributions of multiplicity}
\begin{figure}[th]
\centerline{\includegraphics[scale=0.37]{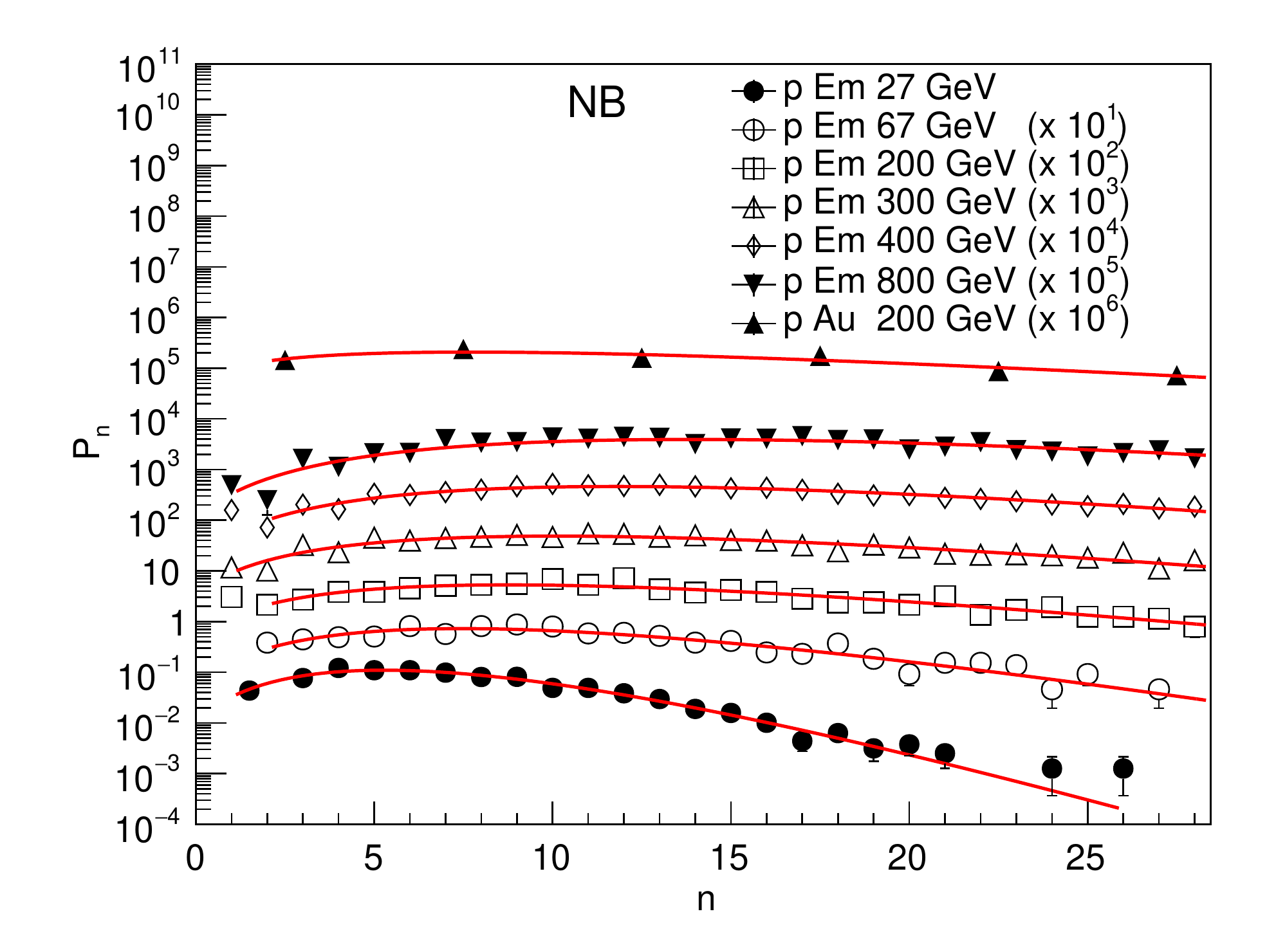}\includegraphics[scale=0.37]{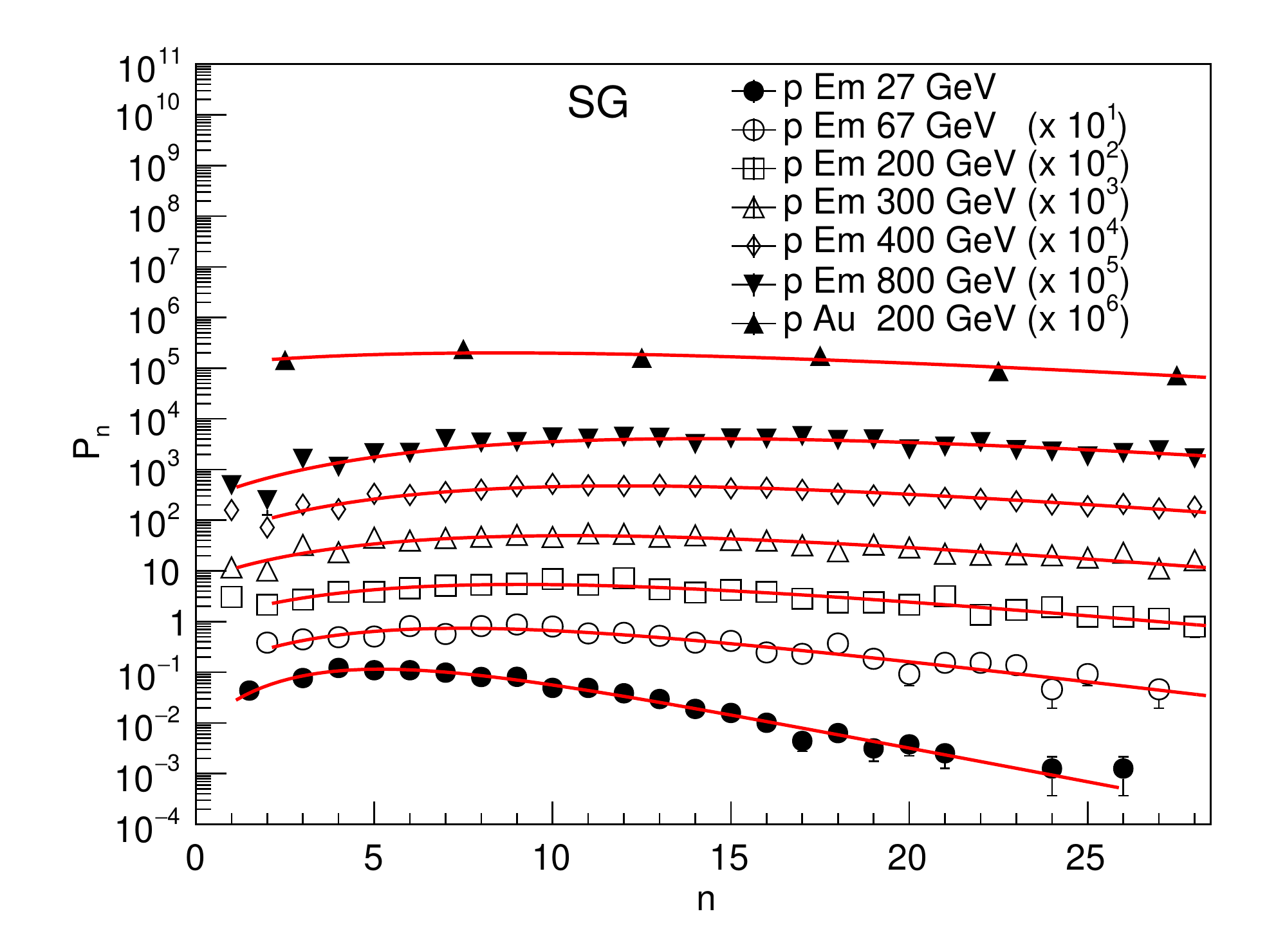}}
\centerline{\includegraphics[scale=0.37]{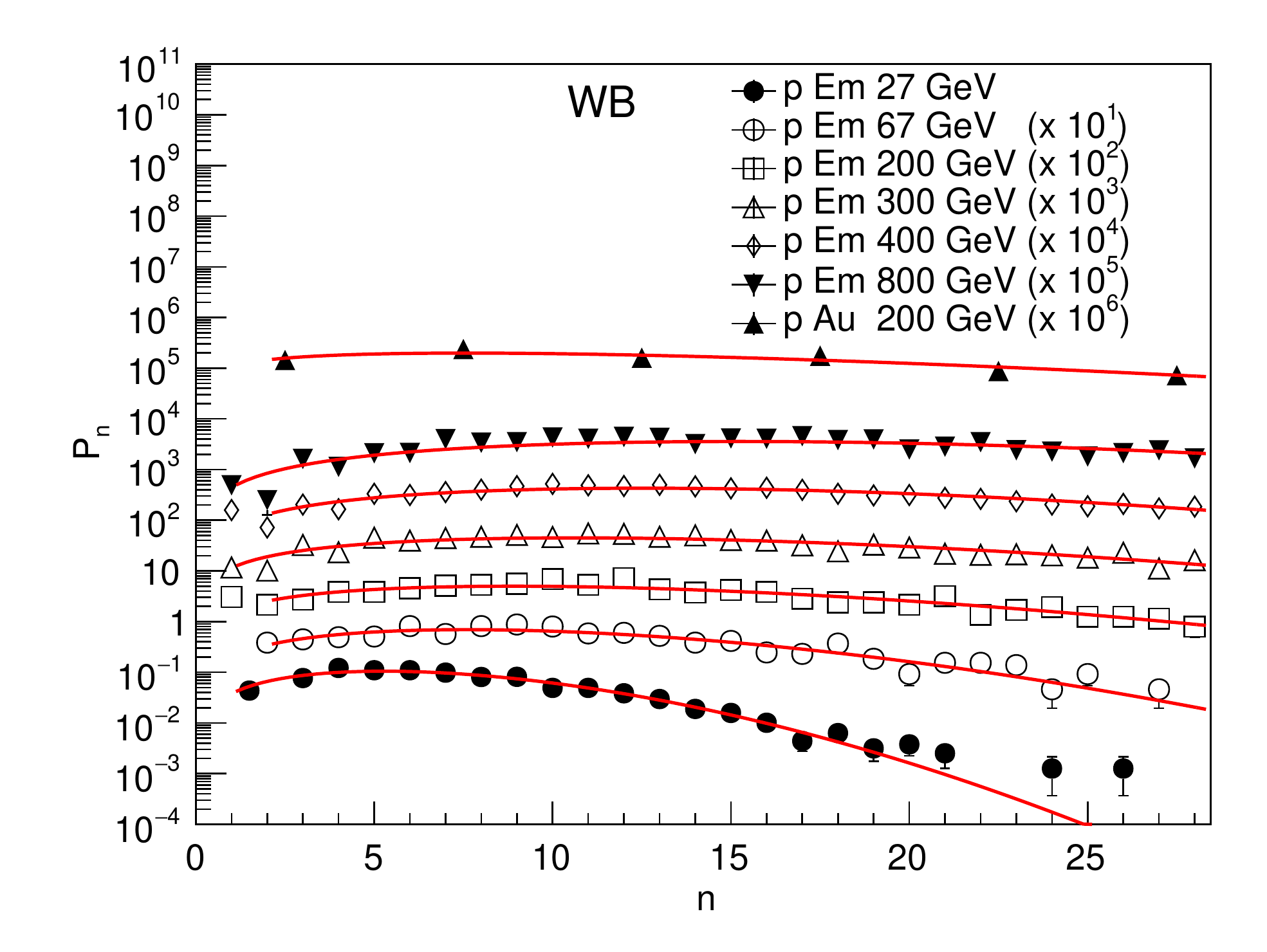}\includegraphics[scale=0.37]{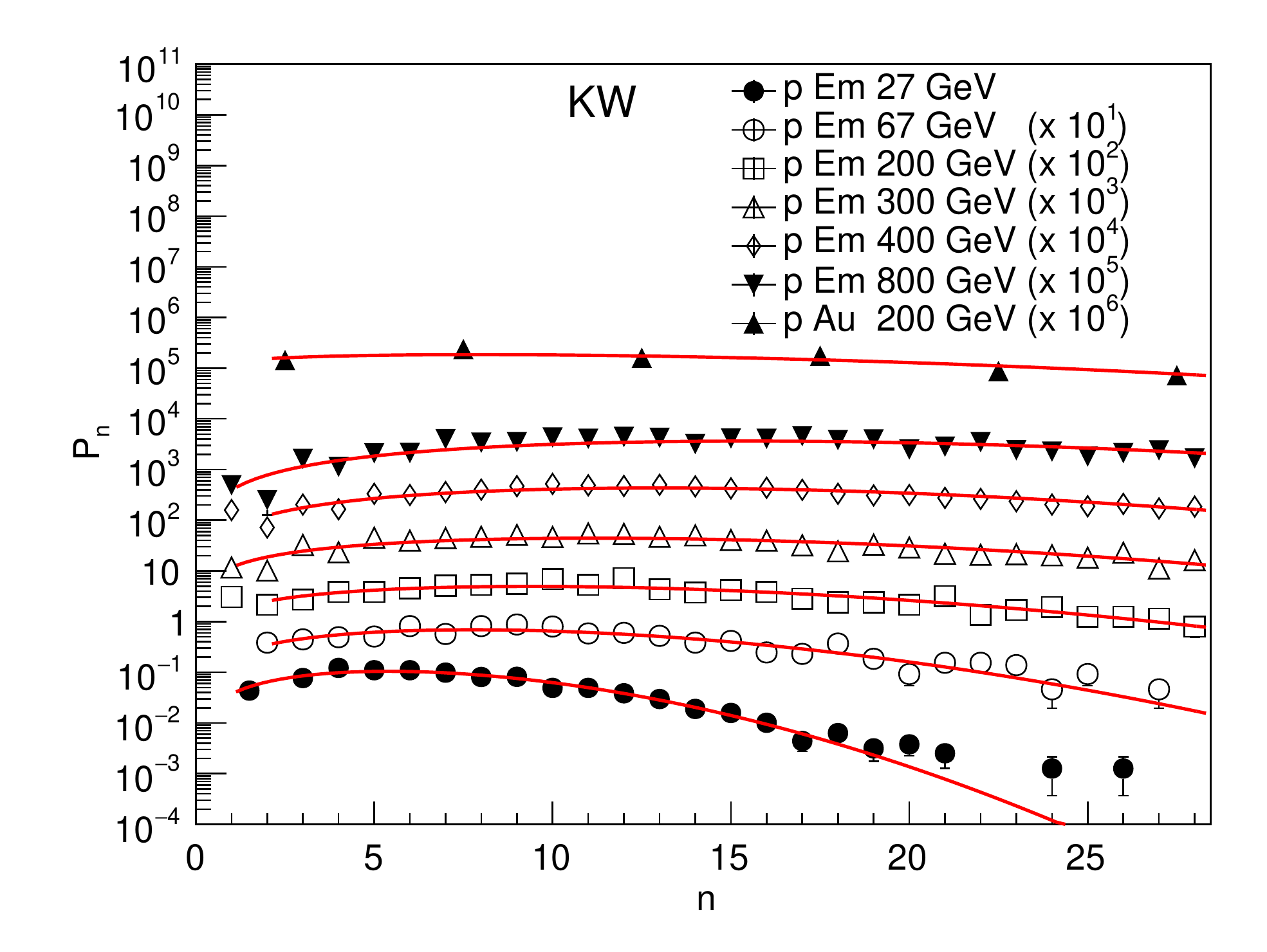}}
\caption{\small{Comparison of data~(points) on charged particle multiplicity in inelastic $p{\text -}Em/Au$ collisions in fixed target experiments, with different projectile energies, fitted with  NB, SG, WB and KW distributions~(solid lines).}}
\label{figpEM}
\end{figure}

The experimental data of hadron-nucleus inelastic collisions mentioned in the previous section, are studied in terms of various distributions: 
 
The PDFs from the Negative Binomial, shifted Gompertz, two parameter Weibull and Krasznovszky-Wagner distributions, are calculated by using equations (\ref{eqNB}), (\ref{eqSG}), (\ref{eqWB}) and (\ref{eqKW}).~Minimum $\chi^{2}$ fits to the $p{\text -}Em/Au$ data are shown in figure~\ref{figpEM}.~Table~1 gives the parameters of the fits for all the distributions and a comparison of corresponding $\chi^{2}/n_{dof}$ and $p$-values is given in table~2.~The minimum $\chi^{2}$ fits were done by using CERN data analysis framework ROOT6.19.~Most of the experimental data points have statistical uncertainties.~Wherever available, the systematic and statistical uncertainties are combined.

For $p{\text -}Em/Au$ interactions, one finds that NB and SG distributions reproduce the data very well at all the energies with SG giving the best results with lowest $\chi^{2}/n_{dof}$ and $p$-values corresponding to $CL>0.1\%$ in each case.~WB and KW distributions reproduce the data well at low energies, but fail completely at the highest energies, 400~GeV and 800~GeV with $p$-values corresponding to $CL<0.01\%$.

\begin{figure}[th]
\centerline{\includegraphics[scale = 0.37]{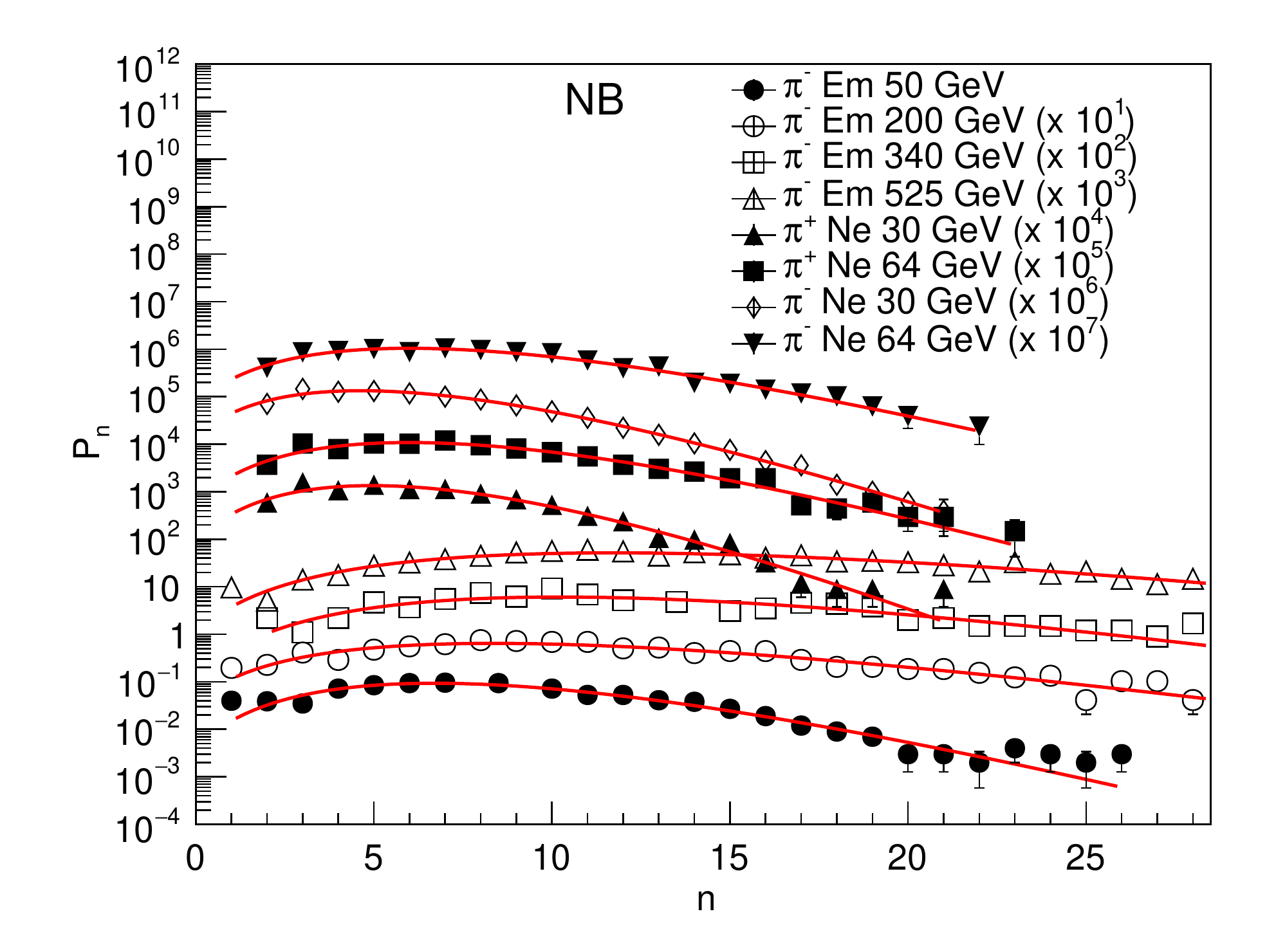}\includegraphics[scale= 0.37]{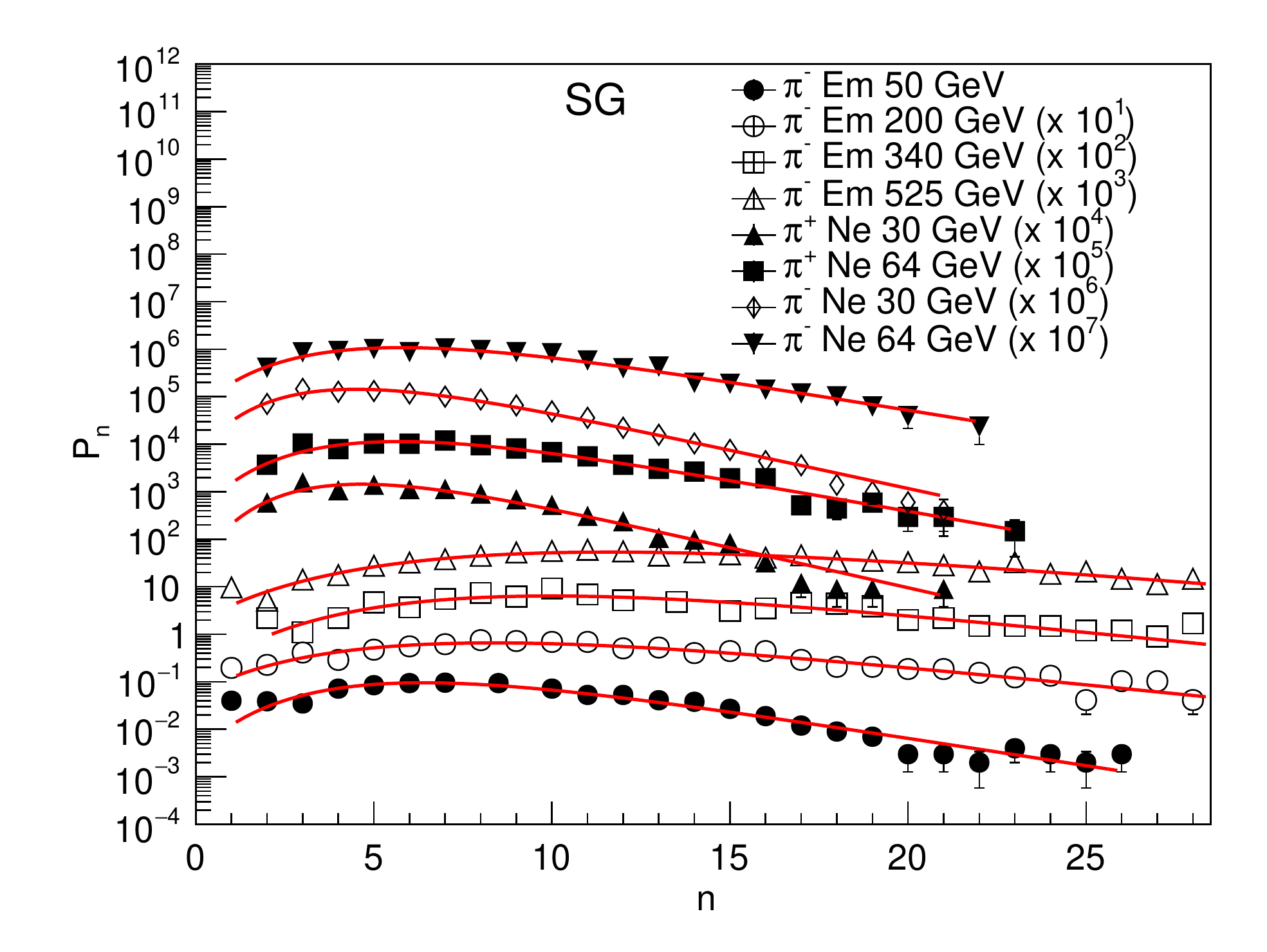}}
\centerline{\includegraphics[scale = 0.37]{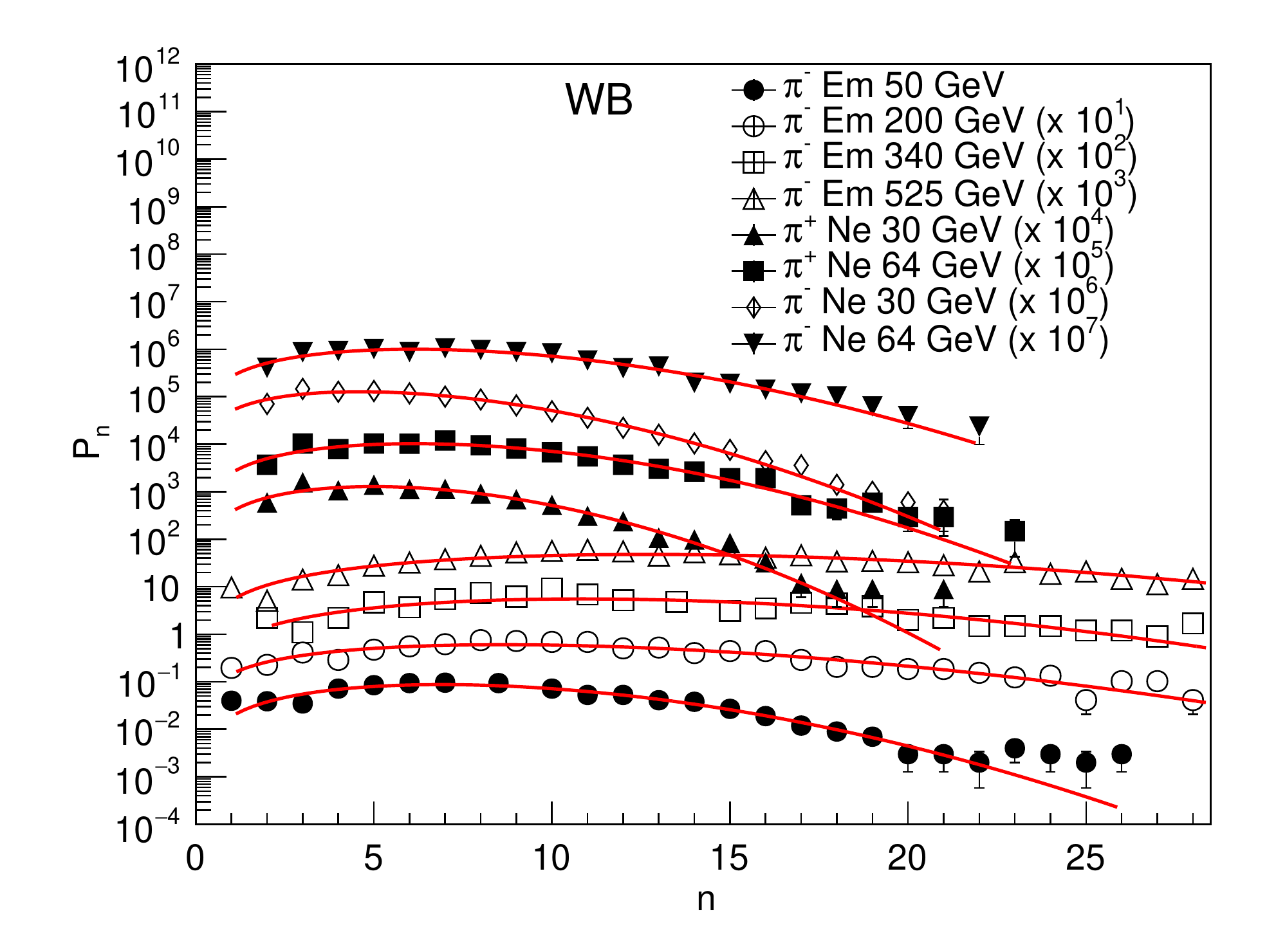}\includegraphics[scale = 0.37]{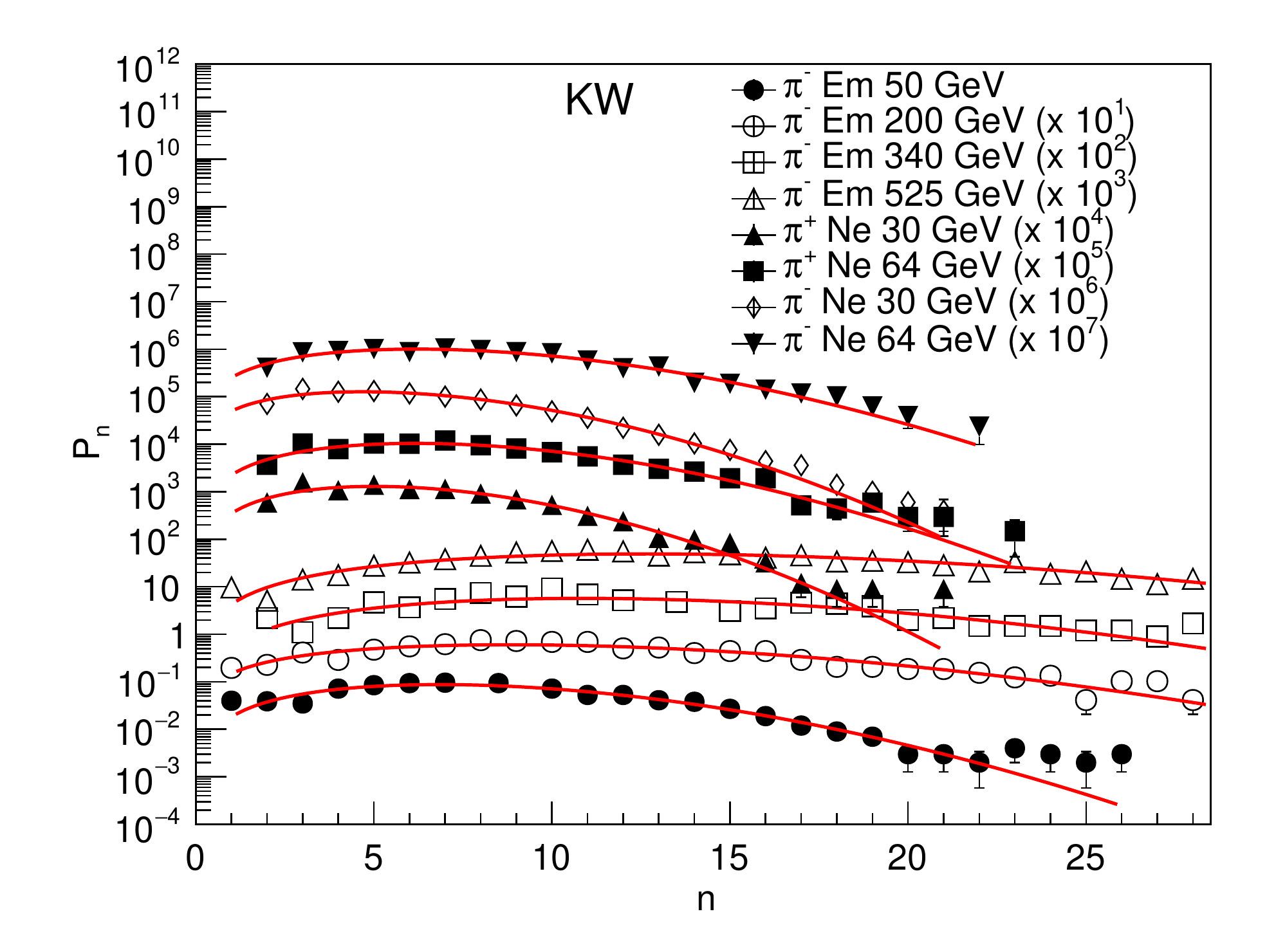}}
\caption{\small{Data on charged particle multiplicity distributions in inelastic $\pi^{-}{\text -}Em$ and $\pi^{\pm}{\text -}Ne$ collisions at different energies with fits from NB, SG, WB and KW distributions.}}
\label{figpiEM}
\end{figure}

Figure~\ref{figpiEM} shows the data and fits to NB, SG, WB, KW distributions for $\pi^{-}{\text -}Em$ and $\pi^{\pm}{\text -}Ne$ interactions at various energies.~Parameters of the fit and $\chi^{2}/n_{dof}$ and $p$-values are given in the tables~1-2.~It is observed that the shape parameter increases slowly with $P_{Lab}$ in all cases.~All distributions reproduce the data at all energies well with the exception of the data for $\pi^{-}{\text -}Em$ interactions at 340~GeV and $\pi^{\pm}{\text -}Ne$ at 30~GeV which are statistically excluded with $p$-values corresponding to $CL<0.01\%$.~Both NB and SG distributions fit the data at most of the energies.

\begin{figure}[th]
\centerline{\includegraphics[scale = 0.38]{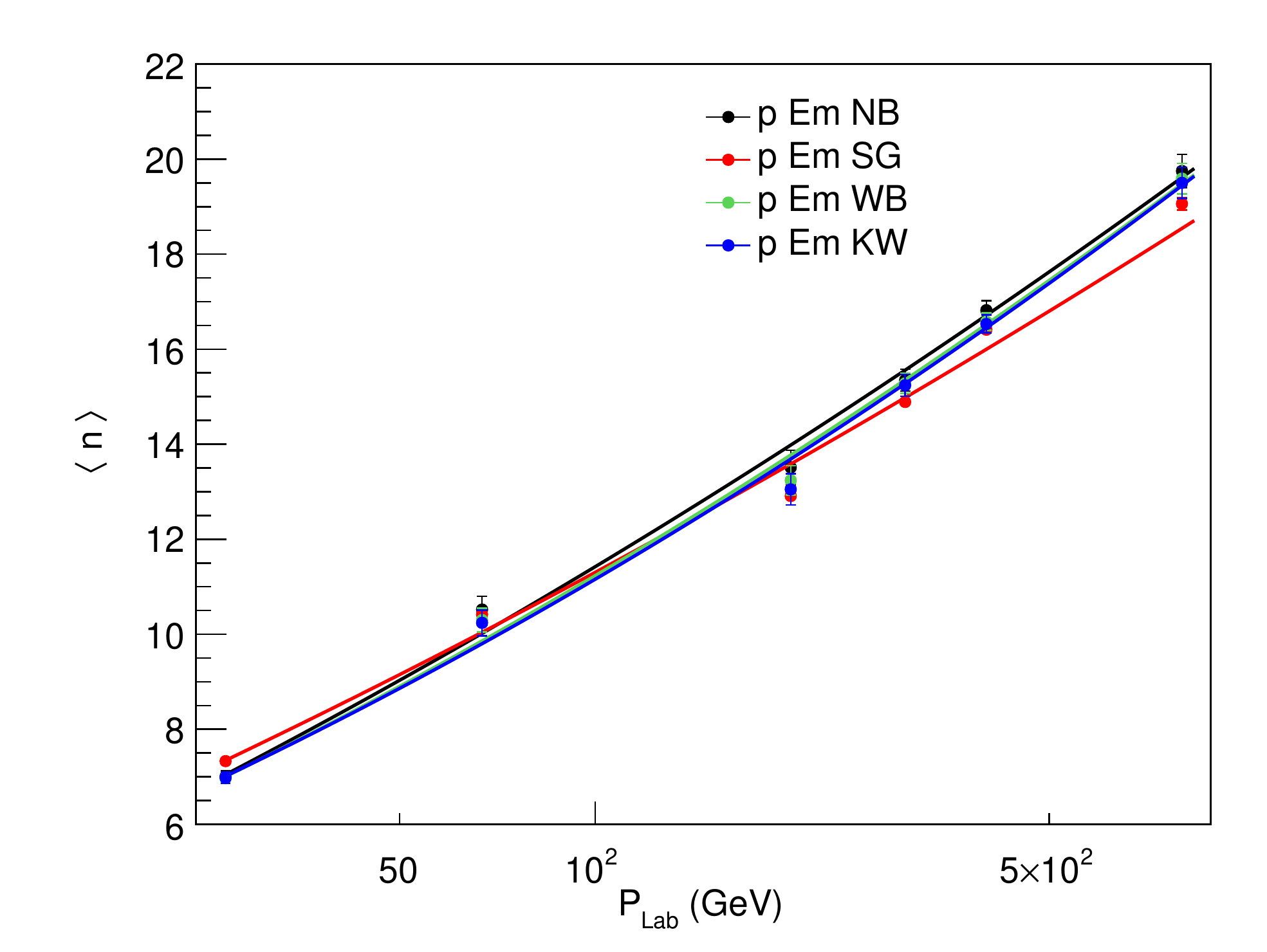}\includegraphics[scale = 0.38]{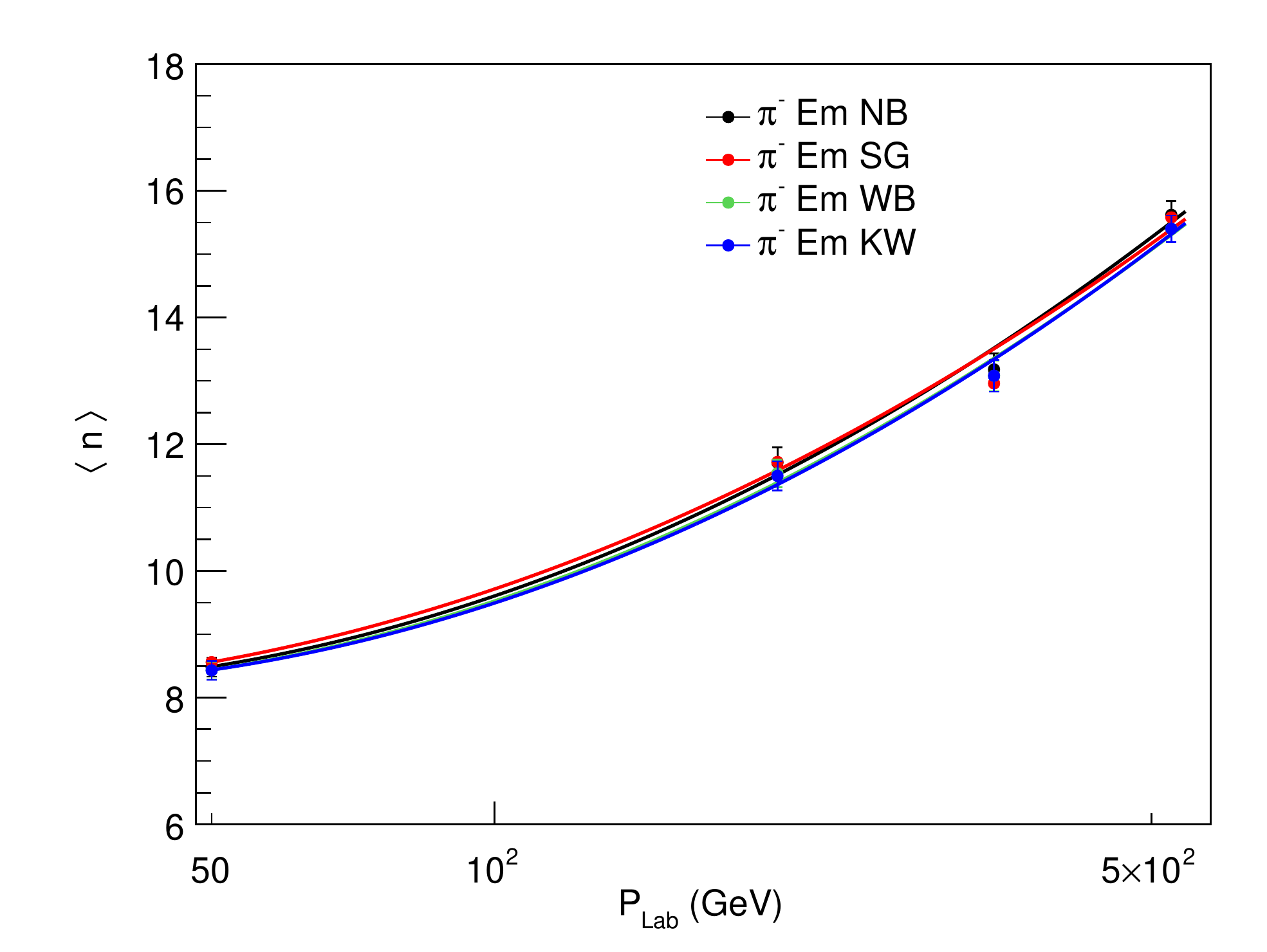}}
\caption{\small{Comparison of experimental average multiplicity (data points) and the values from best fit (solid lines) of NB, SG, WB and KW distributions in inelastic $p{\text -}Em$ (top) and $\pi^{-}{\text -}Em$ (bottom) interactions at different energies.}}
\label{figMult}
\end{figure}
\begin{figure}[th]
\centerline{\includegraphics[scale = 0.60]{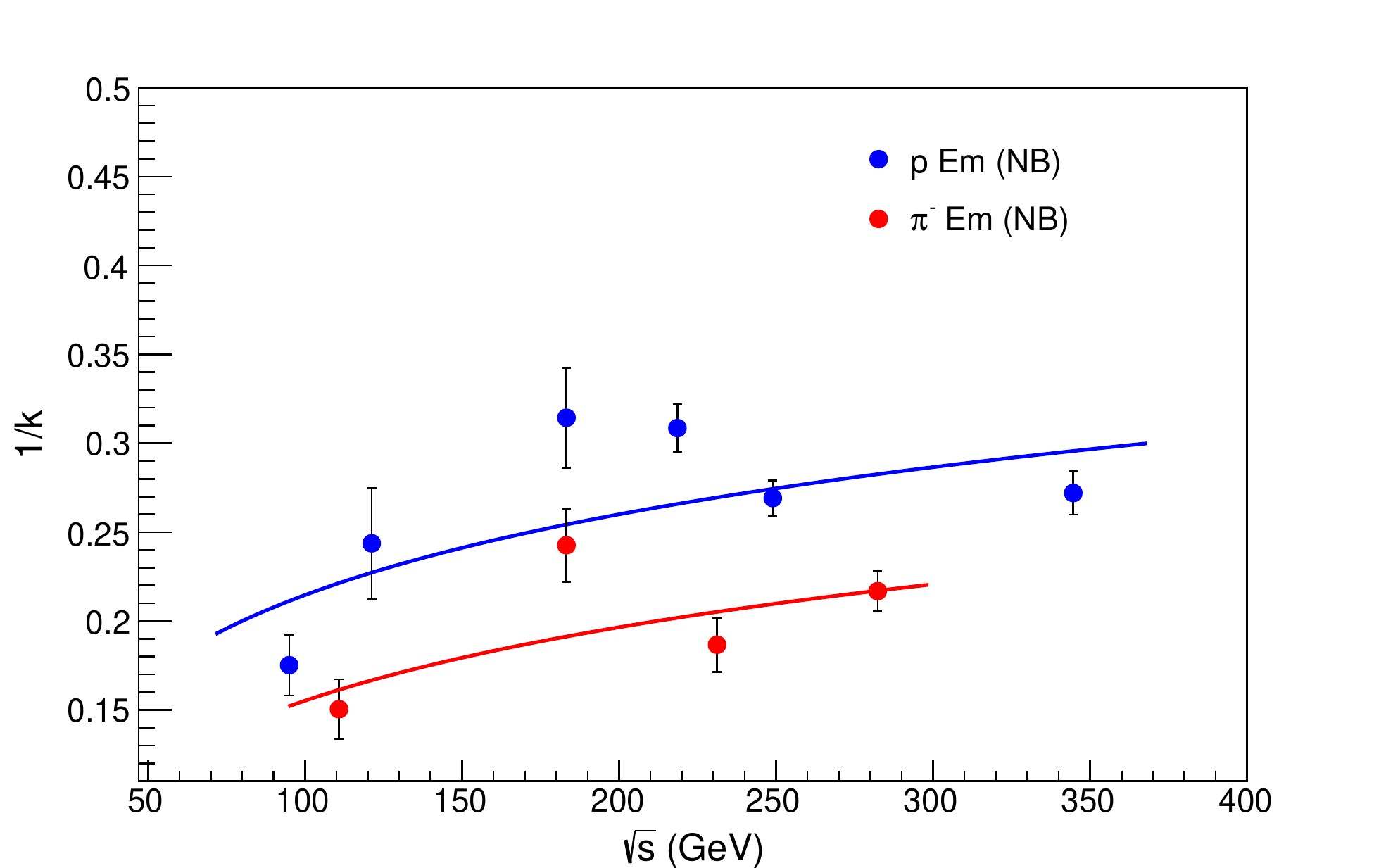}}
\caption{\small{ Shape parameter $k$ dependence on $\sqrt{s}$ for NB distribution in inelastic $p{\text -Em}$ and $\pi^{-}{\text -Em}$ collisions at different energies.}}
\label{figp-piEMshape}
\end{figure}

\begin{table}[tb]
\scalebox{0.8}{ 
\tbl{Scale and shape parameters for NB, SG, WB and KW distributions.}
{\begin{tabular}{|c|c|c||c|c||c|c||c|c|}
\hline
{\bf Energy (GeV)} & \multicolumn{2}{c||}{\bf NB} & \multicolumn{2}{c||}{\bf SG} & \multicolumn{2}{c||}{\bf WB} & \multicolumn{2}{c|}{\bf KW} \rbtrr \\ \hline
 & Scale & Shape & Scale & Shape & Scale & Shape & Scale & Shape \rbtrr \\\hline
{\bf p-Em} & & & & & & & & \rbtrr \\\hline 
27 & 7.009 $\pm$ 0.118 & 5.708 $\pm$ 0.556 &	0.308 $\pm$ 0.009 & 3.964 $\pm$ 0.349 &	7.858 $\pm$ 0.129 & 1.886 $\pm$ 0.054 &	6.977 $\pm$ 0.117 & 0.929 $\pm$ 0.047 \rbtrr \\ \hline		
67 & 10.521 $\pm$ 0.284 & 4.103 $\pm$ 0.526 &	0.193 $\pm$ 0.010 & 3.263 $\pm$ 0.456 &	11.595 $\pm$ 0.300 & 1.844 $\pm$ 0.087 &	10.238 $\pm$ 0.266 & 0.888 $\pm$ 0.070 \rbtrr \\ \hline
200 & 13.507 $\pm$ 0.357 & 3.181 $\pm$ 0.284 &	0.145 $\pm$ 0.007 & 2.812 $\pm$ 0.313 &	14.841 $\pm$ 0.380 & 1.712 $\pm$ 0.067 &	13.048 $\pm$ 0.330 & 0.816 $\pm$ 0.051 \rbtrr \\ \hline
300 & 15.354 $\pm$ 0.232 & 3.241 $\pm$ 0.139 &	0.131 $\pm$ 0.003 & 3.044 $\pm$ 0.160 &	17.159 $\pm$ 0.254 & 1.757 $\pm$ 0.033 &	15.235 $\pm$ 0.230 & 0.845 $\pm$ 0.024 \rbtrr \\ \hline
400 & 16.824 $\pm$ 0.199 & 3.715 $\pm$ 0.137 &	0.127 $\pm$ 0.003 & 3.631 $\pm$ 0.158 &	18.682 $\pm$ 0.211 & 1.896 $\pm$ 0.030 &	16.530 $\pm$ 0.187 & 0.952 $\pm$ 0.024 \rbtrr \\ \hline
800 & 19.746 $\pm$ 0.350 & 3.676 $\pm$ 0.164 &	0.111 $\pm$ 0.003 & 3.892 $\pm$ 0.195 &	22.087 $\pm$ 0.371 & 1.945 $\pm$ 0.037 &	19.500 $\pm$ 0.333 & 0.995 $\pm$ 0.029 \rbtrr \\ \hline
{\bf p-AU} & & & & & & & &  \rbtrr \\\hline 
200 & 16.466 $\pm$ 1.183 & 1.994 $\pm$ 0.315 &	0.097 $\pm$ 0.012 & 1.478 $\pm$ 0.408 &	17.893 $\pm$ 1.177 & 1.419 $\pm$ 0.110 &	15.859 $\pm$ 0.997 & 0.602 $\pm$ 0.073 \rbtrr \\ \hline
{\bf $\pi^{-}$-Em} & & & & & & & &  \rbtrr \\\hline 
50 & 8.481 $\pm$ 0.147 & 6.647 $\pm$ 0.740 &	0.268 $\pm$ 0.009 & 4.573 $\pm$ 0.464 &	9.516 $\pm$ 0.162 & 2.061 $\pm$ 0.067 &	8.435 $\pm$ 0.145 & 1.046 $\pm$ 0.059 \rbtrr \\ \hline
200 & 11.705 $\pm$ 0.239 & 4.121 $\pm$ 0.350 &	0.178 $\pm$ 0.007 & 3.492 $\pm$ 0.318 &	12.997 $\pm$ 0.259 & 1.869 $\pm$ 0.060 &	11.502 $\pm$ 0.231 & 0.915 $\pm$ 0.047 \rbtrr \\ \hline
340 & 13.178 $\pm$ 0.254 & 5.357 $\pm$ 0.436 &	0.179 $\pm$ 0.007 & 4.972 $\pm$ 0.422 &	14.764 $\pm$ 0.276 & 2.103 $\pm$ 0.063 &	13.084 $\pm$ 0.245 & 1.123 $\pm$ 0.056 \rbtrr \\ \hline
525 & 15.620 $\pm$ 0.221 & 4.612 $\pm$ 0.237 &	0.146 $\pm$ 0.004 & 4.498 $\pm$ 0.244 &	17.386 $\pm$ 0.242 & 2.080 $\pm$ 0.044 &	15.397 $\pm$ 0.215 & 1.097 $\pm$ 0.036 \rbtrr \\ \hline
{\bf $\pi^{+}$-Ne} & & & & & & & &  \rbtrr \\\hline 
30 & 6.083 $\pm$ 0.062 & 9.625 $\pm$ 0.796 &	0.391 $\pm$ 0.006 & 5.202 $\pm$ 0.258 &	6.874 $\pm$ 0.068 & 2.066 $\pm$ 0.038 &	6.114 $\pm$ 0.061 & 1.080 $\pm$ 0.035 \rbtrr \\ \hline
64 & 7.546 $\pm$ 0.117 & 7.183 $\pm$ 0.728 &	0.305 $\pm$ 0.008 & 4.755 $\pm$ 0.371 &	8.474 $\pm$ 0.129 & 2.057 $\pm$ 0.057 &	7.531 $\pm$ 0.116 & 1.076 $\pm$ 0.051 \rbtrr \\ \hline
{\bf $\pi^{-}$-Ne} & & & & & & & &  \rbtrr \\\hline 
30 & 5.998 $\pm$ 0.054 & 6.956 $\pm$ 0.404 &	0.373 $\pm$ 0.005 & 4.387 $\pm$ 0.188 &	6.777 $\pm$ 0.060 & 1.905 $\pm$ 0.029 &	6.016 $\pm$ 0.054 & 0.941 $\pm$ 0.027 \rbtrr \\ \hline
64 & 7.772 $\pm$ 0.131 & 6.091 $\pm$ 0.619 &	0.282 $\pm$ 0.009 & 4.136 $\pm$ 0.348 &	8.692 $\pm$ 0.141 & 1.994 $\pm$ 0.061 &	7.715 $\pm$ 0.127 & 1.016 $\pm$ 0.052 \rbtrr \\ \hline
\end{tabular}}
}
\label{tabPar}
\end{table}

\begin{table}[tb]
\scalebox{0.8}{ 
\tbl{Comparison of $\chi^{2}/n_{dof}$ and $p$-values for NB, SG, WB and KW distributions.}
{\begin{tabular}{|c|c|c||c|c||c|c||c|c|}
\hline
{\bf Energy (GeV)} & \multicolumn{2}{c||}{\bf NB} & \multicolumn{2}{c||}{\bf SG} & \multicolumn{2}{c||}{\bf WB} & \multicolumn{2}{c|}{\bf KW} \rbtrr \\ \hline
 & $\chi^2/n_\mathrm{dof}$ & $p$-value & $\chi^2/n_\mathrm{dof}$ & $p$-value & $\chi^2/n_\mathrm{dof}$ & $p$-value & $\chi^2/n_\mathrm{dof}$ & $p$-value \rbtrr \\\hline
{\bf p-Em} & & & & & & & & \rbtrr \\\hline 
27 & 1.00 (19.01/19) & 0.4561	& 0.95 (18.10/19) & 0.5156	& 1.38 (26.24/19) & 0.1238	& 1.49 (28.27/19) & 0.0785 \rbtrr \\ \hline
67 & 1.12 (25.67/23) & 0.3169	& 1.09 (24.96/23) & 0.3521	& 1.19 (27.37/23) & 0.2405	& 1.22 (28.08/23) & 0.2128 \rbtrr \\ \hline
200 & 0.90 (28.80/32) & 0.6295	& 0.83 (26.50/32) & 0.7412	& 1.18 (37.83/32) & 0.2203	& 1.34 (42.94/32) & 0.0938 \rbtrr \\ \hline
300 & 1.23 (49.30/40) & 0.1488	& 1.26 (50.51/40) & 0.1232	& 1.78 (71.30/40) & 0.0017	& 2.11 (84.53/40) & 0.0001 \rbtrr \\ \hline
400 & 1.28 (52.41/41) & 0.1091	& 1.25 (51.30/41) & 0.1300	& 2.38 (97.61/41) & $<$ 0.0001	& 2.56 (105.16/41) & $<$ 0.0001 \rbtrr \\ \hline
800 & 1.60 (75.02/47) & 0.0058	& 1.70 (80.06/47) & 0.0019	& 2.37 (111.35/47) & $<$ 0.0001	& 2.41 (113.35/47) & $<$ 0.0001 \rbtrr \\ \hline
{\bf p-AU} & & & & & & & &  \rbtrr \\\hline 
200 & 0.67 (4.67/7) & 0.6998	& 0.73 (5.14/7) & 0.6432	& 0.77 (5.36/7) & 0.6159	& 1.03 (7.18/7) & 0.4109 \rbtrr \\ \hline
{\bf $\pi^{-}$-Em} & & & & & & & &  \rbtrr \\\hline 
50 & 1.59 (34.96/22) & 0.0391	& 1.99 (43.73/22) & 0.0038	& 1.65 (36.29/22) & 0.0283	& 1.66 (36.51/22) & 0.0268 \rbtrr \\ \hline
200 & 1.06 (32.82/31) & 0.3778	& 0.97 (30.02/31) & 0.5163	& 1.25 (38.85/31) & 0.1572	& 1.30 (40.27/31) & 0.1231 \rbtrr \\ \hline
340 & 2.55 (71.48/28) & $<$ 0.0001	& 2.42 (67.65/28) & $<$ 0.0001	& 3.07 (86.03/28) & $<$ 0.0001	& 2.98 (83.37/28) & $<$ 0.0001 \rbtrr \\ \hline
525 & 1.18 (54.10/46) & 0.1927	& 0.99 (45.75/46) & 0.4825	& 1.93 (88.88/46) & 0.0002	& 1.83 (83.97/46) & 0.0005 \rbtrr \\ \hline
{\bf $\pi^{+}$-Ne} & & & & & & & &  \rbtrr \\\hline
30 & 5.57 (89.06/16) & $<$ 0.0001	& 7.42 (118.79/16) & $<$ 0.0001	& 5.98 (95.74/16) & $<$ 0.0001	& 5.84 (93.46/16) & $<$ 0.0001 \rbtrr \\ \hline
64 & 1.74 (31.35/18) & 0.0262	& 1.94 (34.90/18) & 0.0097	& 2.13 (38.35/18) & 0.0035	& 2.06 (37.05/18) & 0.0052 \rbtrr \\ \hline
{\bf $\pi^{-}$-Ne} & & & & & & & &  \rbtrr \\ \hline 
30 & 2.72 (46.31/17) & 0.0002	& 3.62 (61.57/17) & $<$ 0.0001 & 3.78 (64.20/17) & $<$ 0.0001	& 4.09 (69.45/17) & $<$ 0.0001 \rbtrr \\ \hline
64 & 1.13 (19.24/17) & 0.3148	& 1.31 (22.32/17) & 0.1728	& 1.38 (23.45/17) & 0.1350	& 1.37 (23.37/17) & 0.1377 \rbtrr \\ \hline
\end{tabular}}
}
\label{tabChi}
\end{table}

\begin{figure}
\centerline{\includegraphics[scale = 0.38]{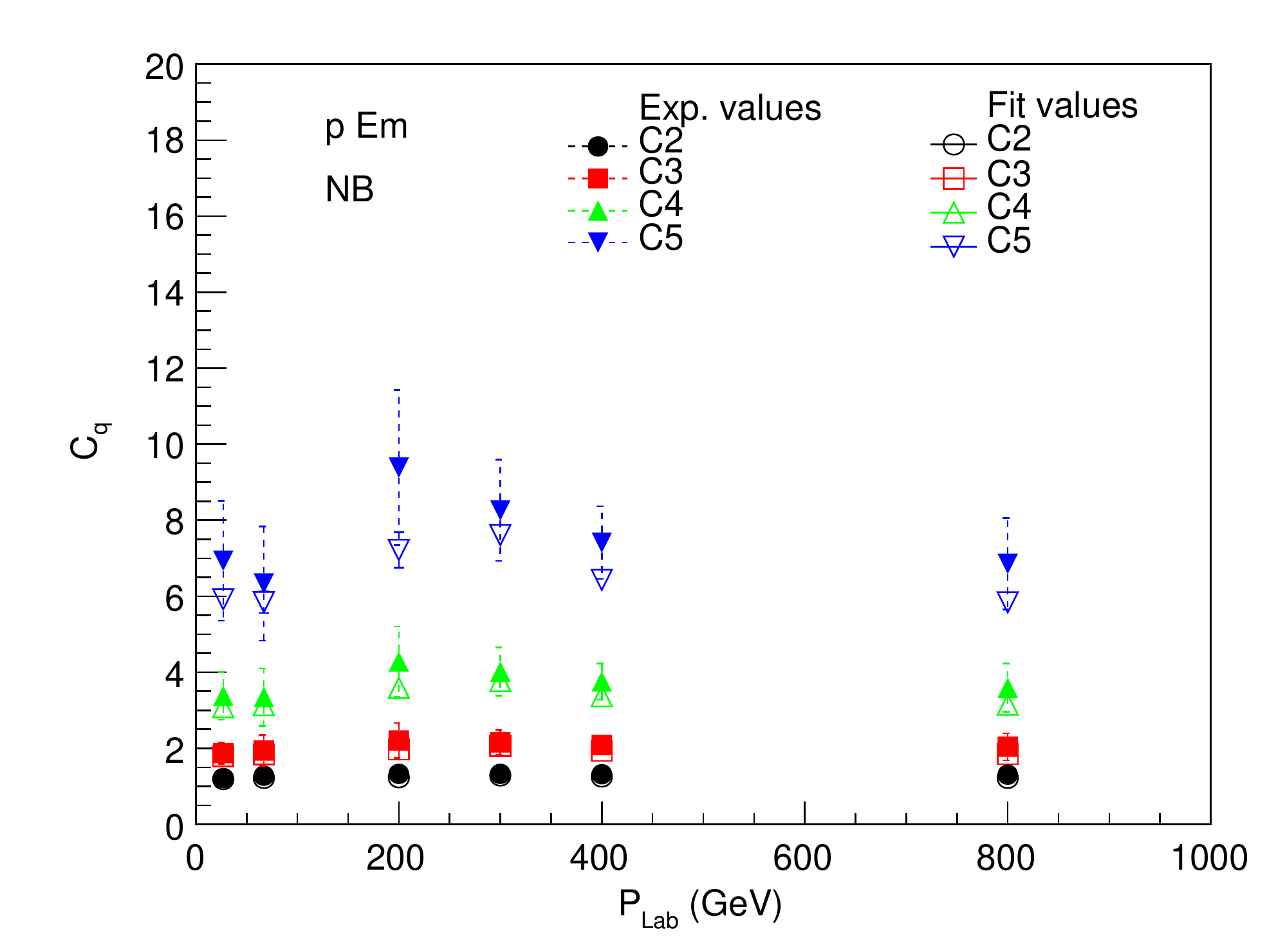}\includegraphics[scale = 0.38]{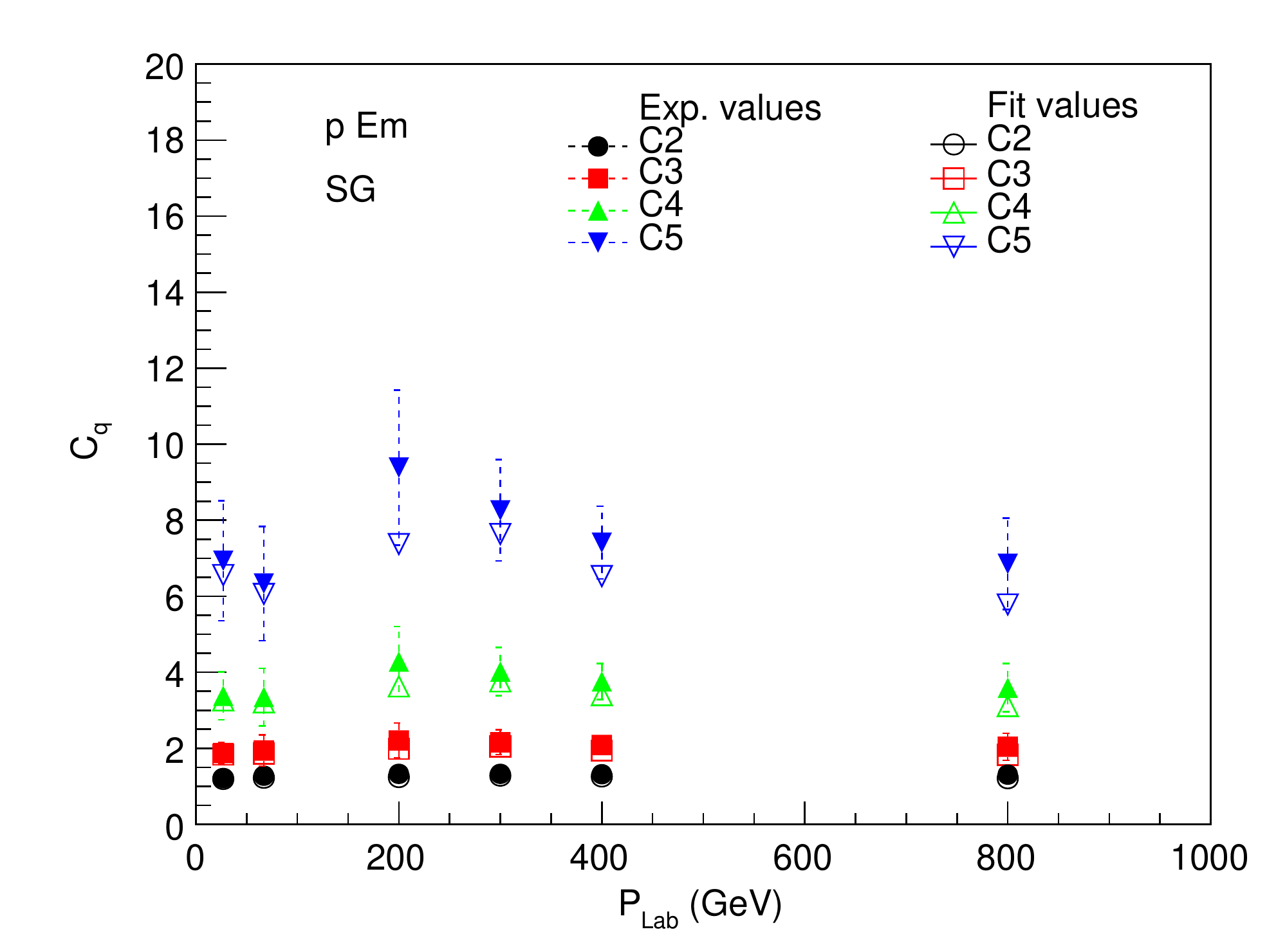}}
\centerline{\includegraphics[scale = 0.38]{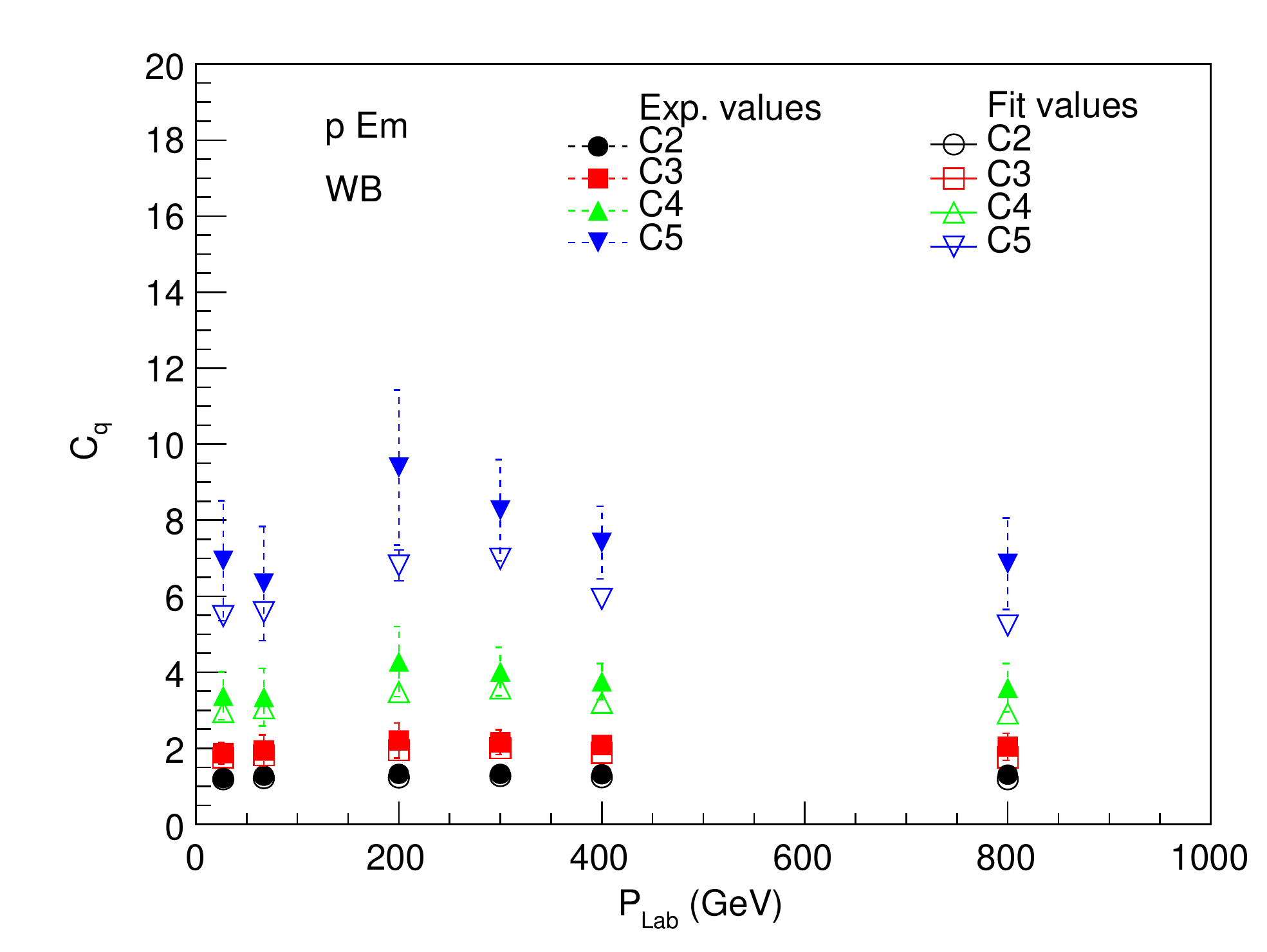}\includegraphics[scale = 0.38]{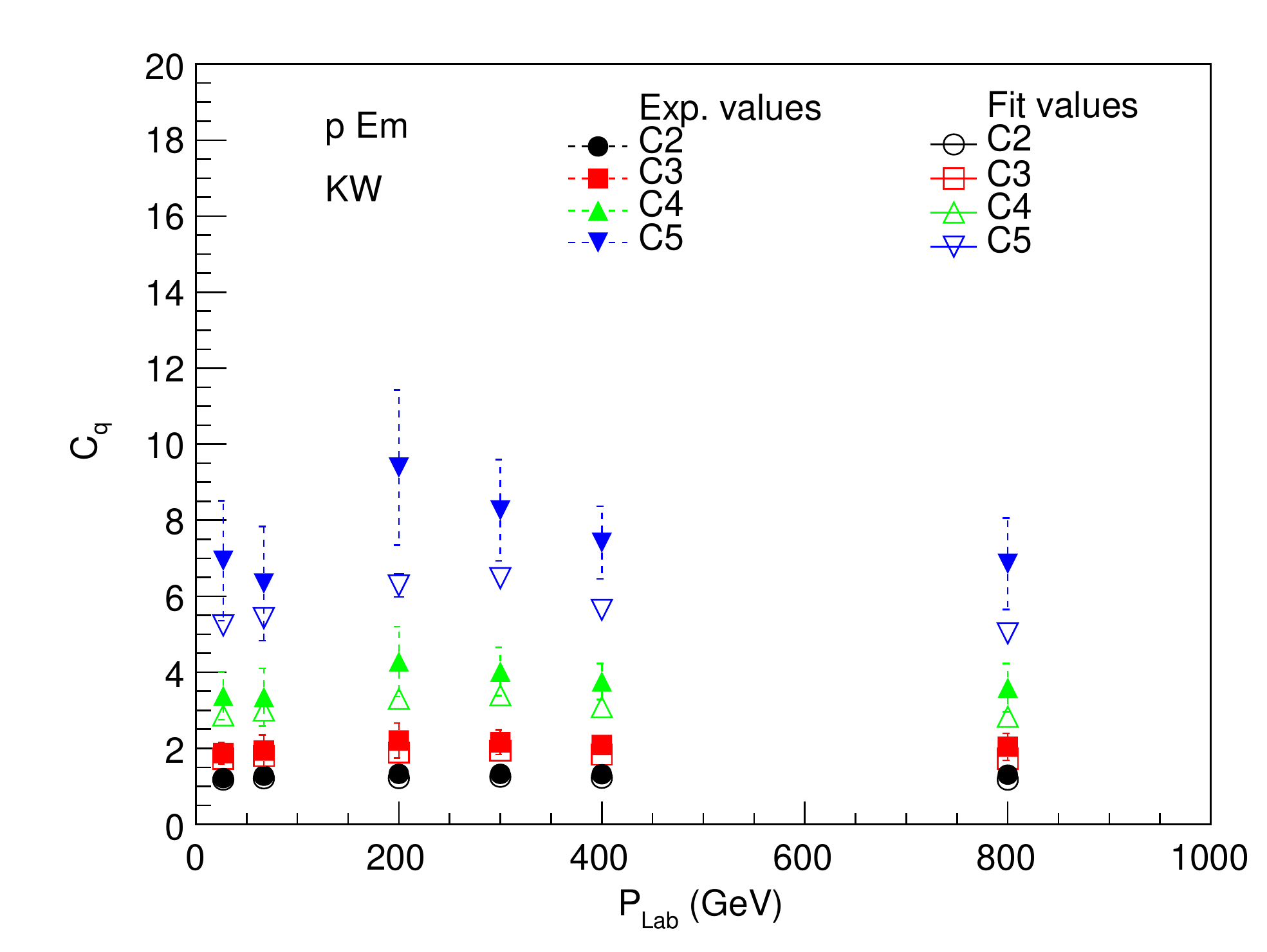}}
\caption{\small{Normalised moments calculated from the data and from NB, SG, WB and KW fit distributions in inelastic $p{\text -}Em$.}}
\label{figMompEM}
\end{figure}

\begin{figure}
\centerline{\includegraphics[scale = 0.38]{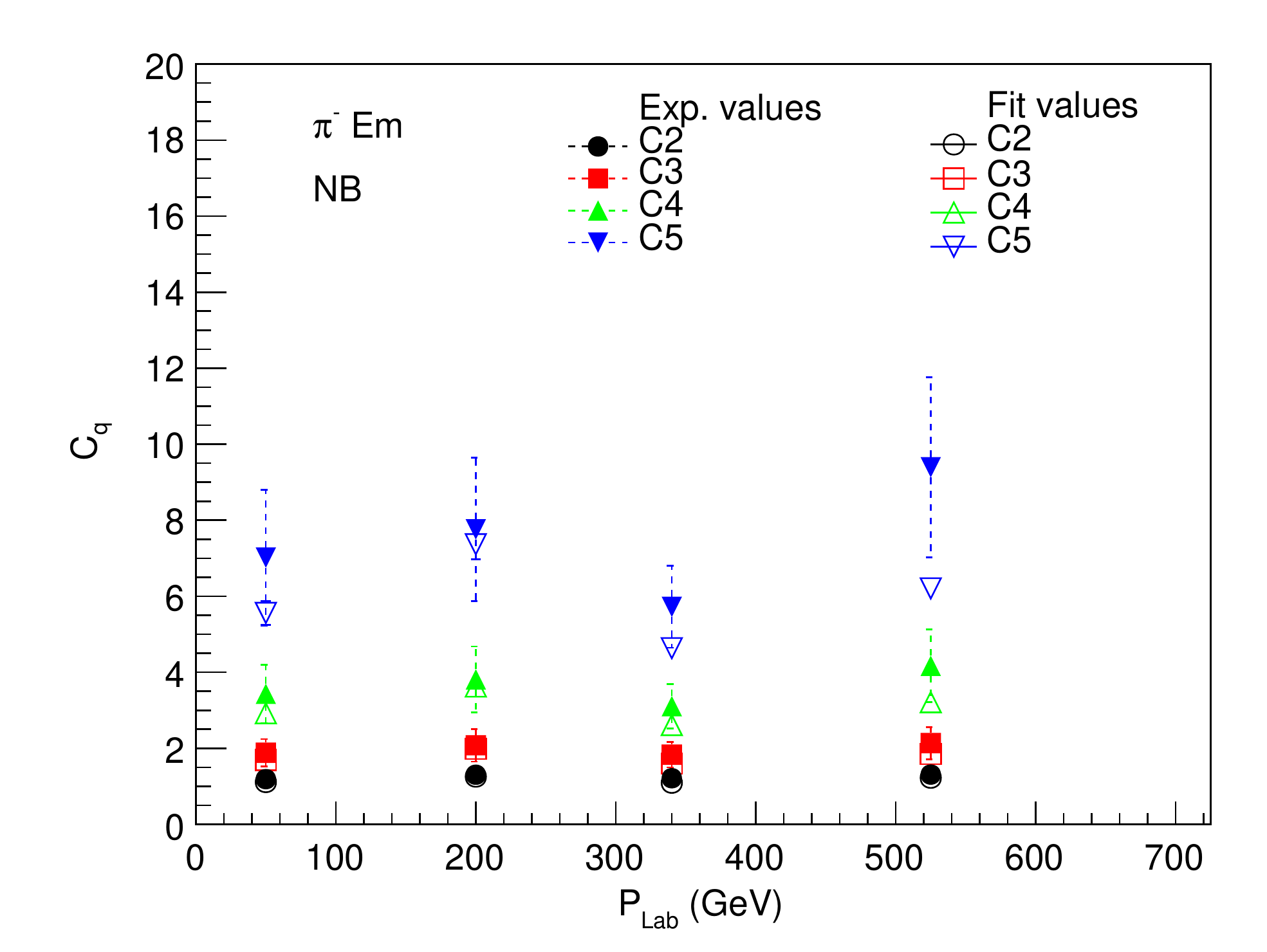}\includegraphics[scale = 0.38]{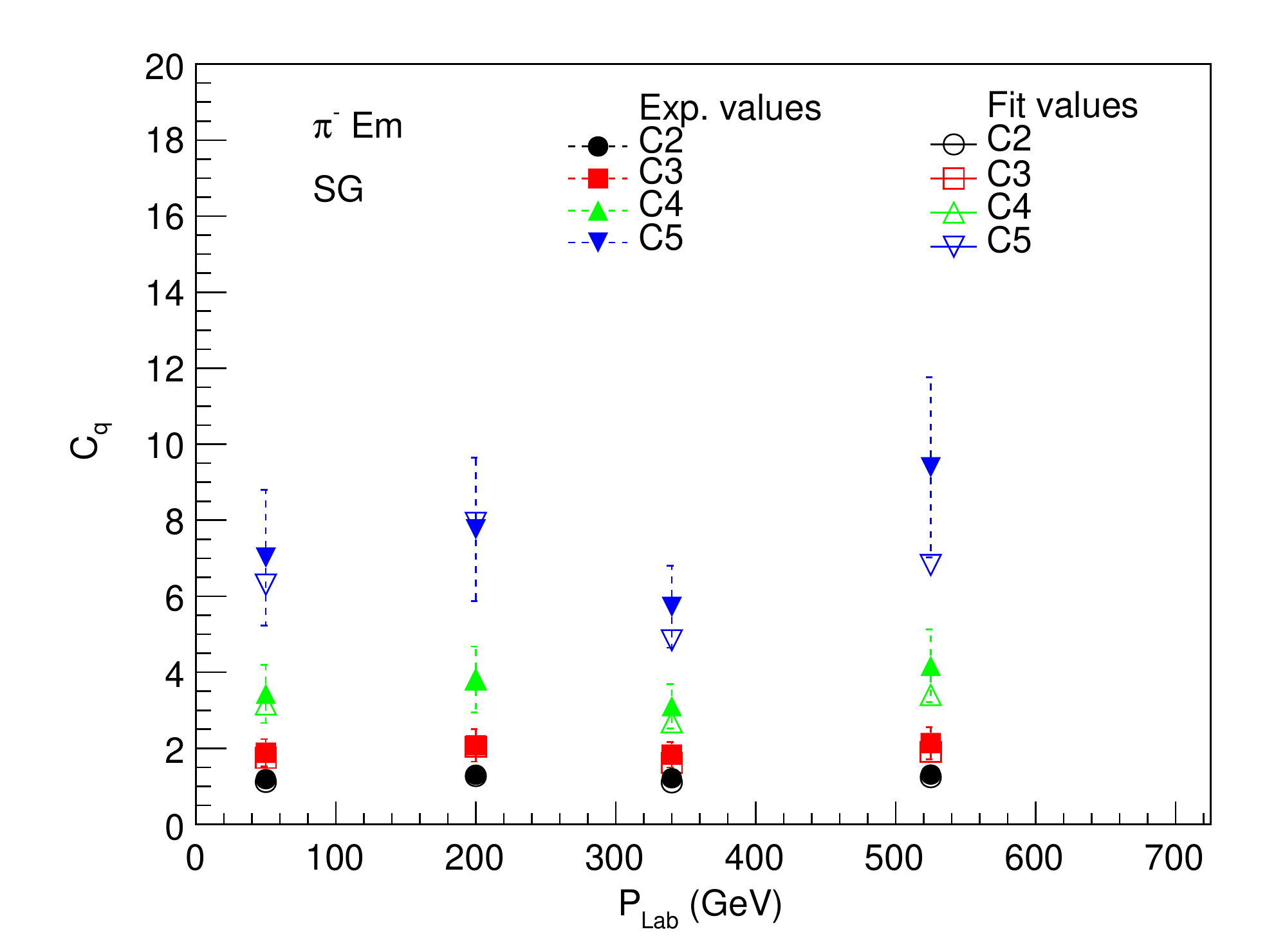}}
\centerline{\includegraphics[scale = 0.38]{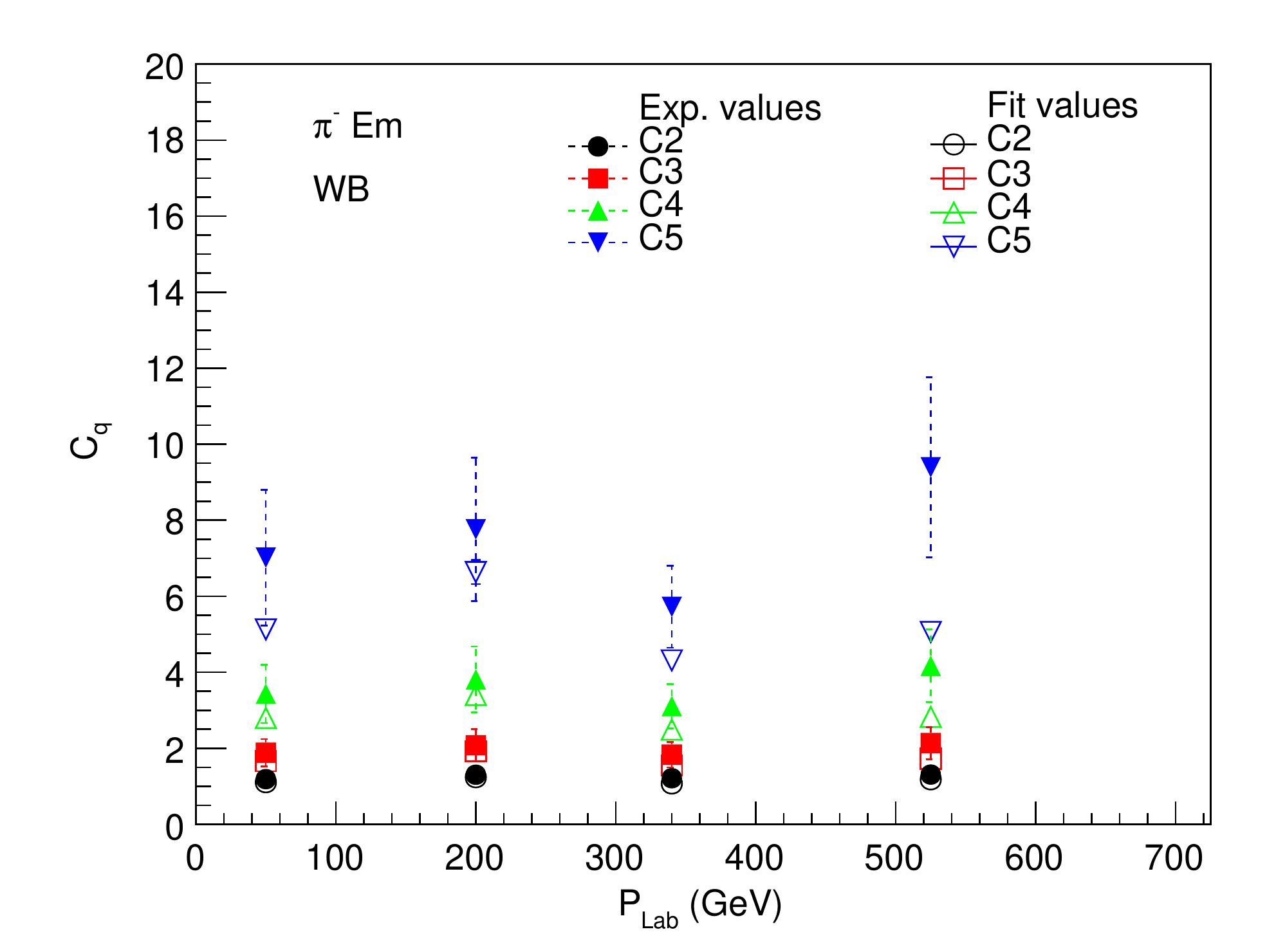}\includegraphics[scale = 0.38]{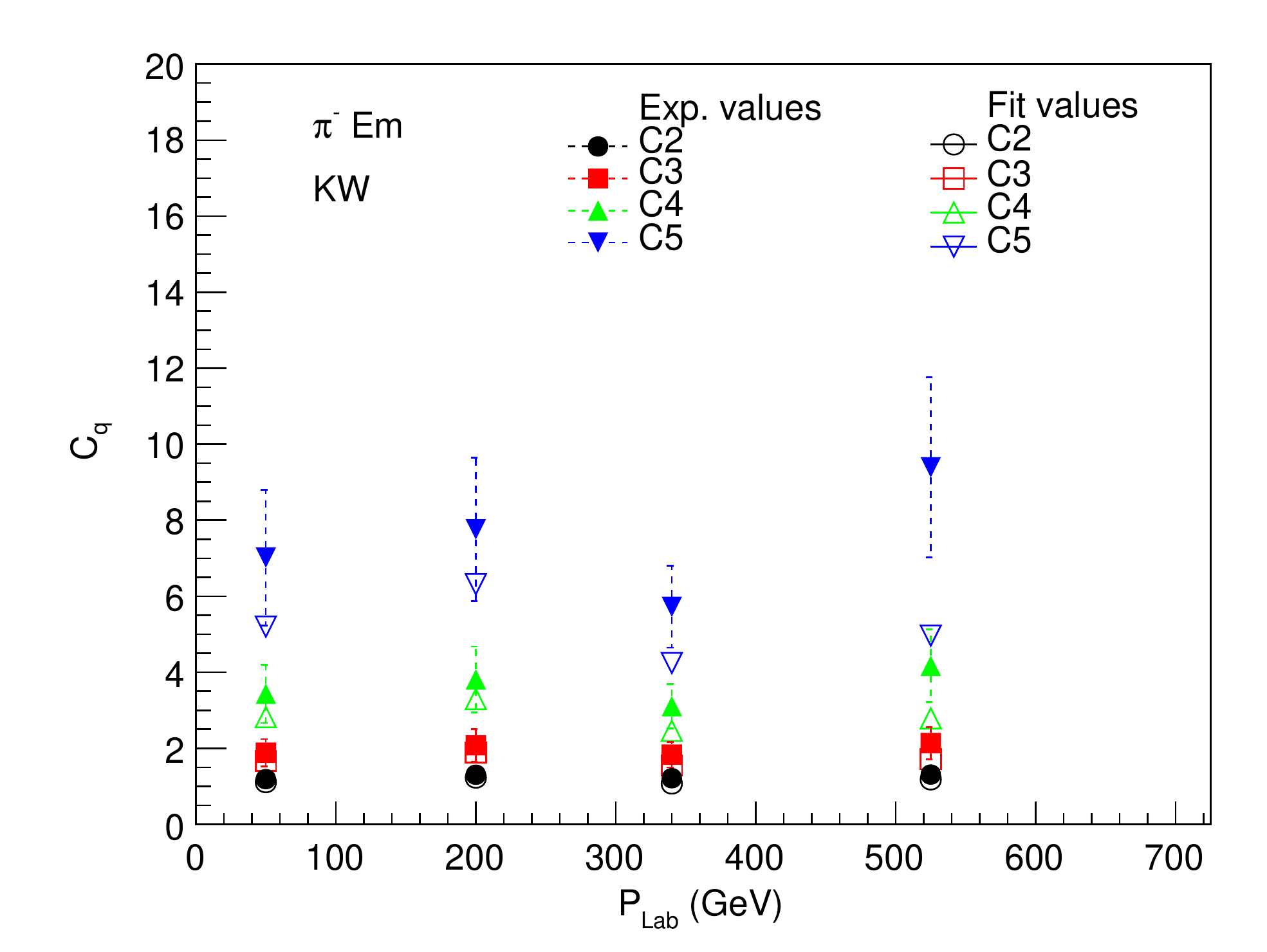}}
\caption{\small{Normalised moments calculated from the data and of NB, SG, WB and KW fit distributions in inelastic $\pi^{-}{\text -}Em$.}}
\label{figMompiEM}
\end{figure}

Table~3 shows the values of $\langle{n}\rangle$, expected from every fitted distribution and the experimentally measured values [\refcite{Barb}-\refcite{Rees}].~It is observed that in most of the cases, the two values are in agreement to each other, within the limits of error.~In case of p-Au data the differences are on account of very few and sparsed data points.~This shows that the distributions represent the data very well.~Figures~\ref{figMult}~(left)-and-(right) show the dependence of average charged particle multiplicity $\langle{n}\rangle$, calculated from different distributions fitted to the data, on the projectile energy $P_{Lab}$ for $p{\text -}Em$ and $\pi^{-}{\text -}Em$ interactions.~The dependence is parameterised as a quadratic equation in $ln\,(P_{Lab})$ as shown in equation (\ref{eqMult}).~The fit parameters are given in table~4.
\begin{equation}
\langle{n}\rangle=a + b *(ln P_{Lab}) + c *(ln P_{Lab})^{2} 
\label{eqMult} 
\end{equation}
For the NB distribution, scale parameter measures the average multiplicity, as shown in equation (\ref{eqNB}).~The shape parameter which determines the width of the distribution also depends on $\sqrt{s}$ according to $k^{-1} = \alpha + \beta (\it {ln\,s})$.~We calculate $\sqrt{s}$ from the $P_{Lab}$ by assuming the target mass of emulsion as 72.~Figure~4 shows the $k^{-1}$ dependence on $\sqrt{s}$ for both $p{\text -}Em$  and $\pi^{-}{\text -}Em$  interactions.~The parameters of fits are:
\begin{gather}
k^{-1} = (-0.1193\pm0.113) + (0.0596 \pm 0.021){ln\sqrt{s}}\hspace*{5mm} p{\text -}Em \\
k^{-1} = (-0.0869\pm0.082) + (0.0655\pm0.015){ln\sqrt{s}}\hspace*{5mm} \pi^{-}{\text -}Em 
\label{eqShape}
\end{gather}
Figures~\ref{figMompEM} and \ref{figMompiEM} show the dependence of normalised moments $C_q$ on $P_{Lab}$ in $p{\text -}Em$ and $\pi^{-}{\text -}Em$ interactions, at different energies.~Comparison between the experimental values calculated from the data and the values calculated from fits is shown for each fitted distribution.~The values of the $C_q$ moments for $q$=2 to 5 are given in tables \ref{tabMomExp}-9.~It may be observed that moments show increase with energy.~The center{\text -}of{\text -}mass energy of the $p{\text -}Em$ interaction for $P_{Lab}$ = 800~GeV is approximately 335~GeV, assuming the average atomic weight of emulsion as 72.~Already the KNO scaling violation was reported at ISR energies in $pp$ collisions [\refcite{ISR}] and also by the UA5 collaboration at 200~GeV [\refcite{UA5}].
\begin{figure}
\centerline{\includegraphics[scale = 0.38]{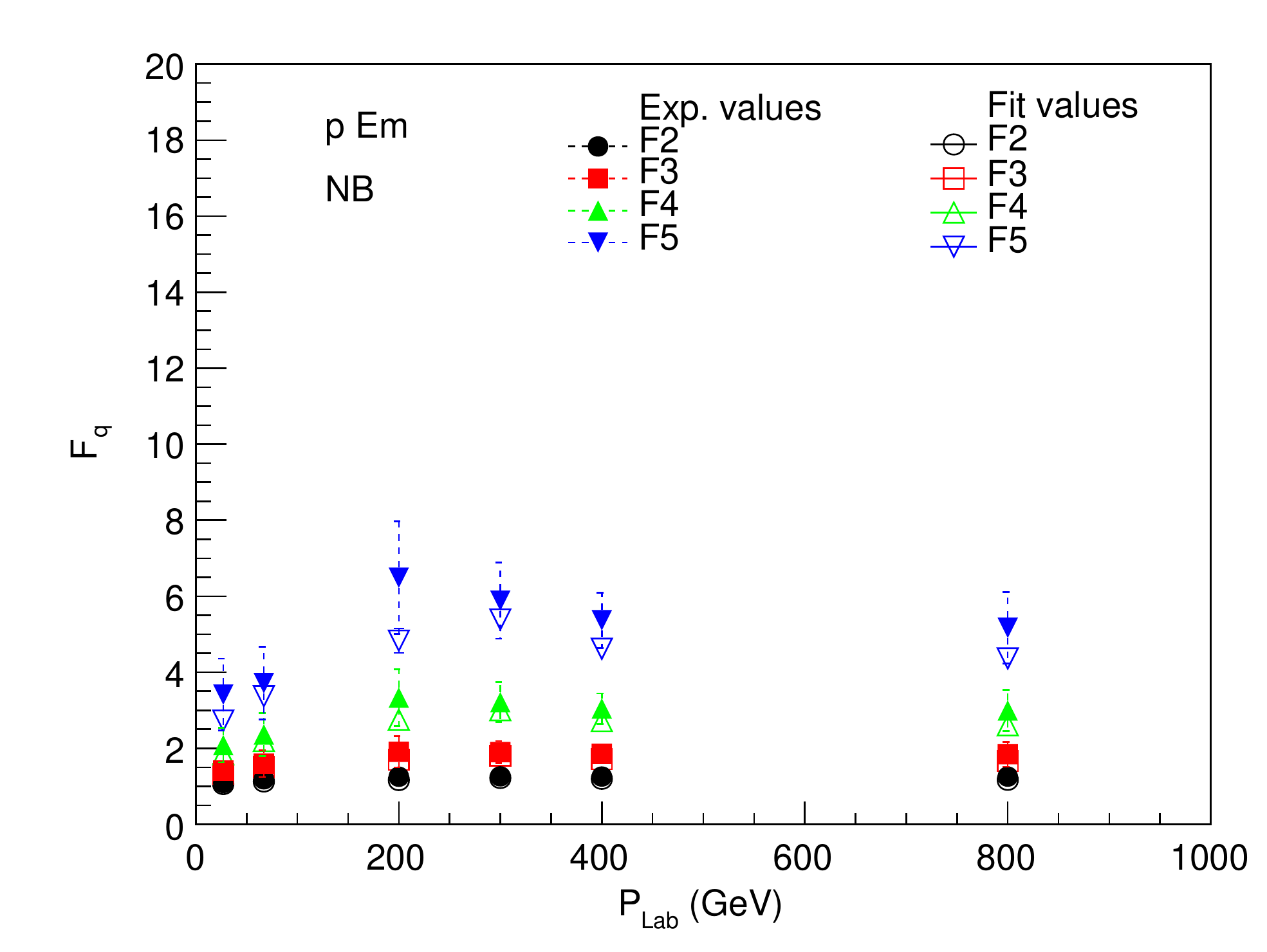}\includegraphics[scale = 0.38]{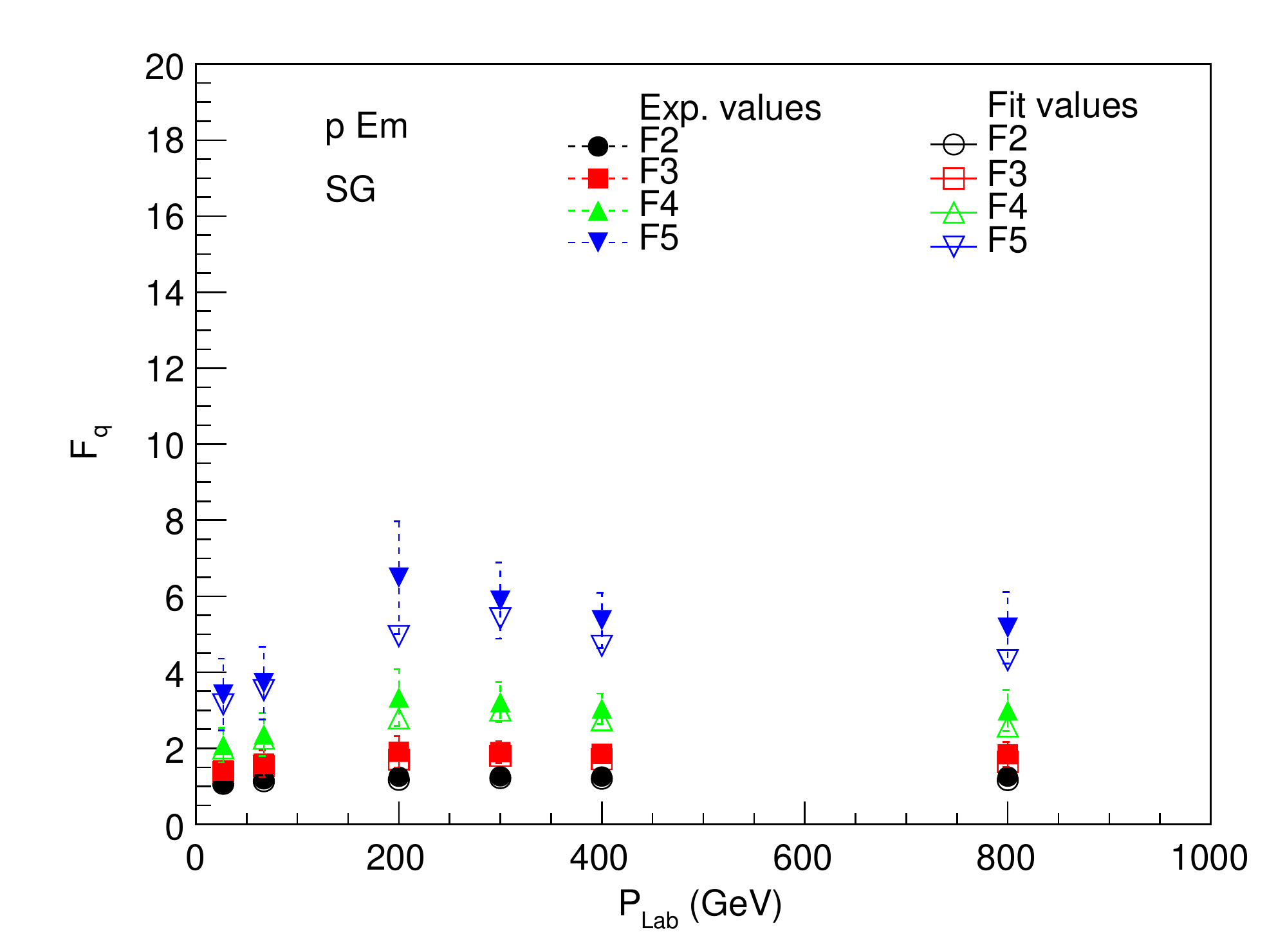}}
\centerline{\includegraphics[scale = 0.38]{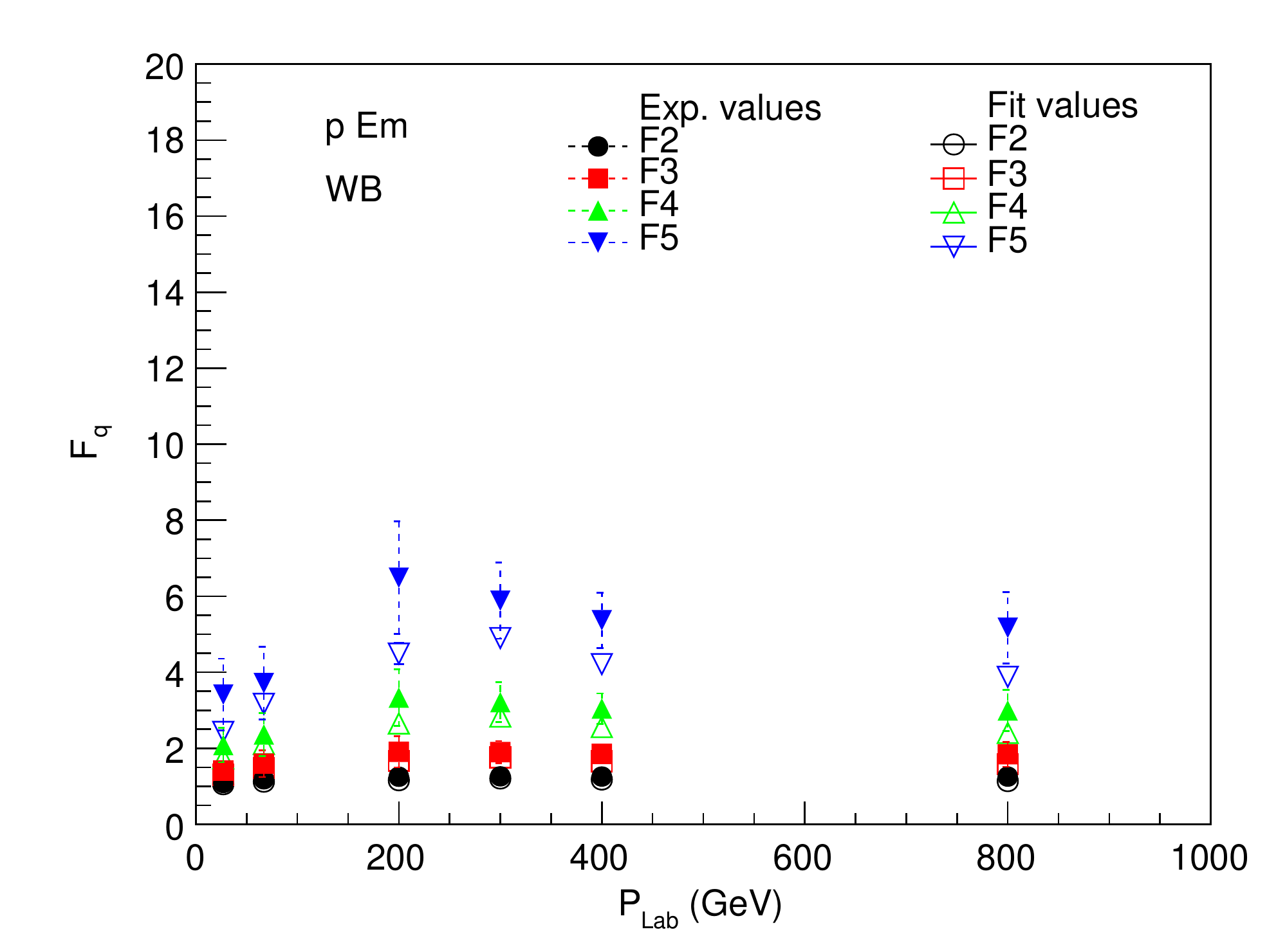}\includegraphics[scale = 0.38]{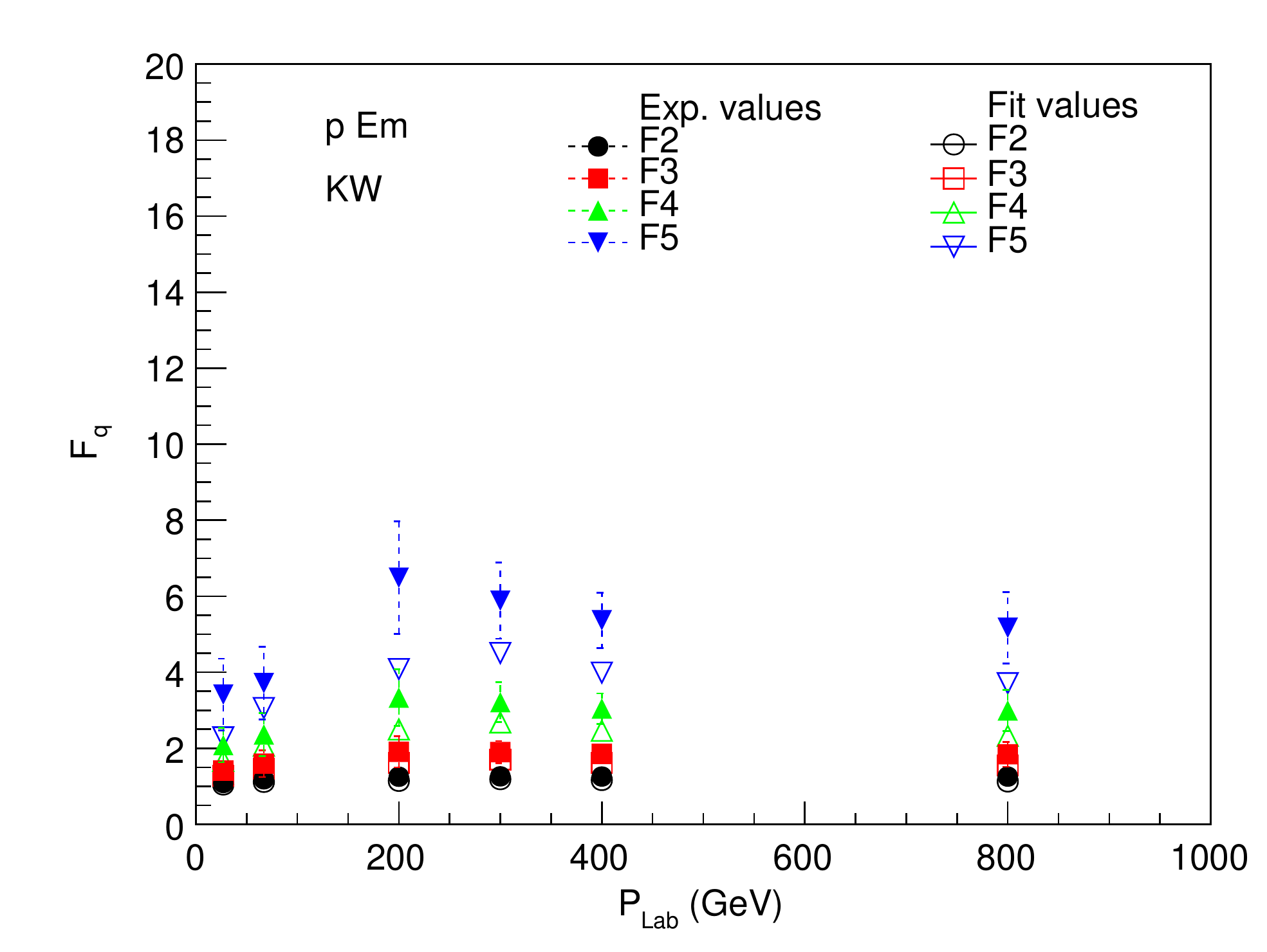}}
\caption{\small{Normalised factorial moments calculated from the data and of NB, SG, WB and KW fit distributions in inelastic $p{\text -}Em$.}}
\label{figFacMompEM}
\end{figure}

\begin{figure}
\centerline{\includegraphics[scale = 0.38]{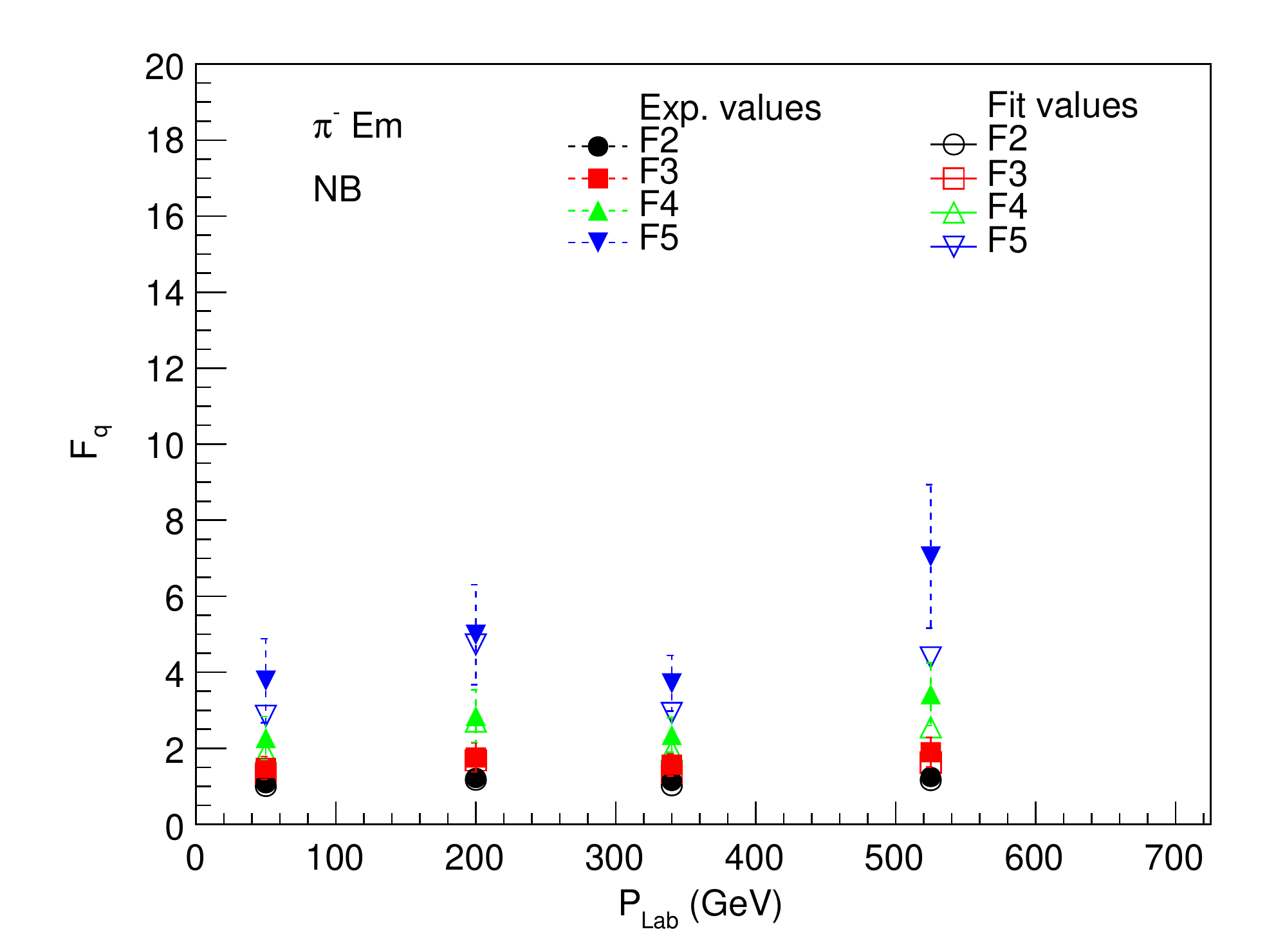}\includegraphics[scale = 0.38]{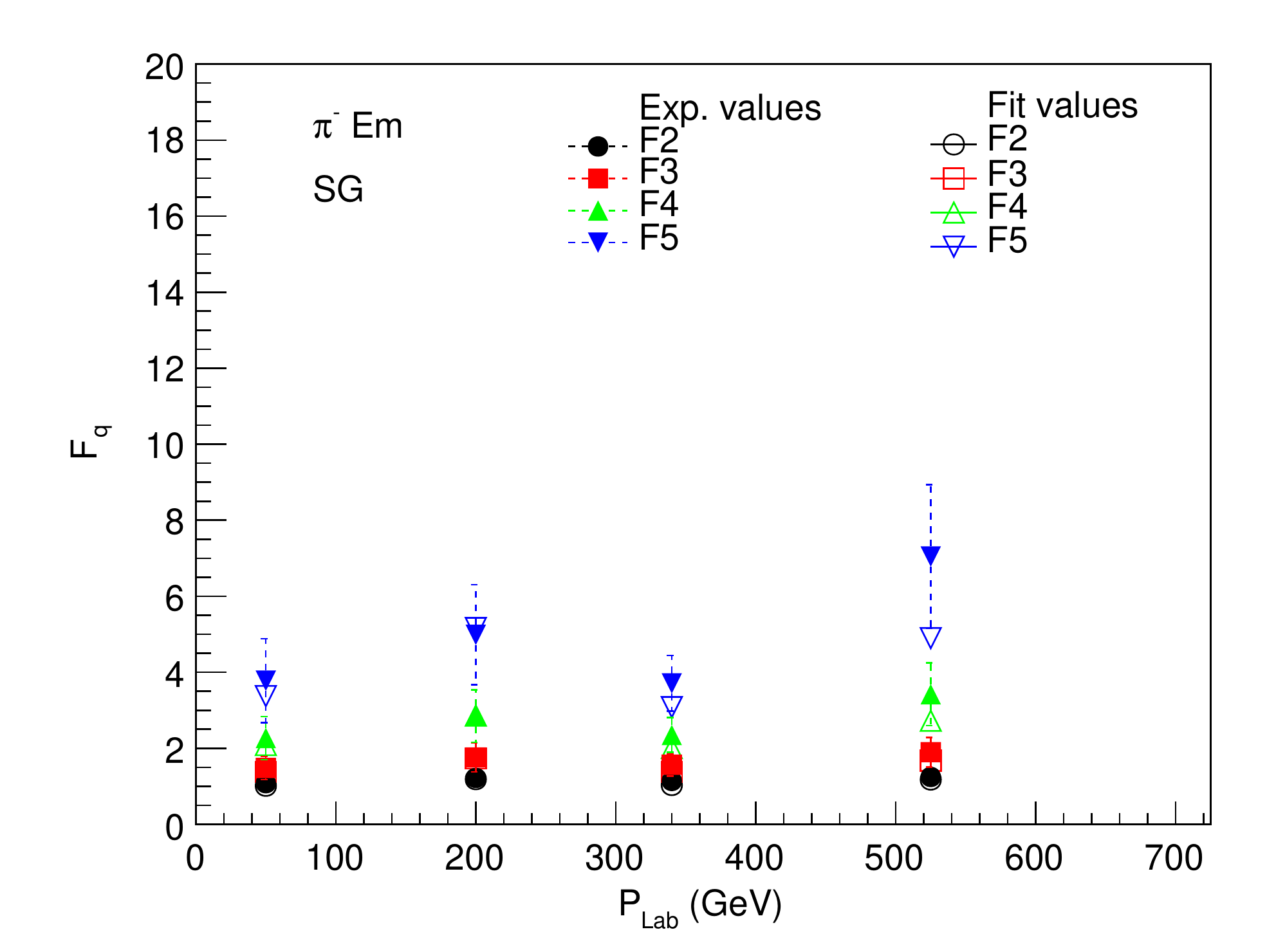}}
\centerline{\includegraphics[scale = 0.38]{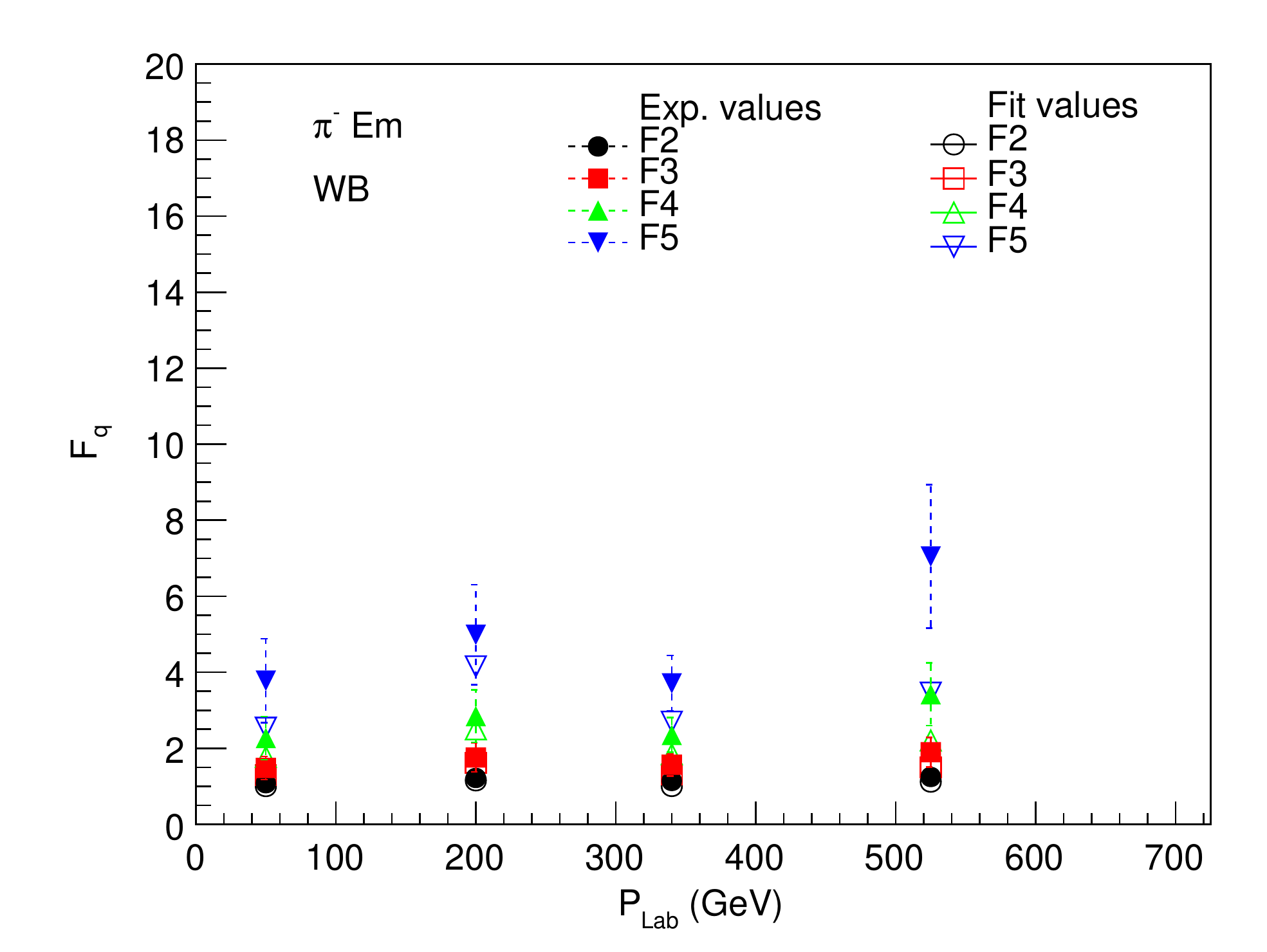}\includegraphics[scale = 0.38]{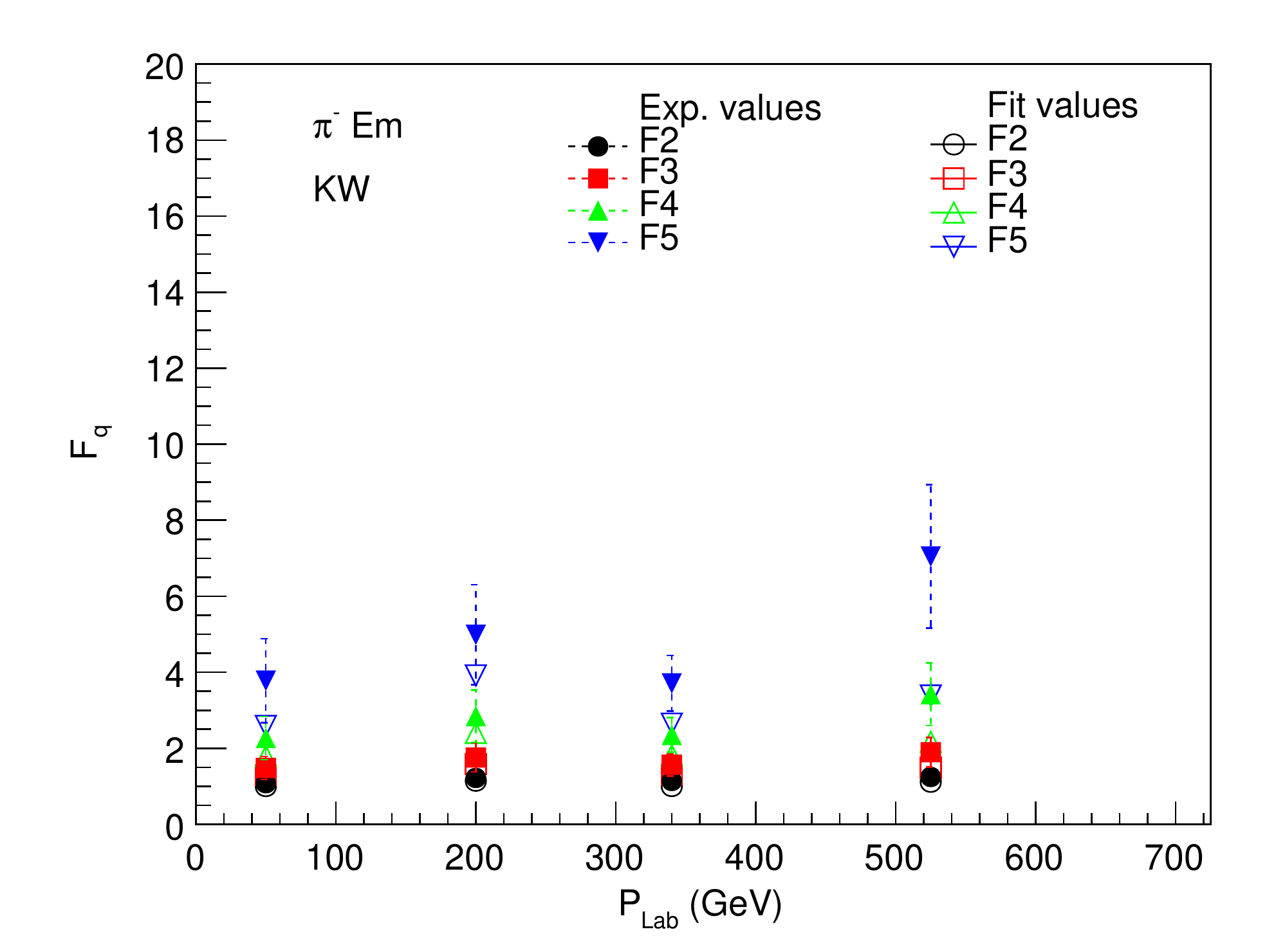}}
\caption{\small{Normalised factorial moments calculated from the data and of NB, SG, WB and KW fit distributions in inelastic $\pi^{-}{\text -}Em$.}}
\label{figFacMompiEM}
\end{figure}

\begin{figure}
\centerline{\includegraphics[scale = 0.38]{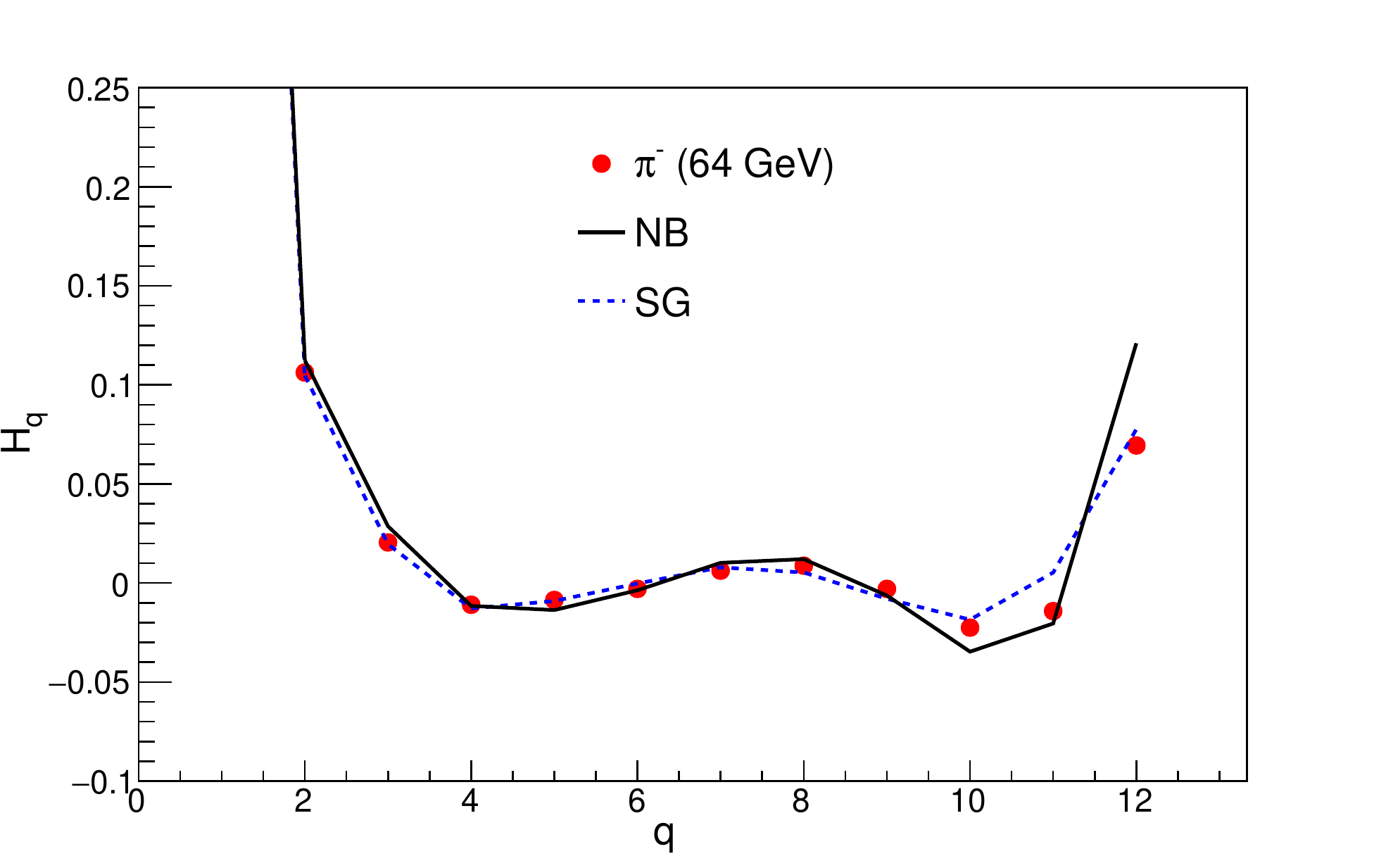}\includegraphics[scale = 0.38]{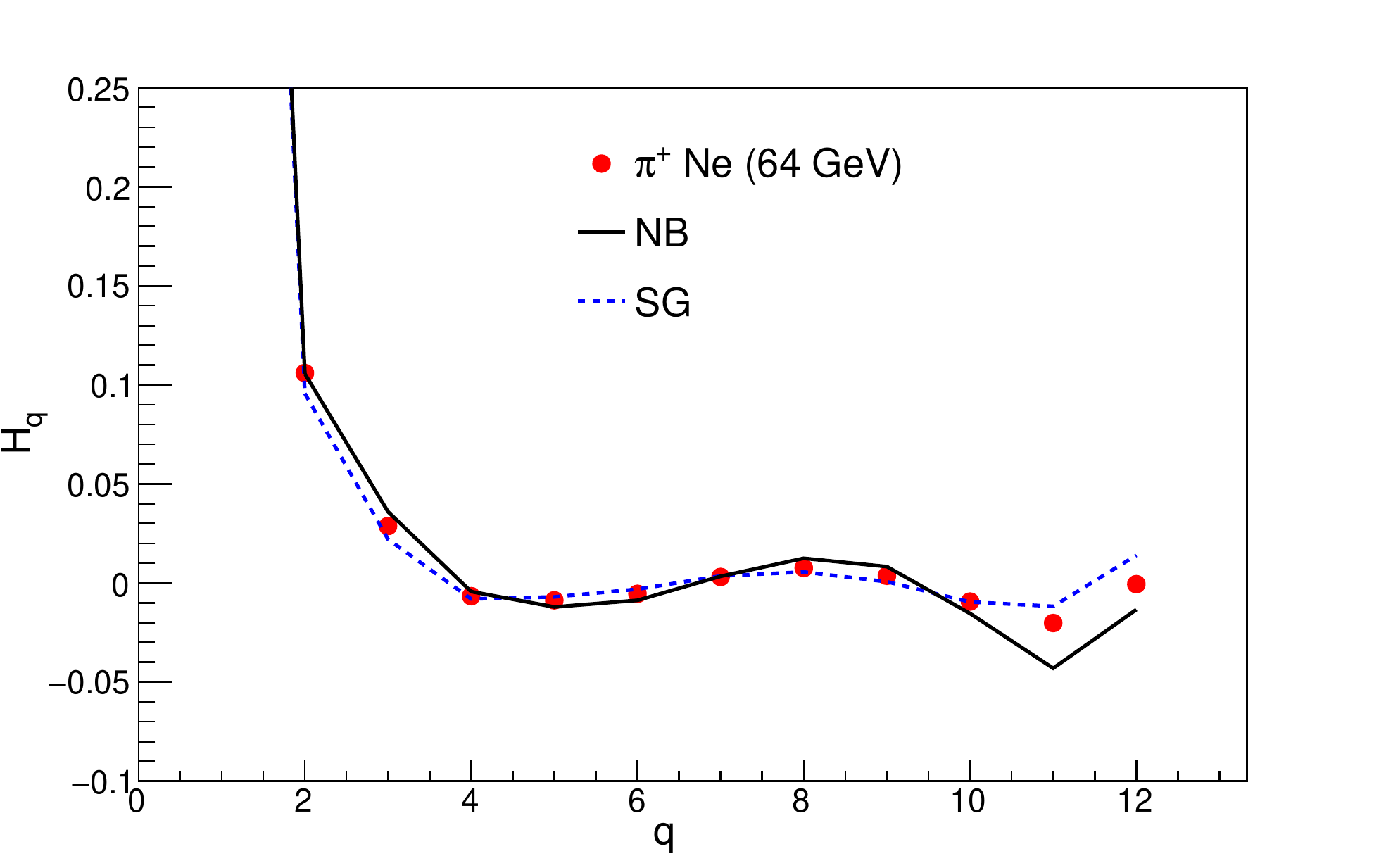}}
\centerline{\includegraphics[scale = 0.38]{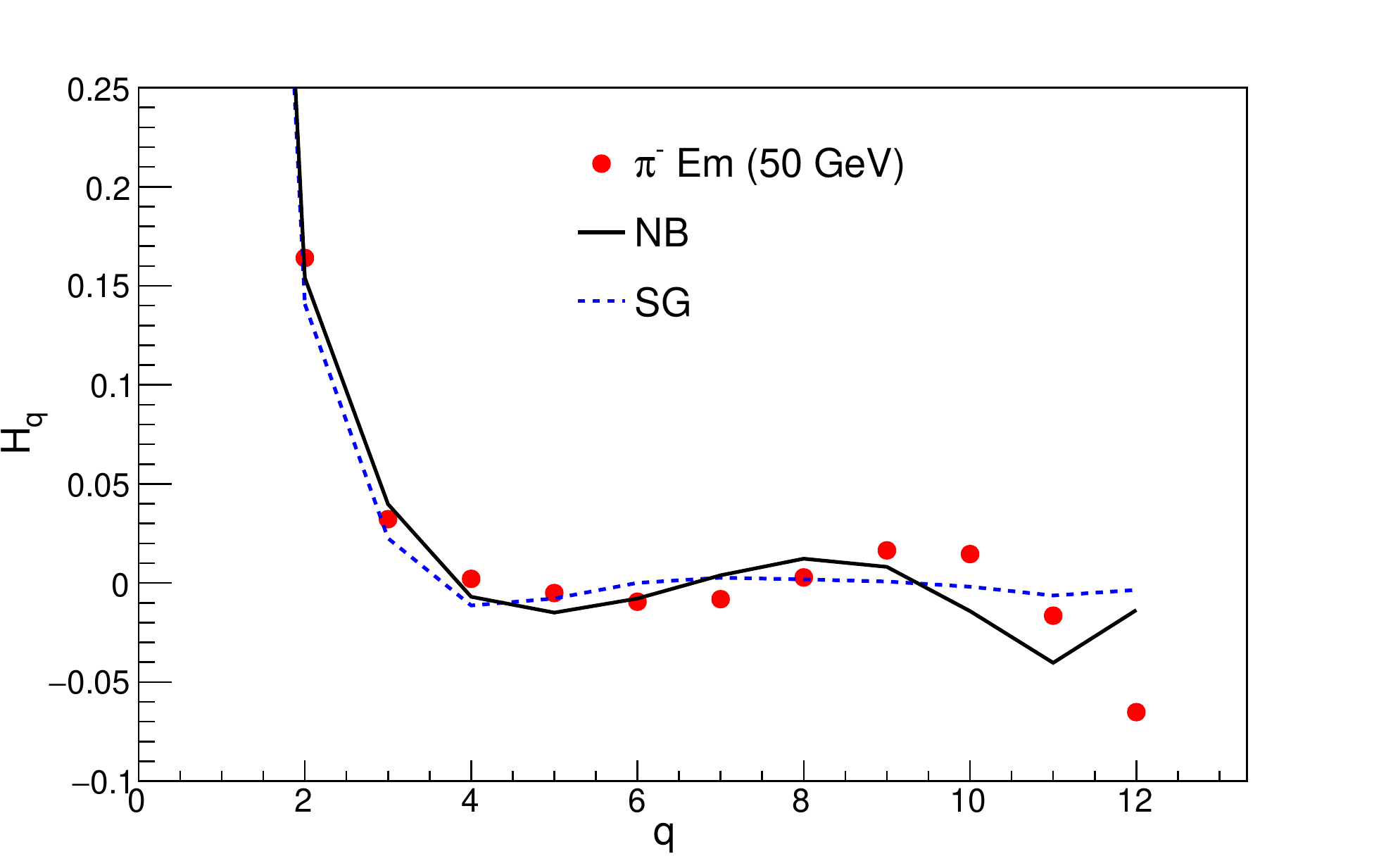}\includegraphics[scale = 0.38]{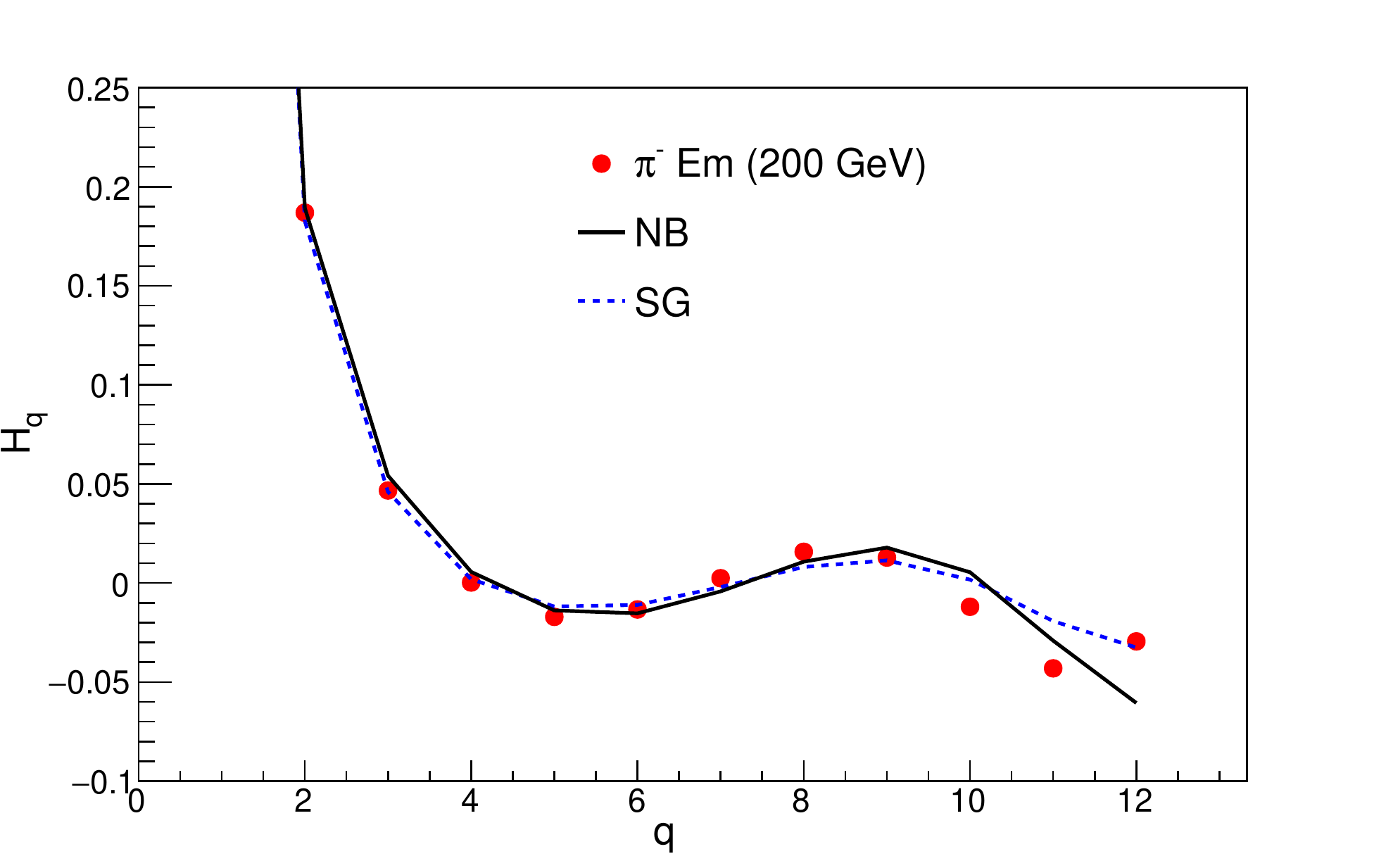}}
\centerline{\includegraphics[scale = 0.38]{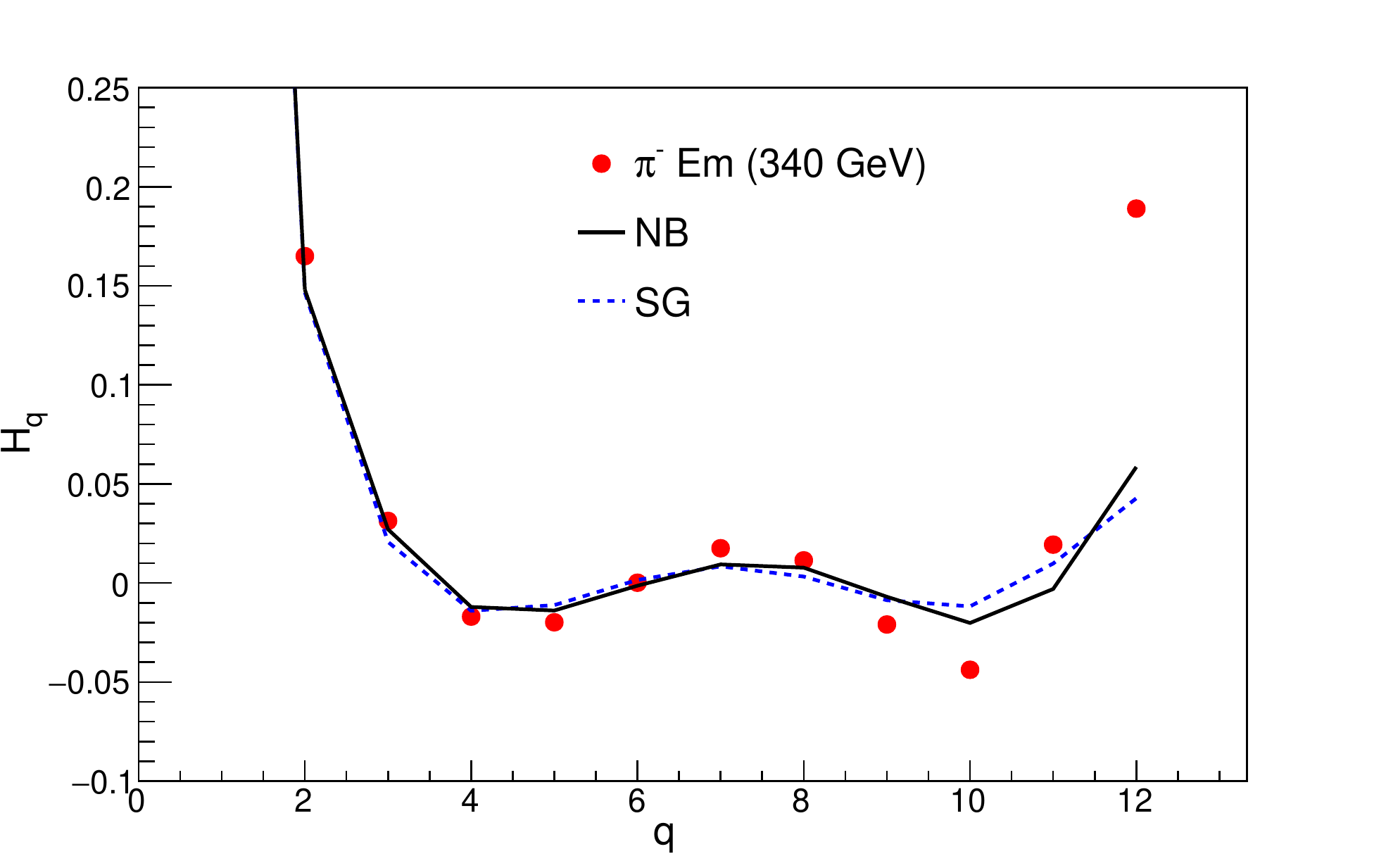}\includegraphics[scale = 0.38]{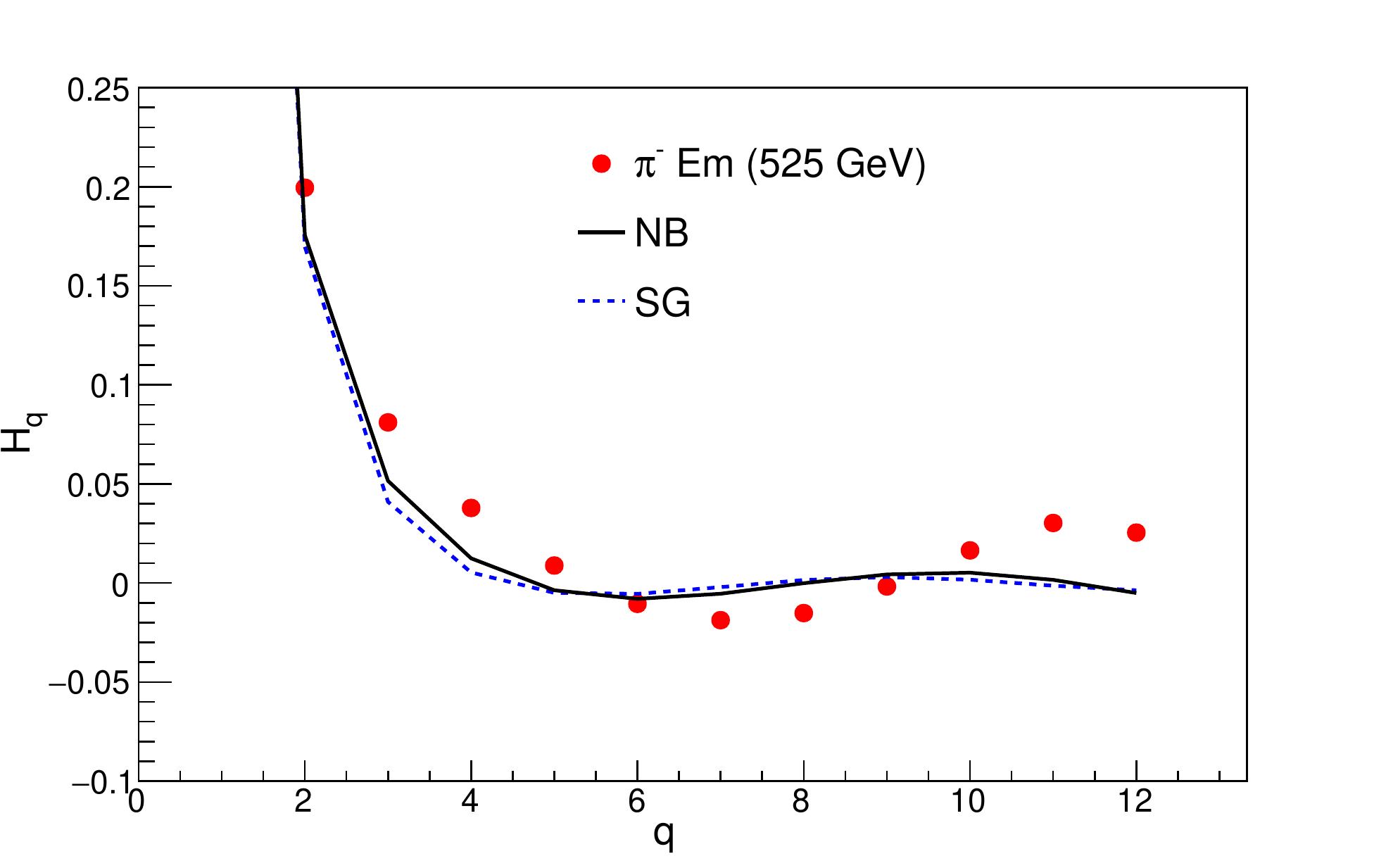}}
\caption{\small{$H_{q}$ moments calculated from the data, NB and SG fit distributions in inelastic $\pi^{-}{\text -}Em$.}}
\label{figHqPi}
\end{figure}

\begin{figure}
\centerline{\includegraphics[scale = 0.38]{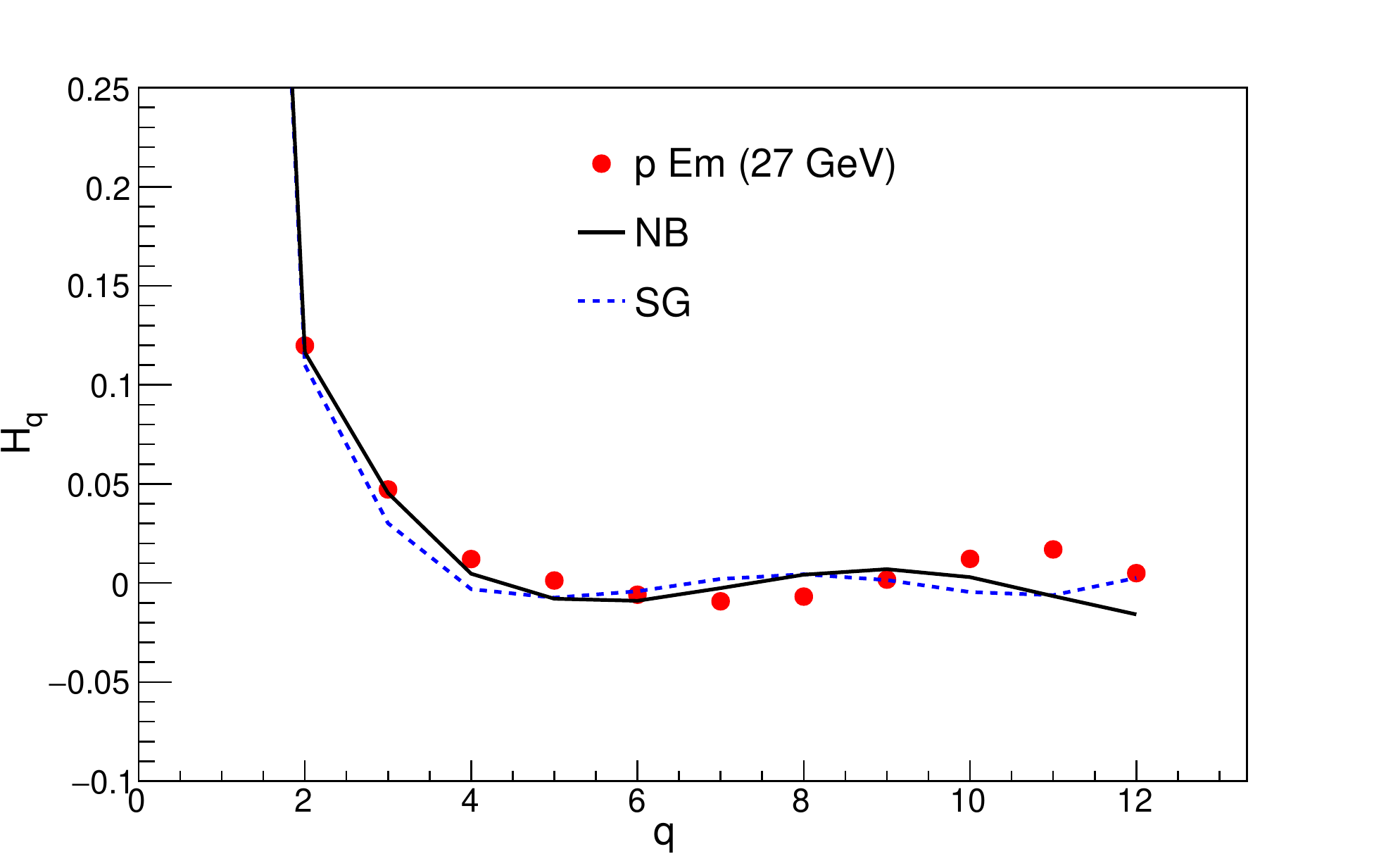}\includegraphics[scale = 0.38]{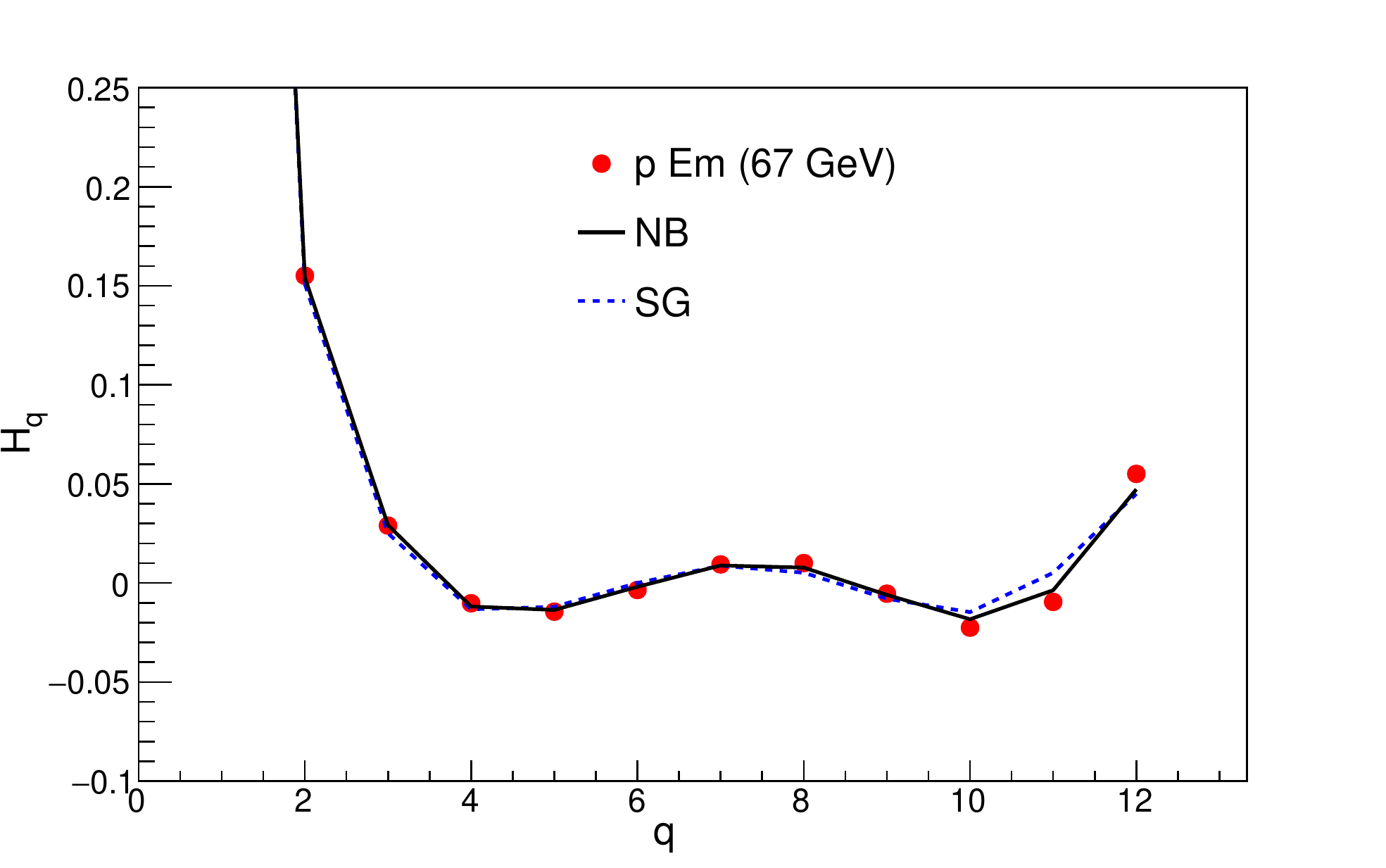}}
\centerline{\includegraphics[scale = 0.38]{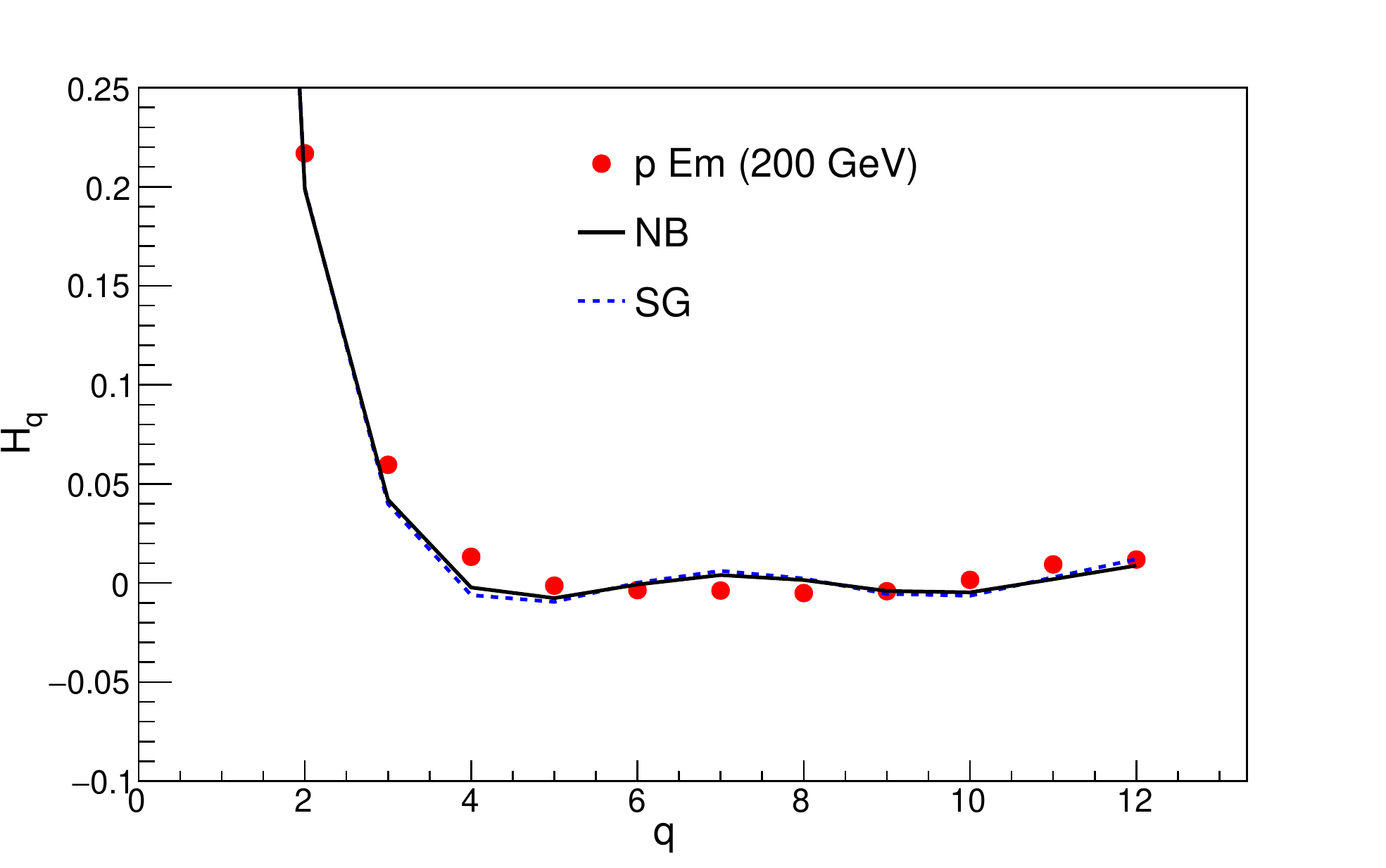}\includegraphics[scale = 0.38]{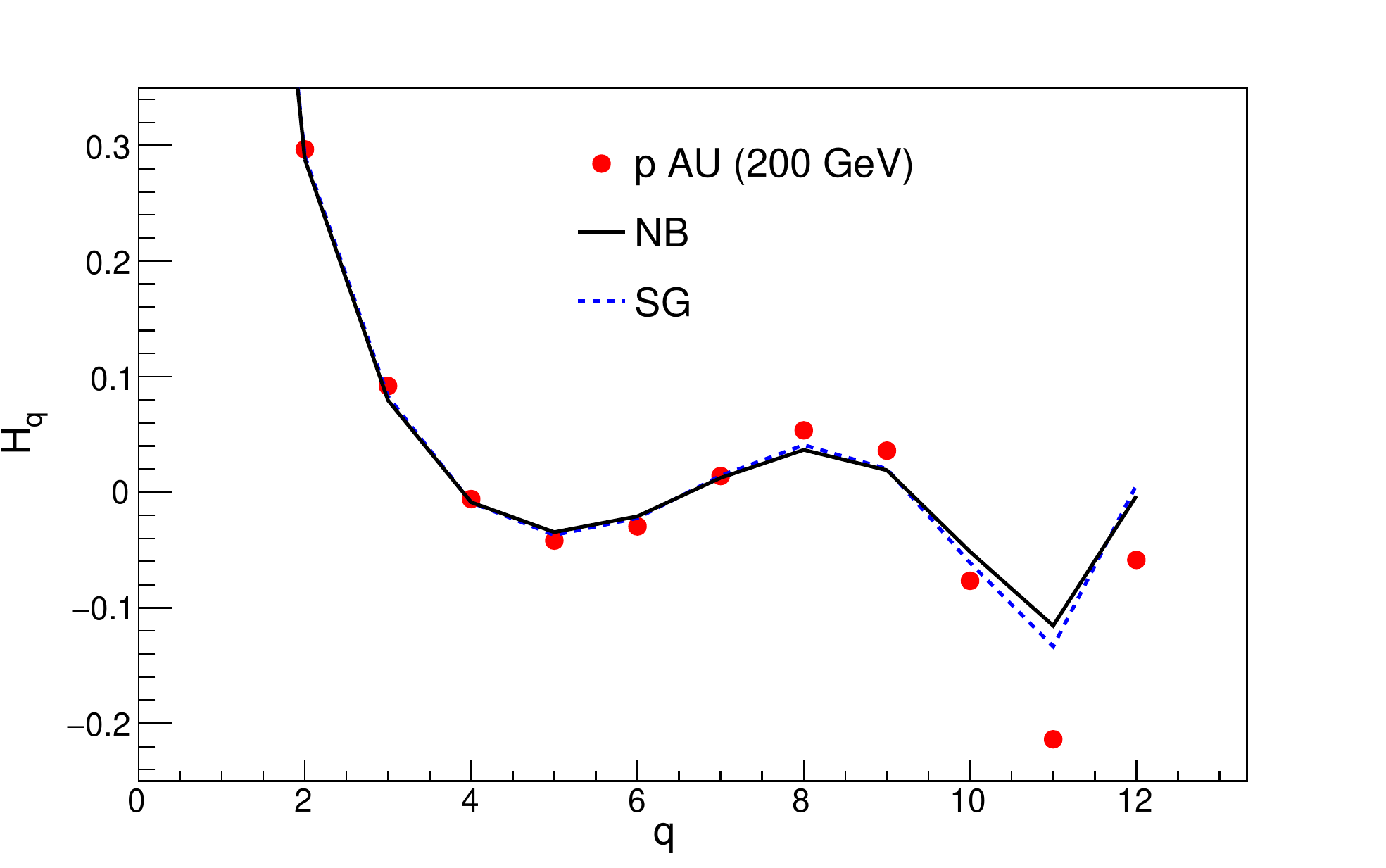}}
\centerline{\includegraphics[scale = 0.38]{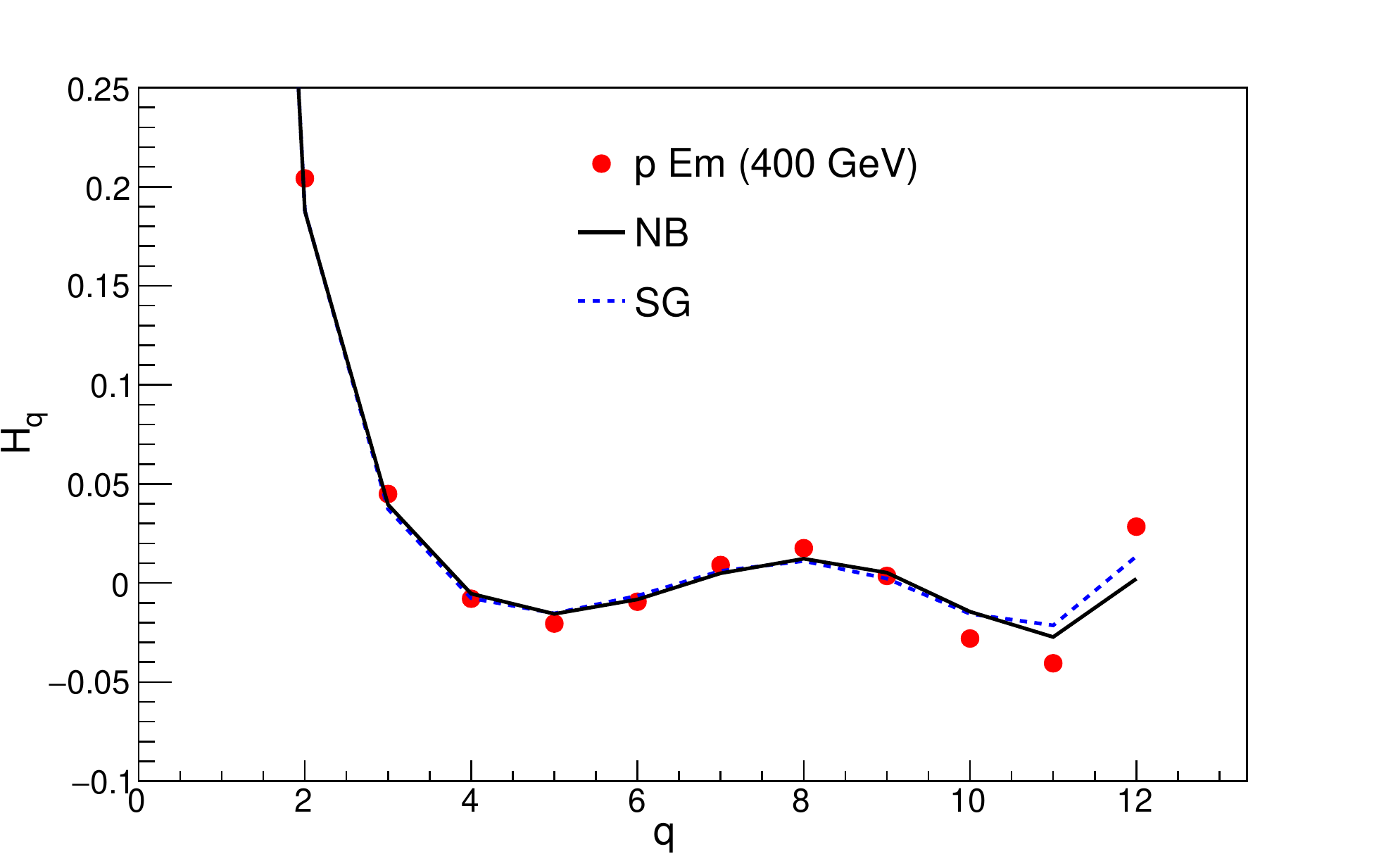}\includegraphics[scale = 0.38]{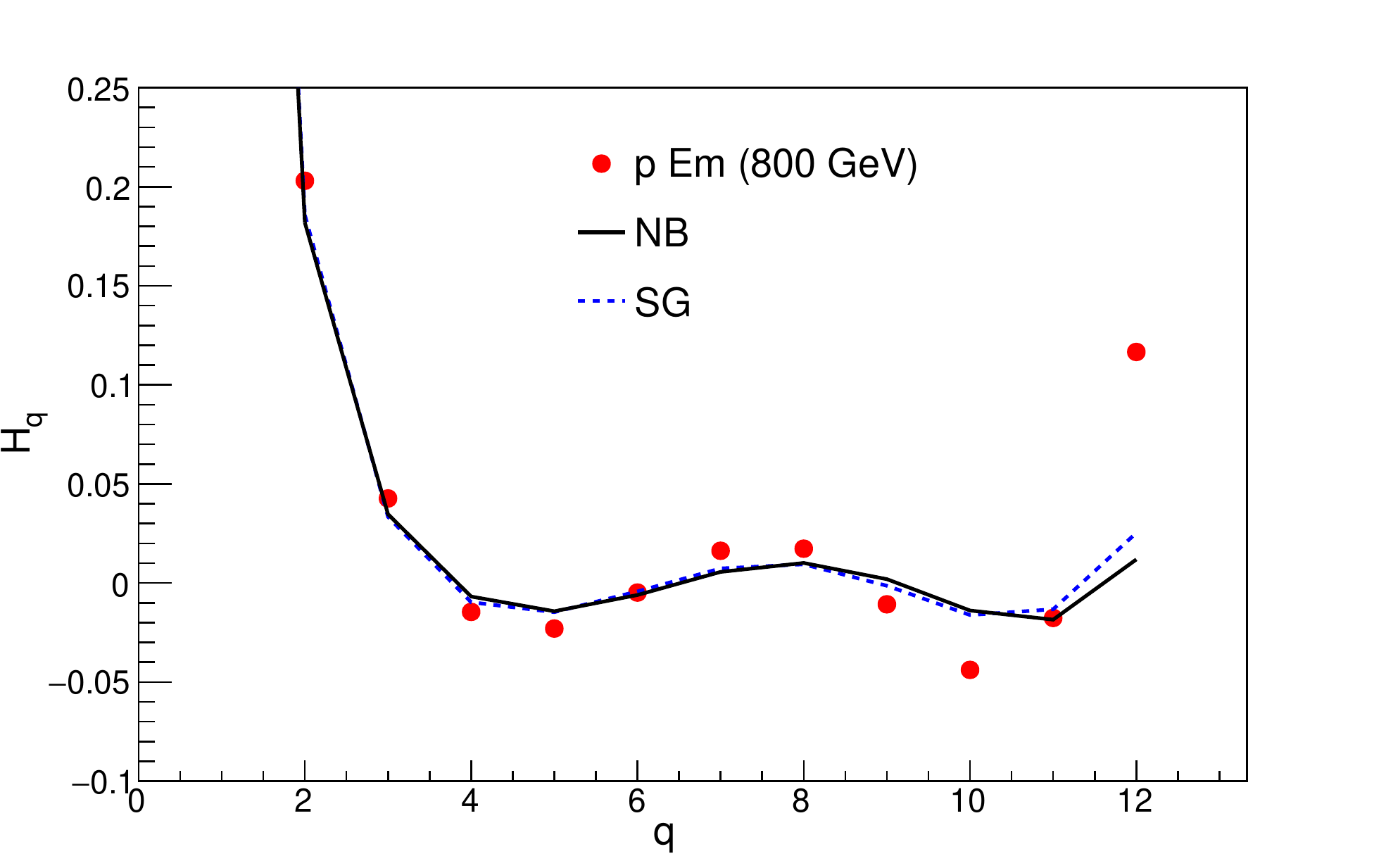}}
\caption{\small{$H_{q}$ moments calculated from the data, NB and SG fit distributions in inelastic $p{\text -}Em$.}}
\label{figHqP}
\end{figure}

Figures~\ref{figFacMompEM} and \ref{figFacMompiEM} show the dependence of normalised factorial moments $F_q$ on $P_{Lab}$ in $p{\text -}Em$ and $\pi^{-}{\text -}Em$ interactions, at different energies.~Comparison between data and fit values for different distributions is shown for each fitted distribution.~The values of the moments for $q$=2 to 5 are given in tables \ref{tabMomExp}-9.~The fact that $1/k$ and hence the factorial moments, are increasing with energy indicates the violation of KNO scaling, as also inferred in the reference [\refcite{ZA}].~Following the description from Ref. [\refcite{ZA}], this also indicates that the particles produced are correlated and the interaction dynamics is influenced by these correlations.~No distribution has $F_q$ smaller than unity, prohibiting the negative correlations.

Figures~9-10 show the $H_{q}$ moments (equation~($\ref{eqHq}$)) as a function of the rank $q$ for the data, NB and SG distributions for the $p{\text -}Em$ and $\pi^{-}{\text -}Em$ interactions.~It may be observed that the dependence of $H_{q}$ on $q$ is very similar at all energies for both the cases, with a steep descent to a minimum value around $q_{min}\sim$5-6.~The agreement between the data and the fit values is very good for both NB and SG distributions.~The shape of the charged-particle multiplicity distribution analysed in terms of the $H_{q}$ shows quasi-oscillations.~The special oscillation pattern of the ratio of cumulants to factorial moments, $H_{q}$  with a first minimum occuring around $q_{min}\sim$5 and quasi-oscillations about zero are expected for larger values of $q$ as predicted by I.M. Dremin et al [\refcite{DR}].~These predictions also come from the next-to-next-to-leading-log-approximation~(NNLLA) of perturbative QCD.
  
\section{Conclusions}
A detailed analysis to compare the multiplicity spectra of charged particles produced in $p/\pi{\text -}$nucleus interactions in the fixed target experiments, with projectile energies ranging from 27~GeV to 800~GeV, has been done in the framework of the negative binomial, the shifted-Gompertz, the Weibull distribution and the Krasznovszky-Wagner distribution.~Out of these, SG and WB distributions are very new and proposed to test the new data from collider experiments at higher energies.~The relevance of the comparison is on account of the similar nature of these distributions, whereby each of these distributions has two free parameters.~It is interesting to revisit the old data collected with the visual detectors which offers an opportunity to test and study the particle production with new concepts in terms of these distributions.~During the times when these data were collected, the particle production at different energies was studied in terms of KNO scaling [\refcite{Kob},~\refcite{KNO}].~The understanding in terms of NB distribution came into picture much later, and was used mostly for the data from colliders.~Often, owing to the success of NB, it became a benchmark with respect to which the performance of other distributions were studied.~Our analysis of the data in terms of Negative Binomial, shifted Gompertz, the Weibull and the Krasznovszky-Wagner distributions shows that out of these, the NB and SG distributions reproduce data at most of the energies, while WB and KW do fail at some energies.~The average charged multiplicity $\langle{n}\rangle$ for both $p/\pi-Em$ increases with the c.m.s energy in a $(lns)^{2}$ dependence.~The experimental values of $\langle{n}\rangle$ for $p/\pi-A$ are approximated well with the values calculated from the fitted distributions.~The normalised and factorial moments have been calculated for the first time in these data.~The analysis shows that $C_{q}$ and $F_{q}$ depend upon projectile energy, indicating the KNO scaling violation.~For the NB distribution, the fact that $1/k$, where $k$ is the shape parameter, and hence the factorial moments increasing with energy, indicates the violation of KNO scaling.~In addition the normalised factorial moments $F_{q}$ are all greater than unity, thereby indicating that the particles produced have positive correlations which leads to the interaction dynamics being influenced by these correlations.~Study of $H_{q}$ moments shows that the variation of $H_{q}$ as a function of $q$ is very similar in both $p/\pi-Em$ interactions at different energies. ~All distributions steeply descend to a negative minimum around $q\sim5$.~Quasi oscillations about zero are expected for larger values of $q$.~These observations support the predictions of perturbative QCD [\refcite{DR},~\refcite{Cap}] in the next-to-next-to-leading logarithmic approximation.~The SG distribution, proposed recently provides the description of data from a very low energy to high energy as good as the more established negative binomial distribution.~This also serves as a good test of the validity of this proposed distribution.


\newpage
\begin{table*}[tb]
\scalebox{0.8}{ 
\tbl{Comparison of average charged multiplicity $\langle{n}\rangle$ derived from different distributions with the experimentally measured values.}
{\begin{tabular}{|c|c|c|c|c|c|}
\hline
{\bf Energy (GeV)} & $\mathrm{\langle{n}\rangle_{exp}}$ & $\mathrm{\langle{n}\rangle_{NB}}$ & $\mathrm{\langle{n}\rangle_{SG}}$ & $\mathrm{\langle{n}\rangle_{WB}}$ & $\mathrm{\langle{n}\rangle_{KW}}$ \rbtrr \\ \hline
{\bf p-Em} & & & & & \rbtrr \\ \hline 
27 & 6.23 $\pm$ 0.20 & 	7.01 $\pm$ 0.12 &	7.33 $\pm$ 0.01 &	6.97 $\pm$ 0.11 &	6.98 $\pm$ 0.12 \rbtrr \\ \hline
67 & 9.73 $\pm$ 0.23 & 	10.52 $\pm$ 0.28 &	10.42 $\pm$ 0.03 &	10.30 $\pm$ 0.25 &	10.24 $\pm$ 0.27 \rbtrr \\ \hline
200 & 13.2 $\pm$ 0.2 & 	13.51 $\pm$ 0.36 &	12.91 $\pm$ 0.04 &	13.24 $\pm$ 0.31 &	13.05 $\pm$ 0.33 \rbtrr \\ \hline
300 & 15.1 $\pm$ 0.2 & 	15.35 $\pm$ 0.23 &	14.89 $\pm$ 0.05 &	15.28 $\pm$ 0.21 &	15.24 $\pm$ 0.23 \rbtrr \\ \hline
400 & 16.8 $\pm$ 0.4 & 	16.82 $\pm$ 0.20 &	16.41 $\pm$ 0.05 &	16.58 $\pm$ 0.18 &	16.53 $\pm$ 0.19 \rbtrr \\ \hline
800 & 19.37 $\pm$ 0.37 & 19.75 $\pm$ 0.35 &	19.06 $\pm$ 0.13 &	19.59 $\pm$ 0.32 &	19.50 $\pm$ 0.33 \rbtrr \\ \hline
{\bf p-AU} & & & & & \rbtrr \\ \hline
200 & 21.6 $\pm$ 1.20 &	16.47 $\pm$ 1.18 &	15.51 $\pm$ 0.24 &	16.27 $\pm$ 0.88 &	15.86 $\pm$ 1.00 \rbtrr \\ \hline
{\bf $\pi^{-}$-Em} & & & & & \rbtrr \\ \hline
50 & 8.39 $\pm$ 0.25 & 	8.48 $\pm$ 0.15 &	8.56 $\pm$ 0.01 &	8.43 $\pm$ 0.14 &	8.43 $\pm$ 0.15 \rbtrr \\ \hline
200 & 11.94 $\pm$ 0.34 & 	11.71 $\pm$ 0.24 &	11.72 $\pm$ 0.05 &	11.54 $\pm$ 0.22 &	11.50 $\pm$ 0.23 \rbtrr \\ \hline
340 & 13.34 $\pm$ 0.59 & 	13.18 $\pm$ 0.25 &	12.96 $\pm$ 0.08 &	13.08 $\pm$ 0.24 &	13.08 $\pm$ 0.25 \rbtrr \\ \hline
525 & 15.93 $\pm$ 0.22 &	15.62 $\pm$ 0.22 &	15.58 $\pm$ 0.07 &	15.40 $\pm$ 0.21 &	15.40 $\pm$ 0.21 \rbtrr \\ \hline
{\bf $\pi^{+}$-Ne} & & & & & \rbtrr \\ \hline 
30 & 6.27 $\pm$ 0.06 & 	6.08 $\pm$ 0.06 &	6.27 $\pm$ 0.01 &	6.09 $\pm$ 0.06 &	6.11 $\pm$ 0.06 \rbtrr \\ \hline
64 & 7.72 $\pm$ 0.11 &	7.55 $\pm$ 0.12 &	7.70 $\pm$ 0.02 &	7.51 $\pm$ 0.11 &	7.53 $\pm$ 0.12 \rbtrr \\ \hline
{\bf $\pi^{-}$-Ne} & & & & & \rbtrr \\ \hline
30 & 6.28 $\pm$ 0.05 & 	6.00 $\pm$ 0.05 &	6.26 $\pm$ 0.01 &	6.01 $\pm$ 0.05 &	6.02 $\pm$ 0.05 \rbtrr \\ \hline
64 & 7.93 $\pm$ 0.11 &	7.77 $\pm$ 0.13 &	7.89 $\pm$ 0.02 &	7.70 $\pm$ 0.12 &	7.71 $\pm$ 0.13 \rbtrr \\ \hline
\end{tabular}}
}
\label{tabMult}
\end{table*}

\begin{table}[tb]
\scalebox{0.8}{ 
\tbl{Fit parameters, equation~(\ref{eqMult}) giving dependence of average multiplicity on $P_{Lab}$ for different distributions.}
{\begin{tabular}{|c|c|c|c|c||c|c|c|c|}
\hline
 & \multicolumn{4}{c||}{\bf p-Em} & \multicolumn{4}{c|}{\bf $\pi^{-}$-Em} \rbtrr \rbtrr \\ \hline
Distribution & a & b & c & $\chi^2/n_\mathrm{dof}$ & a & b & c & $\chi^2/n_\mathrm{dof}$ \rbtrr \rbtrr \\\hline
{\bf NB} & -1.2529 & 1.9289 & 0.1786 & 2.08 (6.23/3) &	17.1065 & -5.4421 & 0.8280 & 2.69 (2.69/1) \rbtrr \rbtrr \\ \hline
{\bf SG} &  0.7924 & 1.1187 & 0.2257 & 2.71 (8.13/3) &	16.5492 & -5.7465 & 0.8703 & 18.23 (18.23/1) \rbtrr \rbtrr \\ \hline
{\bf WB} & -0.3299 & 1.5108 & 0.2172 & 2.24 (6.73/3) &	16.8992 & -5.3383 & 0.8116 & 1.99 (1.99/1) \rbtrr \rbtrr \\ \hline
{\bf KW} &  0.1983 & 1.2767 & 0.2396 & 2.26 (6.78/3) &	17.6899 & -5.6651 & 0.8438 & 1.61 (1.61/1) \rbtrr \rbtrr \\ \hline
\end{tabular}}
}
\label{tabMultFit}
\end{table}

\begin{table}[tb]
\scalebox{0.8}{ 
\tbl{Normalised moments $C_{q}$ and normalised factorial moments $F_{q}$ of experimental multiplicity distributions at different energies with different projectiles.}
{\begin{tabular}{|c|c|c|c|c||c|c|c|c|}
\hline
& \multicolumn{4}{c||}{\bf Normalised moments (Exp)} & \multicolumn{4}{c|}{\bf Normalised factorial moments (Exp)}\rbtrr \\\hline
{\bf Energy (GeV)} & $C_{2}$ & $C_{3}$ & $C_{4}$ & $C_{5}$ & $F_{2}$ & $F_{3}$ & $F_{4}$ & $F_{5}$ \rbtrr \\ \hline
{\bf p-Em} & & & & & & & & \rbtrr \\ \hline 
27 & 1.217 $\pm$ 0.150 & 1.874 $\pm$ 0.282 & 3.378 $\pm$ 0.630 & 6.935 $\pm$ 1.581	& 1.087 $\pm$ 0.139 & 1.414 $\pm$ 0.233 & 2.083 $\pm$ 0.452 & 3.413 $\pm$ 0.943 \rbtrr \\ \hline
67 & 1.279 $\pm$ 0.256 & 1.943 $\pm$ 0.414 & 3.350 $\pm$ 0.757 & 6.336 $\pm$ 1.503	& 1.184 $\pm$ 0.240 & 1.597 $\pm$ 0.353 & 2.363 $\pm$ 0.569 & 3.714 $\pm$ 0.960 \rbtrr \\ \hline
200 & 1.328 $\pm$ 0.264 & 2.207 $\pm$ 0.461 & 4.280 $\pm$ 0.922 & 9.383 $\pm$ 2.037	& 1.252 $\pm$ 0.252 & 1.909 $\pm$ 0.410 & 3.335 $\pm$ 0.748 & 6.489 $\pm$ 1.481 \rbtrr \\ \hline
300 & 1.328 $\pm$ 0.188 & 2.159 $\pm$ 0.324 & 4.019 $\pm$ 0.630 & 8.260 $\pm$ 1.332	& 1.262 $\pm$ 0.181 & 1.900 $\pm$ 0.293 & 3.213 $\pm$ 0.525 & 5.883 $\pm$ 1.003 \rbtrr \\ \hline
400 & 1.318 $\pm$ 0.152 & 2.087 $\pm$ 0.255 & 3.759 $\pm$ 0.477 & 7.410 $\pm$ 0.957	& 1.257 $\pm$ 0.146 & 1.853 $\pm$ 0.231 & 3.046 $\pm$ 0.401 & 5.369 $\pm$ 0.727 \rbtrr \\ \hline
800 & 1.307 $\pm$ 0.221 & 2.041 $\pm$ 0.356 & 3.593 $\pm$ 0.634 & 6.854 $\pm$ 1.201	& 1.255 $\pm$ 0.213 & 1.843 $\pm$ 0.327 & 2.994 $\pm$ 0.542 & 5.170 $\pm$ 0.936 \rbtrr \\ \hline
{\bf p-Au} & & & & & & & & \rbtrr \\\hline
200 & 1.485 $\pm$ 0.339 & 2.770 $\pm$ 0.628 & 5.943 $\pm$ 1.285 & 13.958 $\pm$ 2.756	& 1.422 $\pm$ 0.328 & 2.494 $\pm$ 0.579 & 4.951 $\pm$ 1.106 & 10.554 $\pm$ 2.174 \rbtrr \\ \hline
{\bf $\pi^{-}$-Em} & & & & & & & & \rbtrr \\\hline
50 & 1.189 $\pm$ 0.196 & 1.882 $\pm$ 0.360 & 3.433 $\pm$ 0.766 & 7.017 $\pm$ 1.784	& 1.082 $\pm$ 0.183 & 1.484 $\pm$ 0.305 & 2.270 $\pm$ 0.567 & 3.778 $\pm$ 1.106 \rbtrr \\ \hline
200 & 1.305 $\pm$ 0.245 & 2.079 $\pm$ 0.431 & 3.813 $\pm$ 0.866 & 7.759 $\pm$ 1.888	& 1.220 $\pm$ 0.233 & 1.758 $\pm$ 0.380 & 2.846 $\pm$ 0.689 & 4.986 $\pm$ 1.318 \rbtrr \\ \hline
340 & 1.213 $\pm$ 0.215 & 1.832 $\pm$ 0.337 & 3.109 $\pm$ 0.585 & 5.721 $\pm$ 1.077	& 1.141 $\pm$ 0.204 & 1.567 $\pm$ 0.296 & 2.351 $\pm$ 0.459 & 3.709 $\pm$ 0.734 \rbtrr \\ \hline
525 & 1.309 $\pm$ 0.220 & 2.137 $\pm$ 0.423 & 4.172 $\pm$ 0.954 & 9.392 $\pm$ 2.371	& 1.246 $\pm$ 0.213 & 1.898 $\pm$ 0.390 & 3.421 $\pm$ 0.825 & 7.046 $\pm$ 1.888 \rbtrr \\ \hline
{\bf $\pi^{+}$-Ne} & & & & & & & & \rbtrr \\\hline
30 & 1.236 $\pm$ 0.090 & 1.810 $\pm$ 0.159 & 3.042 $\pm$ 0.337 & 5.732 $\pm$ 0.803	& 1.076 $\pm$ 0.082 & 1.269 $\pm$ 0.126 & 1.630 $\pm$ 0.222 & 2.253 $\pm$ 0.420 \rbtrr \\ \hline
64 & 1.249 $\pm$ 0.153 & 1.849 $\pm$ 0.257 & 3.124 $\pm$ 0.500 & 5.856 $\pm$ 1.068	& 1.119 $\pm$ 0.141 & 1.396 $\pm$ 0.210 & 1.901 $\pm$ 0.346 & 2.758 $\pm$ 0.599 \rbtrr \\ \hline
{\bf $\pi^{-}$-Ne} & & & & & & & & \rbtrr \\\hline
30 & 1.254 $\pm$ 0.076 & 1.889 $\pm$ 0.139 & 3.282 $\pm$ 0.301 & 6.389 $\pm$ 0.728	& 1.094 $\pm$ 0.070 & 1.339 $\pm$ 0.111 & 1.800 $\pm$ 0.202 & 2.596 $\pm$ 0.388 \rbtrr \\ \hline
64 & 1.246 $\pm$ 0.155 & 1.826 $\pm$ 0.249 & 3.029 $\pm$ 0.456 & 5.525 $\pm$ 0.910	& 1.119 $\pm$ 0.142 & 1.385 $\pm$ 0.202 & 1.849 $\pm$ 0.310 & 2.594 $\pm$ 0.495 \rbtrr \\ \hline
\end{tabular}}
}
\label{tabMomExp}
\end{table}

\begin{table}[tb]
\scalebox{0.8}{ 
\tbl{Normalised moments $C_{q}$ and normalised factorial moments $F_{q}$ of NB fit distribution for data at different energies with different projectiles.}
{\begin{tabular}{|c|c|c|c|c||c|c|c|c|} 
\hline
& \multicolumn{4}{c||}{\bf Normalised moments (NB)} & \multicolumn{4}{c|}{\bf Normalised factorial moments (NB)}\rbtrr \\\hline
{\bf Energy (GeV)} & C2 & C3 & C4 & C5 & F2 & F3 & F4 & F5 \rbtrr \\ \hline
{\bf p-Em} & & & & & & & & \rbtrr \\ \hline 
27 & 1.190 $\pm$ 0.031 & 1.788 $\pm$ 0.020 & 3.082 $\pm$ 0.023 & 5.923 $\pm$ 0.168	& 1.061 $\pm$ 0.028 & 1.335 $\pm$ 0.013 & 1.843 $\pm$ 0.022 & 2.731 $\pm$ 0.099 \rbtrr \\ \hline
67 & 1.220 $\pm$ 0.042 & 1.840 $\pm$ 0.018 & 3.134 $\pm$ 0.058 & 5.844 $\pm$ 0.285	& 1.127 $\pm$ 0.039 & 1.505 $\pm$ 0.015 & 2.191 $\pm$ 0.044 & 3.371 $\pm$ 0.177 \rbtrr \\ \hline
200 & 1.237 $\pm$ 0.026 & 1.972 $\pm$ 0.008 & 3.587 $\pm$ 0.119 & 7.219 $\pm$ 0.467	& 1.165 $\pm$ 0.025 & 1.697 $\pm$ 0.006 & 2.753 $\pm$ 0.093 & 4.834 $\pm$ 0.322 \rbtrr \\ \hline
300 & 1.285 $\pm$ 0.019 & 2.058 $\pm$ 0.004 & 3.766 $\pm$ 0.049 & 7.609 $\pm$ 0.221	& 1.220 $\pm$ 0.018 & 1.808 $\pm$ 0.004 & 3.001 $\pm$ 0.039 & 5.392 $\pm$ 0.158 \rbtrr \\ \hline
400 & 1.257 $\pm$ 0.014 & 1.934 $\pm$ 0.003 & 3.371 $\pm$ 0.035 & 6.431 $\pm$ 0.148	& 1.198 $\pm$ 0.014 & 1.712 $\pm$ 0.003 & 2.715 $\pm$ 0.028 & 4.619 $\pm$ 0.107 \rbtrr \\ \hline
800 & 1.221 $\pm$ 0.020 & 1.848 $\pm$ 0.006 & 3.148 $\pm$ 0.038 & 5.840 $\pm$ 0.169	& 1.171 $\pm$ 0.020 & 1.662 $\pm$ 0.007 & 2.605 $\pm$ 0.030 & 4.365 $\pm$ 0.125 \rbtrr \\ \hline
{\bf p-Au} & & & & & & & & \rbtrr \\\hline
200 & 1.430 $\pm$ 0.049 & 2.628 $\pm$ 0.073 & 5.541 $\pm$ 0.542 & 12.803 $\pm$ 2.198	& 1.368 $\pm$ 0.049 & 2.359 $\pm$ 0.063 & 4.589 $\pm$ 0.446 & 9.595 $\pm$ 1.647 \rbtrr \\ \hline
{\bf $\pi^{-}$-Em} & & & & & & & & \rbtrr \\\hline
50 & 1.112 $\pm$ 0.022 & 1.692 $\pm$ 0.004 & 2.922 $\pm$ 0.082 & 5.562 $\pm$ 0.313	& 1.009 $\pm$ 0.020 & 1.324 $\pm$ 0.004 & 1.888 $\pm$ 0.061 & 2.856 $\pm$ 0.182 \rbtrr \\ \hline
200 & 1.255 $\pm$ 0.026 & 1.988 $\pm$ 0.002 & 3.626 $\pm$ 0.096 & 7.369 $\pm$ 0.393	& 1.173 $\pm$ 0.025 & 1.679 $\pm$ 0.002 & 2.702 $\pm$ 0.076 & 4.733 $\pm$ 0.267 \rbtrr \\ \hline
340 & 1.094 $\pm$ 0.026 & 1.599 $\pm$ 0.011 & 2.612 $\pm$ 0.035 & 4.633 $\pm$ 0.165	& 1.027 $\pm$ 0.025 & 1.356 $\pm$ 0.010 & 1.943 $\pm$ 0.027 & 2.932 $\pm$ 0.107 \rbtrr \\ \hline
525 & 1.225 $\pm$ 0.020 & 1.853 $\pm$ 0.006 & 3.212 $\pm$ 0.042 & 6.207 $\pm$ 0.193	& 1.163 $\pm$ 0.019 & 1.624 $\pm$ 0.005 & 2.551 $\pm$ 0.035 & 4.392 $\pm$ 0.142 \rbtrr \\ \hline
{\bf $\pi^{+}$-Ne} & & & & & & & & \rbtrr \\\hline 
30 & 1.185 $\pm$ 0.021 & 1.705 $\pm$ 0.016 & 2.790 $\pm$ 0.005 & 5.075 $\pm$ 0.075	& 1.032 $\pm$ 0.019 & 1.190 $\pm$ 0.010 & 1.473 $\pm$ 0.008 & 1.930 $\pm$ 0.042 \rbtrr \\ \hline
64 & 1.204 $\pm$ 0.030 & 1.749 $\pm$ 0.020 & 2.880 $\pm$ 0.016 & 5.238 $\pm$ 0.128	& 1.077 $\pm$ 0.028 & 1.312 $\pm$ 0.014 & 1.727 $\pm$ 0.015 & 2.402 $\pm$ 0.073 \rbtrr \\ \hline
{\bf $\pi^{-}$-Ne} & & & & & & & & \rbtrr \\\hline
30 & 1.234 $\pm$ 0.019 & 1.844 $\pm$ 0.015 & 3.174 $\pm$ 0.005 & 6.120 $\pm$ 0.075	& 1.078 $\pm$ 0.017 & 1.306 $\pm$ 0.009 & 1.736 $\pm$ 0.008 & 2.476 $\pm$ 0.044 \rbtrr \\ \hline
64 & 1.223 $\pm$ 0.033 & 1.789 $\pm$ 0.022 & 2.954 $\pm$ 0.014 & 5.353 $\pm$ 0.126	& 1.098 $\pm$ 0.030 & 1.355 $\pm$ 0.016 & 1.797 $\pm$ 0.013 & 2.493 $\pm$ 0.070 \rbtrr \\ \hline
\end{tabular}}
}
\label{tabMomNB}
\end{table}

\begin{table}[tb]
\scalebox{0.8}{ 
\tbl{Normalised moments $C_{q}$ and normalised factorial moments $F_{q}$ of SG distribution for data at different energies with different projectiles.}
{\begin{tabular}{|c|c|c|c|c||c|c|c|c|}
\hline
& \multicolumn{4}{c||}{\bf Normalised moments (SG)} & \multicolumn{4}{c|}{\bf Normalised factorial moments (SG)}\rbtrr \\\hline
{\bf Energy (GeV)} & $C_{2}$ & $C_{3}$ & $C_{4}$ & $C_{5}$ & $F_{2}$ & $F_{3}$ & $F_{4}$ & $F_{5}$ \rbtrr \\ \hline
{\bf p-Em} & & & & & & & & \rbtrr \\ \hline 
27 & 1.194 $\pm$ 0.031 & 1.830 $\pm$ 0.027 & 3.262 $\pm$ 0.009 & 6.558 $\pm$ 0.056	& 1.066 $\pm$ 0.027 & 1.377 $\pm$ 0.015 & 1.995 $\pm$ 0.009 & 3.157 $\pm$ 0.058 \rbtrr \\ \hline
67 & 1.225 $\pm$ 0.051 & 1.863 $\pm$ 0.052 & 3.209 $\pm$ 0.048 & 6.062 $\pm$ 0.022	& 1.133 $\pm$ 0.046 & 1.528 $\pm$ 0.038 & 2.255 $\pm$ 0.021 & 3.531 $\pm$ 0.015 \rbtrr \\ \hline
200 & 1.243 $\pm$ 0.037 & 1.984 $\pm$ 0.032 & 3.629 $\pm$ 0.007 & 7.374 $\pm$ 0.083	& 1.170 $\pm$ 0.034 & 1.707 $\pm$ 0.023 & 2.786 $\pm$ 0.007 & 4.952 $\pm$ 0.090 \rbtrr \\ \hline
300 & 1.281 $\pm$ 0.025 & 2.049 $\pm$ 0.027 & 3.760 $\pm$ 0.030 & 7.638 $\pm$ 0.027	& 1.216 $\pm$ 0.023 & 1.800 $\pm$ 0.022 & 2.996 $\pm$ 0.017 & 5.418 $\pm$ 0.004 \rbtrr \\ \hline
400 & 1.259 $\pm$ 0.019 & 1.941 $\pm$ 0.020 & 3.399 $\pm$ 0.022 & 6.531 $\pm$ 0.025	& 1.200 $\pm$ 0.018 & 1.718 $\pm$ 0.016 & 2.739 $\pm$ 0.014 & 4.698 $\pm$ 0.009 \rbtrr \\ \hline
800 & 1.212 $\pm$ 0.029 & 1.828 $\pm$ 0.035 & 3.111 $\pm$ 0.052 & 5.784 $\pm$ 0.089	& 1.163 $\pm$ 0.027 & 1.643 $\pm$ 0.030 & 2.572 $\pm$ 0.038 & 4.319 $\pm$ 0.057 \rbtrr \\ \hline
{\bf p-Au} & & & & & & & & \rbtrr \\\hline
200 & 1.420 $\pm$ 0.091 & 2.591 $\pm$ 0.103 & 5.427 $\pm$ 0.117 & 12.467 $\pm$ 0.131	& 1.358 $\pm$ 0.084 & 2.325 $\pm$ 0.082 & 4.487 $\pm$ 0.068 & 9.326 $\pm$ 0.026 \rbtrr \\ \hline
{\bf $\pi^{-}$-Em} & & & & & & & & \rbtrr \\\hline
50 & 1.115 $\pm$ 0.025 & 1.749 $\pm$ 0.012 & 3.153 $\pm$ 0.025 & 6.307 $\pm$ 0.138	& 1.014 $\pm$ 0.021 & 1.382 $\pm$ 0.005 & 2.086 $\pm$ 0.030 & 3.373 $\pm$ 0.102 \rbtrr \\ \hline
200 & 1.270 $\pm$ 0.032 & 2.042 $\pm$ 0.025 & 3.808 $\pm$ 0.001 & 7.931 $\pm$ 0.077	& 1.188 $\pm$ 0.029 & 1.731 $\pm$ 0.016 & 2.861 $\pm$ 0.013 & 5.164 $\pm$ 0.082 \rbtrr \\ \hline
340 & 1.102 $\pm$ 0.033 & 1.622 $\pm$ 0.037 & 2.685 $\pm$ 0.044 & 4.842 $\pm$ 0.057	& 1.034 $\pm$ 0.030 & 1.377 $\pm$ 0.028 & 2.004 $\pm$ 0.025 & 3.084 $\pm$ 0.022 \rbtrr \\ \hline
525 & 1.240 $\pm$ 0.024 & 1.908 $\pm$ 0.023 & 3.403 $\pm$ 0.014 & 6.823 $\pm$ 0.016	& 1.178 $\pm$ 0.022 & 1.677 $\pm$ 0.017 & 2.723 $\pm$ 0.004 & 4.898 $\pm$ 0.030 \rbtrr \\ \hline
{\bf $\pi^{+}$-Ne} & & & & & & & & \rbtrr \\\hline 
30 & 1.190 $\pm$ 0.021 & 1.775 $\pm$ 0.020 & 3.079 $\pm$ 0.014 & 6.037 $\pm$ 0.011	& 1.037 $\pm$ 0.017 & 1.254 $\pm$ 0.011 & 1.690 $\pm$ 0.001 & 2.478 $\pm$ 0.022 \rbtrr \\ \hline
64 & 1.213 $\pm$ 0.032 & 1.805 $\pm$ 0.032 & 3.084 $\pm$ 0.027 & 5.863 $\pm$ 0.005	& 1.086 $\pm$ 0.028 & 1.365 $\pm$ 0.020 & 1.889 $\pm$ 0.005 & 2.797 $\pm$ 0.020 \rbtrr \\ \hline
{\bf $\pi^{-}$-Ne} & & & & & & & & \rbtrr \\\hline
30 & 1.243 $\pm$ 0.018 & 1.909 $\pm$ 0.018 & 3.438 $\pm$ 0.012 & 7.027 $\pm$ 0.012	& 1.086 $\pm$ 0.015 & 1.364 $\pm$ 0.009 & 1.933 $\pm$ 0.002 & 3.004 $\pm$ 0.023 \rbtrr \\ \hline
64 & 1.229 $\pm$ 0.036 & 1.829 $\pm$ 0.038 & 3.096 $\pm$ 0.039 & 5.775 $\pm$ 0.034	& 1.105 $\pm$ 0.031 & 1.393 $\pm$ 0.025 & 1.910 $\pm$ 0.014 & 2.759 $\pm$ 0.004 \rbtrr \\ \hline
\end{tabular}}
}
\label{tabMomSG}
\end{table}

\begin{table}[tb]
\scalebox{0.8}{ 
\tbl{Normalised moments $C_{q}$ and normalised factorial moments $F_{q}$ of WB distribution for data at different energies with different projectiles.}
{\begin{tabular}{|c|c|c|c|c||c|c|c|c|}
\hline
& \multicolumn{4}{c||}{\bf Normalised moments (WB)} & \multicolumn{4}{c|}{\bf Normalised factorial moments (WB)}\rbtrr \\\hline
{\bf Energy (GeV)} & $C_{2}$ & $C_{3}$ & $C_{4}$ & $C_{5}$ & $F_{2}$ & $F_{3}$ & $F_{4}$ & $F_{5}$ \rbtrr \\ \hline
{\bf p-Em} & & & & & & & & \rbtrr \\ \hline 
27 & 1.185 $\pm$ 0.030 & 1.757 $\pm$ 0.020 & 2.954 $\pm$ 0.017 & 5.478 $\pm$ 0.139	& 1.055 $\pm$ 0.027 & 1.304 $\pm$ 0.014 & 1.735 $\pm$ 0.016 & 2.432 $\pm$ 0.081 \rbtrr \\ \hline
67 & 1.214 $\pm$ 0.042 & 1.818 $\pm$ 0.023 & 3.053 $\pm$ 0.040 & 5.583 $\pm$ 0.233	& 1.122 $\pm$ 0.039 & 1.483 $\pm$ 0.019 & 2.117 $\pm$ 0.032 & 3.170 $\pm$ 0.146 \rbtrr \\ \hline
200 & 1.232 $\pm$ 0.027 & 1.949 $\pm$ 0.002 & 3.481 $\pm$ 0.100 & 6.811 $\pm$ 0.404	& 1.160 $\pm$ 0.026 & 1.673 $\pm$ 0.002 & 2.652 $\pm$ 0.079 & 4.492 $\pm$ 0.279 \rbtrr \\ \hline
300 & 1.270 $\pm$ 0.020 & 2.003 $\pm$ 0.008 & 3.574 $\pm$ 0.034 & 6.986 $\pm$ 0.172	& 1.206 $\pm$ 0.019 & 1.759 $\pm$ 0.008 & 2.835 $\pm$ 0.027 & 4.901 $\pm$ 0.122 \rbtrr \\ \hline
400 & 1.237 $\pm$ 0.015 & 1.877 $\pm$ 0.006 & 3.196 $\pm$ 0.023 & 5.931 $\pm$ 0.113	& 1.179 $\pm$ 0.015 & 1.658 $\pm$ 0.006 & 2.560 $\pm$ 0.019 & 4.216 $\pm$ 0.081 \rbtrr \\ \hline
800 & 1.185 $\pm$ 0.022 & 1.759 $\pm$ 0.013 & 2.914 $\pm$ 0.017 & 5.230 $\pm$ 0.108	& 1.137 $\pm$ 0.021 & 1.581 $\pm$ 0.013 & 2.402 $\pm$ 0.013 & 3.880 $\pm$ 0.078 \rbtrr \\ \hline
{\bf p-Au} & & & & & & & & \rbtrr \\\hline
200 & 1.420 $\pm$ 0.055 & 2.586 $\pm$ 0.045 & 5.390 $\pm$ 0.435 & 12.302 $\pm$ 1.809	& 1.358 $\pm$ 0.054 & 2.320 $\pm$ 0.039 & 4.454 $\pm$ 0.361 & 9.185 $\pm$ 1.364 \rbtrr \\ \hline
{\bf $\pi^{-}$-Em} & & & & & & & & \rbtrr \\\hline
50 & 1.107 $\pm$ 0.023 & 1.663 $\pm$ 0.002 & 2.797 $\pm$ 0.063 & 5.127 $\pm$ 0.252	& 1.005 $\pm$ 0.022 & 1.294 $\pm$ 0.001 & 1.776 $\pm$ 0.046 & 2.542 $\pm$ 0.144 \rbtrr \\ \hline
200 & 1.239 $\pm$ 0.027 & 1.925 $\pm$ 0.004 & 3.401 $\pm$ 0.073 & 6.637 $\pm$ 0.317	& 1.156 $\pm$ 0.026 & 1.617 $\pm$ 0.003 & 2.499 $\pm$ 0.058 & 4.156 $\pm$ 0.214 \rbtrr \\ \hline
340 & 1.074 $\pm$ 0.028 & 1.554 $\pm$ 0.017 & 2.492 $\pm$ 0.018 & 4.313 $\pm$ 0.117	& 1.008 $\pm$ 0.027 & 1.317 $\pm$ 0.015 & 1.845 $\pm$ 0.013 & 2.700 $\pm$ 0.075 \rbtrr \\ \hline
525 & 1.183 $\pm$ 0.021 & 1.727 $\pm$ 0.010 & 2.826 $\pm$ 0.024 & 5.062 $\pm$ 0.127	& 1.122 $\pm$ 0.020 & 1.504 $\pm$ 0.009 & 2.206 $\pm$ 0.020 & 3.465 $\pm$ 0.091 \rbtrr \\ \hline
{\bf $\pi^{+}$-Ne} & & & & & & & & \rbtrr \\\hline 
30 & 1.177 $\pm$ 0.021 & 1.671 $\pm$ 0.017 & 2.665 $\pm$ 0.001 & 4.667 $\pm$ 0.059	& 1.024 $\pm$ 0.019 & 1.159 $\pm$ 0.011 & 1.378 $\pm$ 0.005 & 1.691 $\pm$ 0.033 \rbtrr \\ \hline
64 & 1.192 $\pm$ 0.030 & 1.708 $\pm$ 0.023 & 2.744 $\pm$ 0.007 & 4.822 $\pm$ 0.098	& 1.066 $\pm$ 0.028 & 1.275 $\pm$ 0.017 & 1.618 $\pm$ 0.008 & 2.137 $\pm$ 0.056 \rbtrr \\ \hline
{\bf $\pi^{-}$-Ne} & & & & & & & & \rbtrr \\\hline
30 & 1.224 $\pm$ 0.018 & 1.806 $\pm$ 0.014 & 3.032 $\pm$ 0.004 & 5.637 $\pm$ 0.067	& 1.069 $\pm$ 0.016 & 1.273 $\pm$ 0.009 & 1.627 $\pm$ 0.007 & 2.187 $\pm$ 0.039 \rbtrr \\ \hline
64 & 1.214 $\pm$ 0.032 & 1.754 $\pm$ 0.023 & 2.840 $\pm$ 0.008 & 5.013 $\pm$ 0.105	& 1.089 $\pm$ 0.029 & 1.323 $\pm$ 0.017 & 1.705 $\pm$ 0.009 & 2.275 $\pm$ 0.059 \rbtrr \\ \hline
\end{tabular}}
}
\label{tabMomWB}
\end{table}

\begin{table}[tb]
\scalebox{0.8}{ 
\tbl{Normalised moments $C_{q}$ and normalised factorial moments $F_{q}$ of KW distribution for data at different energies with different projectiles.}
{\begin{tabular}{|c|c|c|c|c||c|c|c|c|}
\hline
& \multicolumn{4}{c||}{\bf Normalised moments (KW)} & \multicolumn{4}{c|}{\bf Normalised factorial moments (KW)}\rbtrr \\\hline
{\bf Energy (GeV)} & $C_{2}$ & $C_{3}$ & $C_{4}$ & $C_{5}$ & $F_{2}$ & $F_{3}$ & $F_{4}$ & $F_{5}$ \rbtrr \\ \hline
{\bf p-Em} & & & & & & & & \rbtrr \\ \hline 
27 & 1.176 $\pm$ 0.032 & 1.728 $\pm$ 0.026 & 2.866 $\pm$ 0.001 & 5.232 $\pm$ 0.082	& 1.047 $\pm$ 0.029 & 1.278 $\pm$ 0.019 & 1.671 $\pm$ 0.003 & 2.289 $\pm$ 0.046 \rbtrr \\ \hline
67 & 1.209 $\pm$ 0.044 & 1.799 $\pm$ 0.030 & 2.995 $\pm$ 0.018 & 5.425 $\pm$ 0.166	& 1.117 $\pm$ 0.041 & 1.465 $\pm$ 0.025 & 2.069 $\pm$ 0.015 & 3.060 $\pm$ 0.101 \rbtrr \\ \hline
200 & 1.216 $\pm$ 0.027 & 1.891 $\pm$ 0.003 & 3.301 $\pm$ 0.074 & 6.283 $\pm$ 0.303	& 1.144 $\pm$ 0.026 & 1.618 $\pm$ 0.003 & 2.496 $\pm$ 0.056 & 4.085 $\pm$ 0.200 \rbtrr \\ \hline
300 & 1.252 $\pm$ 0.020 & 1.943 $\pm$ 0.011 & 3.394 $\pm$ 0.021 & 6.476 $\pm$ 0.123	& 1.188 $\pm$ 0.019 & 1.702 $\pm$ 0.011 & 2.680 $\pm$ 0.015 & 4.507 $\pm$ 0.083 \rbtrr \\ \hline
400 & 1.224 $\pm$ 0.015 & 1.836 $\pm$ 0.008 & 3.084 $\pm$ 0.016 & 5.643 $\pm$ 0.087	& 1.166 $\pm$ 0.015 & 1.619 $\pm$ 0.008 & 2.462 $\pm$ 0.012 & 3.991 $\pm$ 0.060 \rbtrr \\ \hline
800 & 1.175 $\pm$ 0.022 & 1.727 $\pm$ 0.015 & 2.830 $\pm$ 0.009 & 5.025 $\pm$ 0.082	& 1.127 $\pm$ 0.021 & 1.550 $\pm$ 0.015 & 2.327 $\pm$ 0.005 & 3.714 $\pm$ 0.056 \rbtrr \\ \hline
{\bf p-Au} & & & & & & & & \rbtrr \\\hline
200 & 1.386 $\pm$ 0.064 & 2.467 $\pm$ 0.008 & 4.994 $\pm$ 0.291 & 11.060 $\pm$ 1.286	& 1.325 $\pm$ 0.062 & 2.207 $\pm$ 0.005 & 4.102 $\pm$ 0.235 & 8.181 $\pm$ 0.948 \rbtrr \\ \hline
{\bf $\pi^{-}$-Em} & & & & & & & & \rbtrr \\\hline
50 & 1.109 $\pm$ 0.023 & 1.670 $\pm$ 0.003 & 2.821 $\pm$ 0.056 & 5.201 $\pm$ 0.225	& 1.006 $\pm$ 0.022 & 1.300 $\pm$ 0.002 & 1.796 $\pm$ 0.039 & 2.590 $\pm$ 0.123 \rbtrr \\ \hline
200 & 1.229 $\pm$ 0.028 & 1.890 $\pm$ 0.011 & 3.292 $\pm$ 0.046 & 6.318 $\pm$ 0.224	& 1.147 $\pm$ 0.027 & 1.584 $\pm$ 0.010 & 2.405 $\pm$ 0.034 & 3.918 $\pm$ 0.143 \rbtrr \\ \hline
340 & 1.073 $\pm$ 0.028 & 1.543 $\pm$ 0.018 & 2.462 $\pm$ 0.012 & 4.243 $\pm$ 0.098	& 1.007 $\pm$ 0.027 & 1.307 $\pm$ 0.016 & 1.820 $\pm$ 0.008 & 2.650 $\pm$ 0.061 \rbtrr \\ \hline
525 & 1.181 $\pm$ 0.022 & 1.713 $\pm$ 0.014 & 2.786 $\pm$ 0.010 & 4.964 $\pm$ 0.084	& 1.119 $\pm$ 0.021 & 1.490 $\pm$ 0.013 & 2.171 $\pm$ 0.007 & 3.391 $\pm$ 0.056 \rbtrr \\ \hline
{\bf $\pi^{+}$-Ne} & & & & & & & & \rbtrr \\\hline 
30 & 1.176 $\pm$ 0.022 & 1.666 $\pm$ 0.020 & 2.652 $\pm$ 0.009 & 4.638 $\pm$ 0.031	& 1.023 $\pm$ 0.020 & 1.155 $\pm$ 0.013 & 1.370 $\pm$ 0.002 & 1.680 $\pm$ 0.018 \rbtrr \\ \hline
64 & 1.190 $\pm$ 0.032 & 1.698 $\pm$ 0.026 & 2.716 $\pm$ 0.006 & 4.756 $\pm$ 0.060	& 1.063 $\pm$ 0.029 & 1.266 $\pm$ 0.020 & 1.599 $\pm$ 0.002 & 2.102 $\pm$ 0.033 \rbtrr \\ \hline
{\bf $\pi^{-}$-Ne} & & & & & & & & \rbtrr \\\hline
30 & 1.217 $\pm$ 0.020 & 1.780 $\pm$ 0.018 & 2.950 $\pm$ 0.008 & 5.401 $\pm$ 0.032	& 1.062 $\pm$ 0.018 & 1.250 $\pm$ 0.012 & 1.569 $\pm$ 0.001 & 2.061 $\pm$ 0.021 \rbtrr \\ \hline
64 & 1.211 $\pm$ 0.033 & 1.742 $\pm$ 0.028 & 2.805 $\pm$ 0.006 & 4.924 $\pm$ 0.065	& 1.086 $\pm$ 0.031 & 1.312 $\pm$ 0.021 & 1.679 $\pm$ 0.002 & 2.225 $\pm$ 0.035 \rbtrr \\ \hline
\end{tabular}}
}
\label{tabMomKW}
\end{table}
\end{document}